\documentclass[aps,prl,reprint,superscriptaddress,floatfix,nobalancelastpage]{revtex4-2}
\usepackage[utf8]{inputenc}
\usepackage{amsthm}
\usepackage{amsmath}
\usepackage{amssymb}
\usepackage{physics}
\usepackage{color}
\usepackage[english]{babel}
\usepackage{graphicx,xcolor}
\usepackage{overpic}
\usepackage{wrapfig}
\usepackage{url}
\usepackage{hyperref}

\begin{document}

\title{Dissipative Time Crystals as Passively Protected Oscillating Qubits}

\author{Mert Esencan}
\affiliation{Clarendon Laboratory, University of Oxford, Parks Road, OX1 3PU, Oxford, United Kingdom}

\author{A. I. Lvovsky}
\affiliation{Clarendon Laboratory, University of Oxford, Parks Road, OX1 3PU, Oxford, United Kingdom}

\author{\fbox{Berislav Bu\v ca}}
\affiliation{Universit\' e Paris-Saclay, CNRS, LPTMS, 91405, Orsay, France}
\affiliation{Niels Bohr Institute, University of Copenhagen, Blegdamsvej 17, 2100, Copenhagen, Denmark}
\affiliation{Clarendon Laboratory, University of Oxford, Parks Road, OX1 3PU, Oxford, United Kingdom}

\begin{abstract}
Passive error correction schemes have so far been limited to static memories rather than dynamical qubits. We demonstrate that a driven-dissipative bosonic system, the Bose--Hubbard dimer (BHD), can encode a persistently oscillating qubit within a dynamical noiseless subsystem (DNS). The strong parity symmetry of the system leads to degenerate stationary states, whose mutual coherence breaks the time translation symmetry, giving rise to a time crystal in the thermodynamic limit. This time crystal undergoes a dissipative phase transition controlled by the driving force, with the DNS present above the transition point. This noiseless subsystem, and hence the qubit, is resilient to collective loss and at most quadratically affected when the strong symmetry condition is weakened. The decoherence from collective dephasing also cancels in the leading order, with the cancellation being exact at a specific drive strength, set by the photon-number difference between the two logical states. Tuning the drive extends this cancellation to unbalanced dephasing. These results establish dissipative time crystals as a medium for autonomously stabilized oscillating qubits.
\end{abstract}

\maketitle

\paragraph{Introduction.---}\!\!\!\!\!\!\! Protecting quantum information from decay in open systems remains a central challenge for practical quantum computers \cite{shor_scheme_1995, unruh_maintaining_1995, chuang_quantum_1995, ekert_dfs, knill_resilient_1998, viola_dynamical_1999, knill_theory_2000, zurek_decoherence_2003, suter_colloquium_2016, egan_fault-tolerant_2021, ding_universally_2025, nielsen_chuang}. \emph{ Active} error correction \cite{shor_scheme_1995, steane_error_1996, gottesman_theory_1998, terhal_quantum_2015}, exemplified by surface codes \cite{dennis_topological_2002, fowler_surface_2012, krinner_realizing_2022} and demonstrated experimentally \cite{barends_superconducting_2014, kelly_state_2015, ofek_extending_2016, chen_exponential_2021, sivak_real-time_2023}, diagnoses and corrects errors through redundant encoding with ancillas, at the price of substantial qubit overhead.

An alternative is \emph{passive} error correction, where dissipative dynamics correct errors without active feedback \cite{verstraete_quantum_2008}, as explored in open many-body systems \cite{simon, simon_new, liu_dissipative_2024, suri_uniformly_2024} and realized in cat qubits \cite{lvovsky_cat, leghtas_confining_2015, lescanne_exponential_2020, grimm_stabilization_2020, alice_bob_cat, amazon_cat}. However, existing passive schemes protect static memories, not coherent evolution. Here we investigate \emph{time crystals}, systems that spontaneously break time-translation symmetry and oscillate persistently
\cite{wilczek_quantum_2012,khemani_phase_2016, heugel_role_2023, li_quantum_2023, zhao_quantum_2025}, as a medium whose oscillation enables passively protected coherent evolution. We focus on the Bose-Hubbard dimer (BHD), a nonlinearly coupled, driven pair of harmonic oscillators which, in the presence of dissipation, is known to enable a continuous time crystal state \cite{pudlik_dynamics_2013,cristobal_strong, cristobal, liang_statistical_2024, pi_dynamics_2024}. We demonstrate that this time crystal hosts a noiseless subsystem: a qubit that can be separated from the remainder of the dimer's Hilbert space. The two degrees of freedom of the qubit can be arbitrarily controlled without affecting the integrity of the time crystal.
We show that this qubit is robust with respect to linear loss and dephasing, provided that these errors affect both modes in a collective and balanced manner. This requirement is soft: the detrimental effect on the qubit scales quadratically with respect to non-collective or unbalanced loss.

To our knowledge, this is the first description of a dynamical noiseless subsystem (DNS) as a dissipative phase of matter. In other words, a remarkable paradox is manifest: dissipation being not a nuisance that destroys the qubit state, but a resource that stabilizes it.  
  
\begin{figure*}[!t]
\centering
\includegraphics[width=1\linewidth]{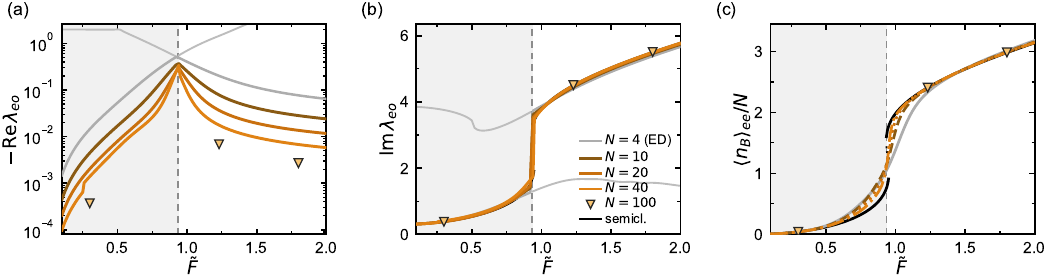}
\caption{(a) Dissipative gap $-\mathrm{Re}\,\lambda_{eo}$ (log scale), (b) frequency $\mathrm{Im}\,\lambda_{eo}$ corresponding to the eigenvalue $\lambda_{eo}$, (c) expected photon number $\langle n_B\rangle/N$ in eigenstate $r_{ee}$, vs drive $\tilde F$, calculated numerically for different scaling parameters $N$, as well as the semiclassical solution. A phase transition is visible at $\tilde F\approx 0.9325$. For $N=4$, exact diagonalization (ED) yields the full eigenspectrum, and the eigenvalue plot corresponding to the slowest-decaying state for each phase is extended beyond the phase transition point, exhibiting a level crossing in (a), see SM for details. For $N=100$, select points are shown due to computational limitations.}
\label{fig:fig1}
\end{figure*}

\paragraph{The model and symmetries.---}\!\!\!\!\!\!\! The driven-dissipative BHD consists of two identically driven, coupled Kerr modes \(a_1\) and \(a_2\) subject to loss. We define the \emph{bonding} and \emph{antibonding} modes
    \(a_B=(a_1+a_2)/\sqrt{2}\) and \(a_A=(a_1-a_2)/\sqrt{2}\) as swapping $a_1$ and $a_2$ leaves $a_B$ unchanged while negating $a_A$. In this basis, the Hamiltonian is
\begin{equation}\label{eq:bhd}
\begin{aligned}
 {H} &= \sqrt{2}F({a}^\dagger_B + {a}_B) + (-\Delta - J) {a}^\dagger_B{a}_B + (-\Delta + J) {a}^\dagger_A{a}_A \\
 & +\frac{U}{2}({a}^\dagger_B {a}^\dagger_B {a}_B {a}_B + {a}^\dagger_A {a}^\dagger_A {a}_A {a}_A +
 {a}^\dagger_B {a}^\dagger_B {a}_A {a}_A \\
 & + {a}_B {a}_B {a}^\dagger_A {a}^\dagger_A + 4{a}^\dagger_B a_B \hat{a}^\dagger_A {a}_A
).
\end{aligned}
\end{equation}
$F$ is the drive amplitude, $\Delta$ the detuning of the drive from the mode resonance, $J$ the intermode coupling, and $U$ the Kerr nonlinearity.

To account for dissipative dynamics, we model the open system with the Lindblad master equation assuming Markovian conditions \cite{open_quantum, dutta_introduction_2025, yoshida_theory_2026}:
\begin{equation}\label{eq:liouvillian}
 \hat{\mathcal{L}}{\rho}(t) = \frac{\partial{\rho}}{\partial t} = -i[{H}, {\rho}] + \gamma\hat{\mathcal{D}}[L]{\rho}.
\end{equation}
The dissipator with strength $\gamma$ is $\hat{\mathcal{D}}[L]\rho = L {\rho} L^\dagger - \frac{1}{2} \{L^\dagger L,{\rho}\}$ with the jump operator $L=a_1+a_2=\sqrt2\,a_B$: only the symmetric combination couples to the reservoir, while \(a_A\) is dark to this loss channel.

     In the standard doubled-Hilbert-space (Fock-Liouville) notation for open systems, an operator \(A\) is a vector \(|A\rangle\rangle\) on which the Liouvillian acts as a matrix, with inner product \(\langle\!\langle A|B\rangle\!\rangle=\mathrm{Tr}(A^\dagger B)\). The density operator evolves as
\begin{equation}
|\rho(t)\rangle\rangle = e^{\hat{\mathcal{L}}t}|\rho(0)\rangle\rangle= \sum^D_{j\geq{0}}e^{\lambda_jt}|r_j\rangle\rangle \langle\!\langle l_j|\rho(0)\rangle\rangle,
\end{equation}
where \(\lambda_j\) are the eigenvalues of the Liouvillian \(\hat{\mathcal L}\), with right and left eigenoperators \(|r_j\rangle\rangle\) and \(\langle\!\langle l_j|\), respectively \cite{open_quantum, albert_lindbladians_2018,lindblad_review, stefanini_is_2025}.  
    Generally, barring special symmetries, an open system will decay into a unique steady eigenstate $|r_0\rangle\rangle \equiv |\rho(\infty)\rangle\rangle$ corresponding to $\lambda_0 = 0$ while all other $\lambda_j$ have negative real parts \cite{open_quantum}. The BHD, however, does host degenerate steady states due to a $\mathbb{Z}_2$ strong parity symmetry for the antibonding mode \cite{beri_symmetry}: its parity operator $P=\sum_{n_B,n_A \geq 0}(-1)^{n_A}\ket{n_B, n_A}\bra{n_B, n_A}$ (where $n_{B,A}=a^\dagger_{B,A} a_{B,A}$ are the number operators) commutes both with the Hamiltonian and the jump operators: $[H,P]=[L,P]=0$. \mbox{}\hskip0ptThus the loss affects only the bonding mode while the antibonding parity is preserved. Mathematically, if we split the Hilbert space into the odd ($o$) and even ($e$) subspaces, and the space of operators over that space into $ee$, $oo$, $eo$ and $oe$ subspaces accordingly, the Liouvillian superoperator will map each of these four subspaces onto itself. The strong symmetry thus block-diagonalizes the Liouvillian: \(\hat{\mathcal L}=\hat{\mathcal L}_{ee}\oplus\hat{\mathcal L}_{eo}\oplus\hat{\mathcal L}_{oe}\oplus\hat{\mathcal L}_{oo}\).%

To investigate the thermodynamic limit, we introduce an effective size $N$ by scaling $F=\tilde{F}\sqrt{N}$ and $U=\tilde{U}/N$, keeping $F\sqrt{U}$ fixed \cite{N_scaling1,N_scaling2}. Increasing $N$ scales the driven $a_B$ amplitude by $\sqrt{N}$ (photon number as $N$), while the decreasing $U$ keeps the Kerr effect finite. 

We numerically compute the Liouvillian eigenoperators and eigenvalues in each symmetry sector \cite{Esencan2026GitHub, johansson2012qutip}. We fix $J=1.1$, $\Delta=0.8$, $\tilde U = 1$, $\gamma=1$ throughout and work in a truncated two-mode Fock basis \(\{|n_B,n_A\rangle\}\). The bonding mode carries the large amplitude whereas the antibonding mode remains weakly populated, so we use an asymmetric truncation $n_A^{\max}\ll n_B^{\max}$, elaborated in Supplemental Material (SM) \cite{supplements}.

For each of the two superoperators $\hat{\mathcal L}_{ee}$ and $\hat{\mathcal L}_{oo}$ we find a zero-eigenvalue normalized eigenoperator, ${r}_{ee}$ and ${r}_{oo}$, respectively, which gives the degenerate steady state in each of the two Hilbert space subspaces. For the other two superoperators, we select the eigenoperators ${r}_{eo}$ and ${r}_{oe}$, with ${r}_{eo}={r}_{oe}^\dagger$, whose eigenvalues ($\lambda_{eo} = \lambda_{oe}^*$) have real parts closest to zero. As seen in Fig.~\ref{fig:fig1}, these real parts approach zero with increasing $N$ at both small and large driving amplitudes, while the imaginary parts remain nonzero.  

To understand the physical implication of this finding, we consider the density operator  
    \begin{equation}\label{eq:four_ss}
\rho(t)= b{r}_{ee} + (1-b){r}_{oo} + ce^{\lambda_{eo}t}{r}_{eo} + c^*e^{\lambda_{oe}t}{r}_{oe},
\end{equation}    
  For any $b\in [0,1]$ and a sufficiently small $|c|$, this density matrix represents a physical state. The values of $b$ and $1-b$ are populations of the even and odd sectors of the Hilbert space, respectively --- so the above state can be used to represent a logical bit.   
  In the thermodynamic limit, $\mathrm{Re}\lambda_{eo}=\mathrm{Re}\lambda_{oe}\to0$, so the coherence $ce^{\lambda_{eo}t}$ is preserved indefinitely, rendering the system a time crystal, except near $\tilde F_c\approx0.9325$, where the real parts of $\lambda_{eo}$ and $\lambda_{oe}$ tend to a constant value that significantly deviates from zero and does not decrease with $N$, forming a level crossing [Fig.~\ref{fig:fig1}(a)] \cite{cristobal_strong, cristobal}. When the drive increases through this critical point, the system undergoes a phase transition: the dominant persistently oscillating state jumps to the branch of higher frequency and higher expected photon number [Fig.~\ref{fig:fig1}(b,c)]. At $\tilde F=1.8$ the bonding occupation is $\langle n_B\rangle\approx 3N$. We call this macroscopically occupied phase mode \emph{bright}, and the phase corresponding to the driving force below $\tilde F_c$ \emph{dim}. We will see next that this transition also captures an abrupt change in quantum information preservation.

\begin{figure}
\centering \includegraphics[width=1\linewidth]{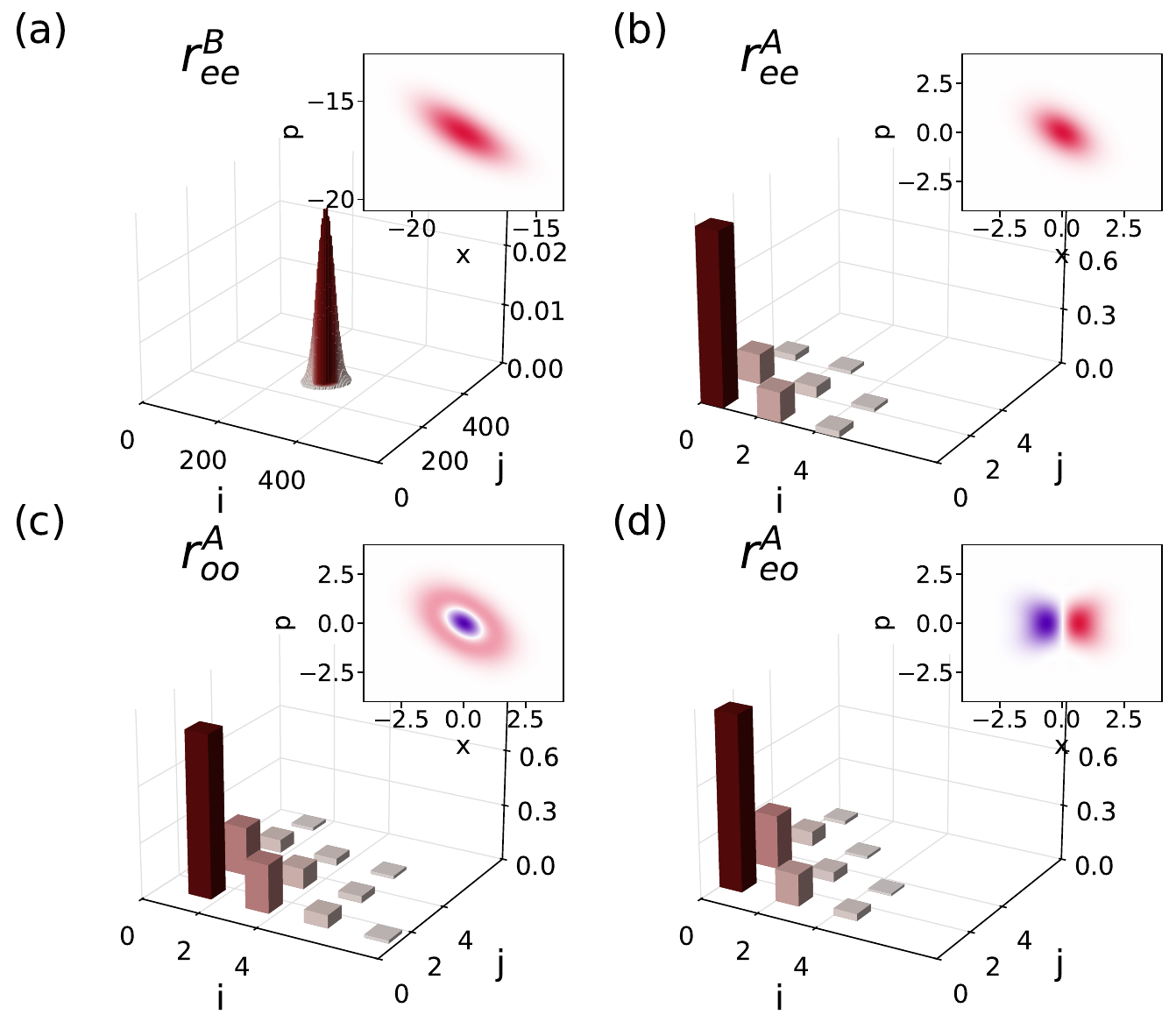}
\caption{Matrices in the Fock basis (absolute values) and Wigner functions (insets) for the partial traces $r_{ee}^B=\Tr_A(r_{ee})$ (a), $r_{ee}^A=\Tr_B(r_{ee})$ (b), $r_{oo}^A=\Tr_B(r_{oo})$ (c), $r_{eo}^A=\Tr_B(r_{eo})$ (d), of the eigenoperators in the bonding (a) and antibonding (b--d) modes at $\tilde F=1.8$ and $N=100$. The reductions $r^B_{oo}$ and $r^B_{eo}$ are indistinguishable from $r^B_{ee}$ ($\langle n_B\rangle\approx300$), so only the latter is displayed in (a). The Wigner function in (d) shows the real part, $i,j$ denote photon numbers for the Fock basis.}
\label{fig:fig2}
\end{figure}

    The semiclassical (mean field) equations of Lled\'o \emph{et al.}~\cite{cristobal_strong} for $\alpha_{A,B}=\langle a_{A,B}\rangle$ support our quantum simulations: for our parameters they yield a single stable bonding solution throughout most drive strengths, while the antibonding mode oscillates persistently with microscopic amplitude at the frequency
\begin{equation}\label{eq:sc_frequency}
\omega_A = \sqrt{(U |\alpha_B|^2 - \Delta + J)\,(3U |\alpha_B|^2 - \Delta + J)}.
\end{equation}
This frequency (derived in SM \cite{supplements}) behaves similarly to the imaginary part of $\lambda_{eo}$ exhibiting bifurcation near the phase transition point $\tilde F=\tilde F_c$ [Fig.~\ref{fig:fig1}(a), SM]. The quantum bonding amplitudes approach the semiclassical solution with increasing $N$ [Fig.~\ref{fig:fig1}(c)]. The eigenoperators themselves behave consistently with these findings: traced over the antibonding mode [Fig.~\ref{fig:fig2}(a)], all are approximated by a macroscopic coherent state of the semiclassically predicted amplitude. Traced over the bonding mode [Fig.~\ref{fig:fig2}(b--d)], the eigenoperators are microscopic, loosely approximated by $\ketbra{0}{0}$, $\ketbra{1}{1}$, $\ketbra{0}{1}$ and $\ketbra{1}{0}$ for ${r}_{ee}$ ${r}_{oo}$, $r_{eo}$ and $r_{oe}$, respectively. The Wigner functions of ${r}_{eo}$ and ${r}_{oe}$ spin in the antibonding phase space at their eigenfrequencies, close to $\omega_A$.

\begin{figure}
 \centering
     \includegraphics[width=0.97\linewidth]{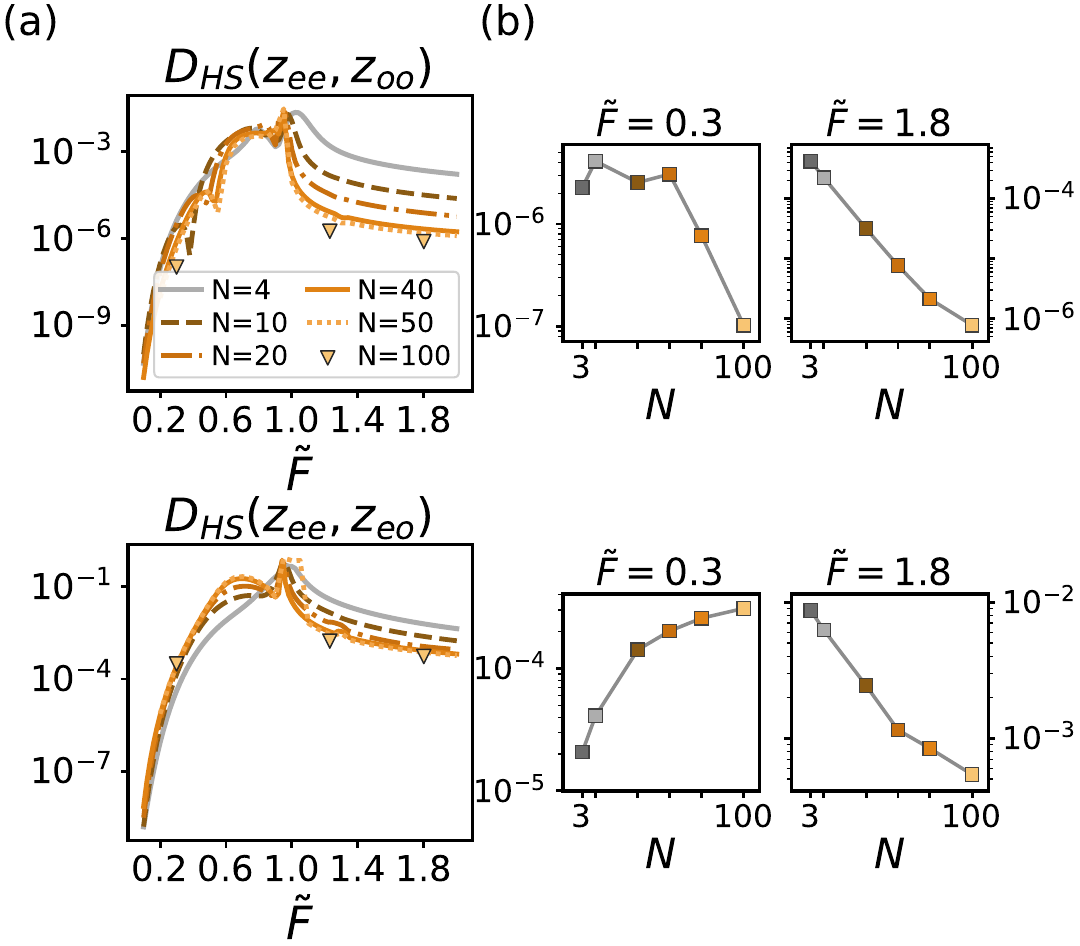}
   \vspace{-5pt}
 \caption{(a) HS distances of the occupied parity blocks $z_{ij}$ vs $\tilde F$: $N=4$ exact line, $N=10,20,40,50,100$ computed via sparse matrix analysis (SM \cite{supplements}). (b) Scaling with $N$ at $\tilde F=0.3$ and $1.8$, points colored by $N$: the monotonic trend of the HS-distance reverses across the phase transition. Only two pairs are shown: the rest follow from $r_{oe}=r_{eo}^{\dagger}$ and the triangle inequality.}
 \label{fig:fig3}
\end{figure}

\paragraph{Noiseless subsystem as a phase.---}\!\!\!\!\!\!\! 
In the bright phase, we find an additional remarkable property of the state \eqref{eq:four_ss}. Writing the four operators in the form \cite{simon}
\begin{equation}
\begingroup
\setlength\arraycolsep{3pt}
r_{ee}=\begin{pmatrix} z_{ee} & 0\\ 0 & 0 \end{pmatrix},\,
r_{oo}=\begin{pmatrix} 0 & 0\\ 0 & z_{oo} \end{pmatrix},\,
r_{eo}=\begin{pmatrix} 0 & z_{eo}\\ 0 & 0 \end{pmatrix},
\endgroup    
\end{equation}
and $r_{oe} = r_{eo}^\dagger$, in a basis that separates the even and odd subspaces and diagonalizes ${r}_{ee}$ and ${r}_{oo}$, we observe that the four blocks $z_{ij}$ are equal up to a coefficient.
To this end, we compute the normalized Hilbert--Schmidt (HS) distance,
\begin{equation}\label{eq:hs_distance}
D_{HS}({z}_{ij},{z}_{i'j'}) = 1 - \left \langle\frac{{z}_{ij}}{\langle {z}_{ij}, {z}_{ij}\rangle}, \frac{{z}_{i'j'}}{\langle {z}_{i'j'}, {z}_{i'j'}\rangle}\right\rangle,
\end{equation}
where $\langle A,B\rangle=\sqrt{\mathrm{Tr}(|A^\dagger B|)}$,
comparing the matrices inside the occupied parity blocks.
As evidenced by Fig.~\ref{fig:fig3}, in the bright phase all pairwise distances tend to zero in the thermodynamic limit. Note that, in contrast, below the transition the distance \emph{grows} with $N$: this flow reversal at $\tilde F_c$ is the operational signature of the quantum-information transition. 

Consequently, in the thermodynamic limit the state \eqref{eq:four_ss} takes the form
    \begin{equation}\label{eq:qubit}
 {\rho}(t) = \rho_Q\otimes {M}, \textrm{ where } \rho_Q=\begin{pmatrix}
b & ce^{\lambda_{eo}t}\\
 c^*e^{\lambda_{oe}t} & 1-b
\end{pmatrix}
\end{equation}  
    with the matrix $M=z_{ij}$. The density matrix $\rho_Q$ is the \emph{noiseless subsystem}  \cite{ekert_dfs,duan_preserving_1997, zanardi_noiseless_1997, lidar_decoherence-free_1998, knill_theory_2000, nielsen_fighting_2010} --- a proper qubit with arbitrary allowable parameters $b$ and $c$ formed by the even (logical 0) and odd (logical 1) subspaces. While the state $M$ of each individual subspace is non-pure due to dissipation, the qubit state itself can be made pure and remain such indefinitely long. 

\paragraph{Passive Dephasing Protection.---}\!\!\!\!\!\!\! An important question is what happens to the noiseless qubit under dephasing \cite{nielsen_chuang, open_quantum, zurek_decoherence_2003}. We investigate collective dephasing, adding to our dissipator the collective Lindblad jump operator $L_d^{\eta} = n_B + \eta\, n_A$. Independently of $\eta$, the dephasing preserves the strong symmetry: $[L_d,P]=0$.   

  \begin{figure}[t]
 \centering
 \includegraphics[width=1\linewidth]{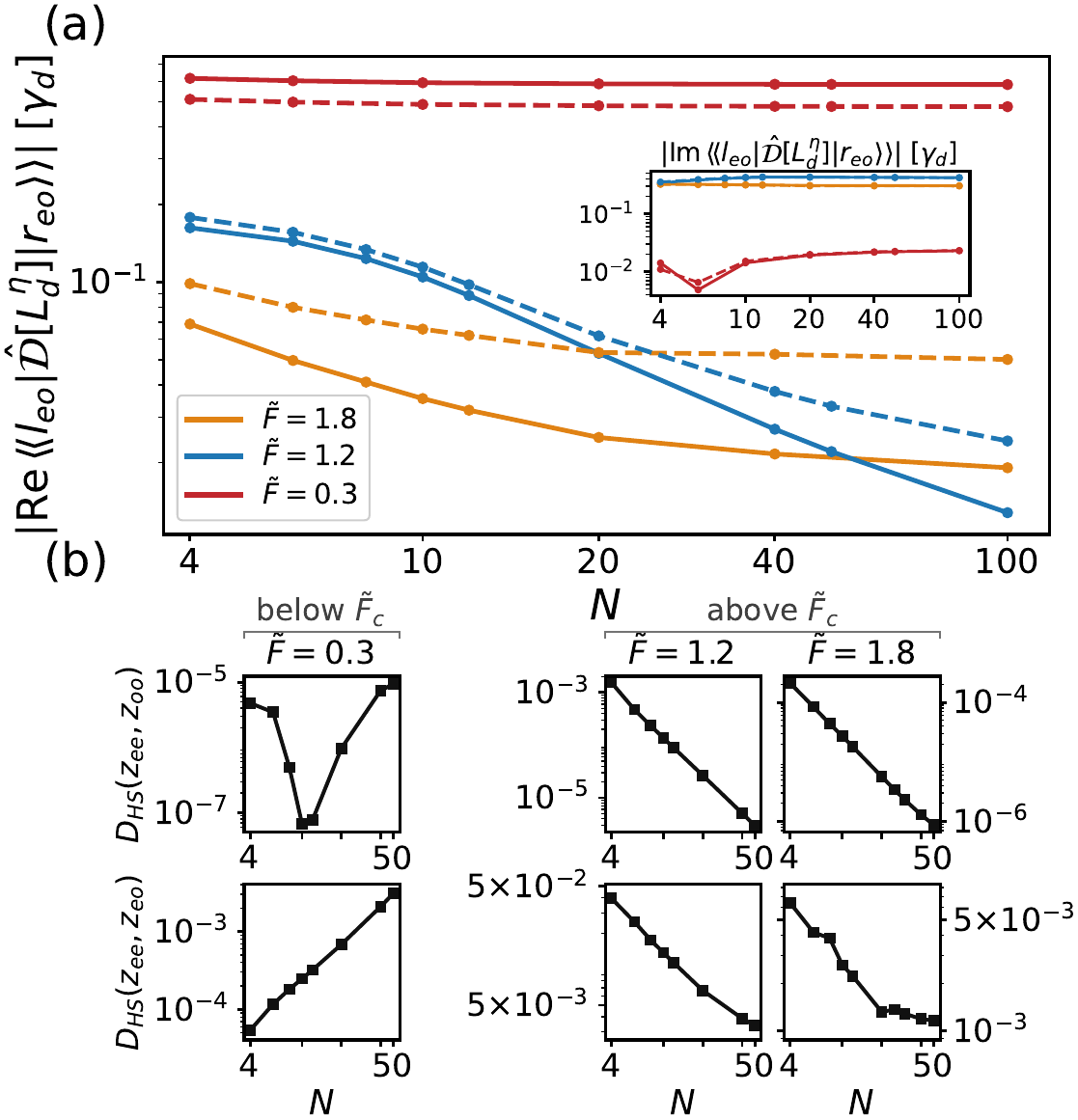}
 \vspace{-12pt}
 \caption{(a) Real part of the leading-order correction to the eigenvalue $\lambda_{eo}$ due to dephasing $\langle\!\langle l_{eo}|\hat{\mathcal{D}}[L_d^{\eta}]|r_{eo}\rangle\!\rangle$ plotted in the units of $\gamma_d$. Solid lines correspond to $\eta=1$, dashed to $\eta=0.9$. Inset: imaginary part. Below the transition ($\tilde F=0.3$), no protection is observed. (b) DNS analysis of the dephased Liouvillian ($\gamma_d=0.01\gamma$): $D_{HS}(z_{ee},z_{oo})$ (top row) and $D_{HS}(z_{ee},z_{eo})$ (bottom row) at $\tilde F=0.3$, $1.2$, $1.8$. The behavior mimics Fig.~\ref{fig:fig3}(b): both distances fall with $N$ above the transition, below it the HS distance \emph{rises}.}
 \label{fig:fig4}
\end{figure}

\begin{figure}[t] \centering
 \includegraphics[width=\linewidth]{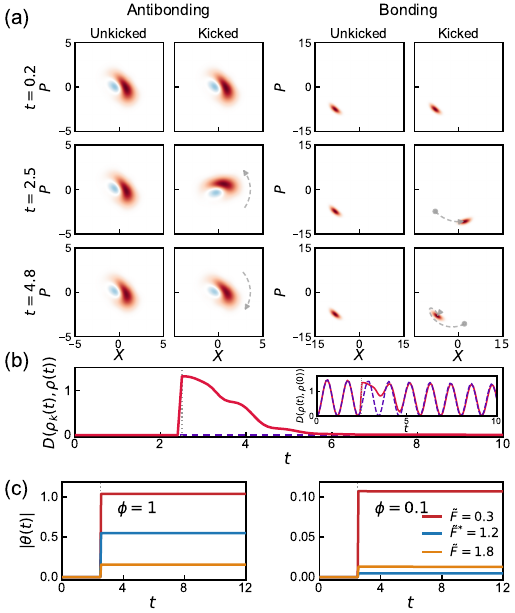}
 \vspace{-10pt}
 \caption{Evolution of state \eqref{eq:four_ss}, initialized with \(b=c=1/2\) at \(\tilde F=1.8\) and \(N=20\), after a 1 rad phase kick at \(t=2.5\), with time $t$ in units of $1/\gamma$. (a) Wigner-function snapshots of bonding and antibonding reductions. Dashed arrows: the kick-induced rotation ($t=2.5$) and return ($t=4.8$). (b) Distance between the kicked and unkicked states, \(D(\rho_k(t),\rho(t))\), restored to zero after the kick. Inset: distances of both states from \(\rho(0)\), showing the oscillation persists. (c) Evolution of the phase of the qubit $\rho_Q(t)$, showing no correction. The effect on the qubit phase for $\tilde F=\tilde F^*$ scales quadratically with the kick. }
 \label{fig:fig5}
\end{figure}

Let us first treat dephasing as a perturbation to the persistently oscillating states, as done for cat-qubit memories \cite{mirrahimi_dynamically_2014}. In Fig.~\ref{fig:fig4}(a), we plot the magnitudes of the real and imaginary parts of the leading-order dephasing perturbation to the unperturbed eigenvalue, $\delta_\lambda = \langle\bra{l_j}\hat{\mathcal{L}}^{\eta}_d\ket{r_j}\rangle$, for $\hat{\mathcal{L}}^{\eta}_d = \gamma_d\hat{\mathcal{D}}[L^{\eta}_d]$, where $\gamma_d$ is the dephasing rate. In the bright phase, the change in the real part of $\lambda_{eo}$, the dephasing-induced decay, is strongly suppressed with $N$. The suppression saturates for large $N$ except for drive strengths approaching the ``magic'' value of $\tilde F^{*}\approx1.217$ \cite{supplements}. On the other hand, the imaginary part of $\delta_\lambda$ stays constant with $N$, implying a small frequency shift.

To understand the physics behind this observation, we specialize to $\eta=1$ and notice that the dephasing mechanism consists in the environment measuring the total photon number $n_A+n_B$ in the system. Comparing the two logical states of the qubit, the odd state holds about one more photon in the antibonding mode but fewer in the bonding mode, so across the bright phase the totals already nearly match. At $\tilde F= \tilde F^{*}$
the imbalances cancel exactly and the dephasing vanishes in the linear order. Below the transition there is no cancellation. The protection is retained for imperfect balance $\eta\ne 1$, albeit with the matching drive $\tilde F^{*}$ dependent on $\eta$, as elaborated further in the End Matter.

We repeat the DNS scaling analysis in the presence of dephasing with $\gamma_d=0.01\gamma$ in Fig. \ref{fig:fig4} (b). The behavior closely mirrors the undephased case [Fig. \ref{fig:fig3} (b)]: the HS distances between the four symmetry-sector eigenoperators again fall with $N$ in the DNS phase. Thus, although dephasing alters the eigenoperators, the dephased Liouvillian hosts a noiseless subsystem of its own, at a slightly shifted frequency [Fig.~\ref{fig:fig4}(b)].

Dephasing can be envisioned as a continuous train of infinitesimally small kicks. Therefore, to gain further insight, we apply macroscopic instantaneous phase kicks of 1 rad magnitude to both modes, and evolve the system under the master equation \eqref{eq:liouvillian} \cite{johansson2012qutip}. As seen in Fig.~\ref{fig:fig5}(b), the kicked state $\rho_k(t)$ returns approximately to the unkicked state $\rho(t)$, as the distance between these states $D(\rho_k(t), \rho(t)) = \operatorname{Tr}\!\left[(\rho(t) - \rho_k(t))^\dagger (\rho(t) - \rho_k(t))\right]$ diminishes during the evolution after the kick, with long-time sinusoidal behavior for both $\hat{\mathcal{L}}_d$ and $\hat{\mathcal{L}}$.

    Wigner functions visualize this evolution in Fig.~\ref{fig:fig5}(a), with an animation in the repository \cite{Esencan2026GitHub}. The bonding mode returns because the dissipative dynamics admit a unique semiclassical bonding solution. More remarkably, the antibonding mode also recovers. This is mainly because, while recovering its phase, the bonding mode temporarily changes its amplitude through the self-phase modulation term ${a}^\dagger_B {a}^\dagger_B {a}_B {a}_B$ in the Hamiltonian \eqref{eq:bhd}. The cross-phase modulation term ${a}^\dagger_B a_B \hat{a}^\dagger_A {a}_A$ then causes this modified amplitude to affect the phase of the antibonding mode, correcting it to nearly its initial value \cite{supplements}. In other words, the classical bonding mode acts as a ``guardian" to the quantum antibonding mode.

It is important to distinguish the effect of the kick on the overall physical state $\rho(t)$ and its qubit part, $\rho_Q(t)$. In both cases, the phase is instantaneously affected by the kick. In contrast to the host state, however, the qubit is not repaired by the subsequent evolution; the phase shift is permanent. Fortunately, in the bright phase this phase shift is much less significant than the kick magnitude and vanishes in the first order for $\tilde F=\tilde F^{*}$ [Fig.~\ref{fig:fig5}(c)].

  \begin{table}[!t]
\caption{The two dissipative phases of the driven BHD.}
\label{tab:phases}

\footnotesize
\setlength{\tabcolsep}{0pt}
\begin{ruledtabular}
\begin{tabular}{
 @{}l
 @{\hspace{5pt}}c
 @{\hspace{5pt}}c
 @{\hspace{10pt}}l
 @{}
}
 & $\tilde F<0.93$ & $\tilde F>0.93$ & signature\\
\colrule
frequency
& low
& high
& jump at $\tilde F_c$ 
 \\

time crystal
& yes
& yes
& gap $\to0$ with $N$ 
 \\

memory
& preserved
& preserved
& zero modes $r_{ee},r_{oo}$\\

DNS qubit
& no
& yes
& $D_{HS}$ flow reverses 
 \\

dephasing
& unprotected
& suppressed
& $n_{ee}-n_{oo}$; cancels at $\tilde F^{*}$ 
 \\
\end{tabular}
\end{ruledtabular}

\end{table}

  \paragraph{Discussion and outlook.---}\!\!\!\!\!\!\! We have demonstrated that, controlled by the drive strength, the BHD undergoes a phase transition   
\cite{minganti2018spectral} into a DNS in the form of a persistently oscillating qubit: the macroscopic coherent amplitude of the bonding mode stabilizes the microscopic quantum coherence carrying the logical information, protected against collective loss and dephasing. Table~\ref{tab:phases} summarizes the two phases and their signatures.

    Such protection makes strong symmetric open systems a platform for controlled phase shifts: the oscillation frequency is tunable through system parameters, so a judicious control of parameters can rotate the qubit around the $Z$ axis of the Bloch sphere, enabling single-qubit phase shift gates including the non-Clifford T gate. Measurement of the qubit in the computational basis is straightforward: it consists in measuring the photon number in the antibonding mode and determining its parity. Single-qubit rotations around other axes and measuring it in bases other than computational, as well as two-qubit gates, require further investigation, which we leave to future work.

We focus on collective, balanced dephasing and loss to preserve the strong-parity symmetry. The effects of small deviations from the balance disappear in the first order (see End Matter). 
Being a passive scheme, our error correction approach requires no ancillas, syndromes, or feedback. Superconducting circuits offer a natural realization: two high-coherence cavity modes mixed by a Josephson coupler supplying the Kerr terms, and only the bonding mode drained through a filtered cold line, which appears experimentally simpler than the engineered two-photon loss of cat qubits \cite{leghtas_confining_2015, lescanne_exponential_2020, alice_bob_cat, amazon_cat}. Two modes are best realized as normal modes of a single cavity, where noise couples to them collectively. Kerr microresonators \cite{taheri_dissipative_2022}, cat-qubit architectures \cite{mamaev_dissipative_2018, amazon_cat}, and collectively dissipative spin ensembles \cite{osullivan_signatures_2020} share this common-reservoir structure.

\paragraph{Acknowledgments.---}\!\!\!\!\!\!\! We thank C. Lledó, S. Lieu, and D. Sels for their discussions. We are also grateful for the conversation with K. Müller that led to new computational insights. B.B. acknowledges funding by the French National Research Agency (ANR) under Project No. ANR-24-CPJ1-0150-01 and the research grant (No. 42085) from Villum Fonden. A.L.'s research is supported by EPSRC Standard Grant EP/Y020596/1 and the EPSRC Impact Acceleration Account Award EP/X525777/1. B.B. would like to dedicate this article to his loved, talented, and lovely wife Vendi who has always loved, supported, and assisted him as his unsung hero, especially during the most difficult times \emph{when the dreams run dry}.

\bibliography{citations}

@misc{supplements,
  note = {See Supplemental Material for the logical structure of the
  encoded qubit, the leading-order perturbative treatment of global dephasing,
  the loss and local-dephasing channel battery, computational methods with
  convergence checks, semiclassical convergence, kick-strength scans, and the
  ab initio theory of the phase-kick response and operating drive}
}

@misc{nielsen_chuang,
	title = {Quantum {Computation} and {Quantum} {Information}: 10th {Anniversary} {Edition}},
	shorttitle = {Quantum {Computation} and {Quantum} {Information}},
	url = {https://www.cambridge.org/highereducation/books/quantum-computation-and-quantum-information/01E10196D0A682A6AEFFEA52D53BE9AE},
	language = {english},
	urldate = {2022-08-23},
	journal = {Higher Education from Cambridge University Press},
	author = {Nielsen, Michael A. and Chuang, Isaac L.},
	month = dec,
	year = {2010},
	doi = {10.1017/CBO9780511976667},
}

@book{open_quantum,
	title = {The {Theory} of {Open} {Quantum} {Systems}},
	url = {https://academic.oup.com/book/27757},
	language = {english},
	urldate = {2022-08-23},
	author = {Breuer, Heinz-Peter and Petruccione, Francesco},
    publisher = {Oxford University Press},
	month = jan,
	year = {2007},
	doi = {10.1093/acprof:oso/9780199213900.001.0001},
}

@article{lindblad_review,
	title = {A short introduction to the {Lindblad} {Master} {Equation}},
	volume = {10},
	issn = {2158-3226},
	url = {http://arxiv.org/abs/1906.04478},
	doi = {10.1063/1.5115323},
	number = {2},
	urldate = {2022-08-23},
	journal = {AIP Advances},
	author = {Manzano, Daniel},
	month = feb,
	year = {2020},
	note = {arXiv:1906.04478 [cond-mat, physics:quant-ph]},
	pages = {025106},
}

@article{cristobal,
	title = {A dissipative time crystal with or without $\mathbb{Z}_2$ symmetry breaking},
	volume = {22},
	issn = {1367-2630},
	url = {http://arxiv.org/abs/2004.02855},
	doi = {10.1088/1367-2630/ab9ae3},
	number = {7},
	urldate = {2022-08-23},
	journal = {New Journal of Physics},
	author = {Lledó, Cristóbal and Szymańska, Marzena H.},
	month = jul,
	year = {2020},
	note = {arXiv:2004.02855 [cond-mat, physics:quant-ph]},
	pages = {075002},
}

@article{beri_symmetry,
	title = {A note on symmetry reductions of the {Lindblad} equation: transport in constrained open spin chains},
	volume = {14},
	issn = {1367-2630},
	shorttitle = {A note on symmetry reductions of the {Lindblad} equation},
	url = {https://iopscience.iop.org/article/10.1088/1367-2630/14/7/073007},
	doi = {10.1088/1367-2630/14/7/073007},
	number = {7},
	urldate = {2022-08-23},
	journal = {New Journal of Physics},
	author = {Buča, Berislav and Prosen, Tomaž},
	month = jul,
	year = {2012},
	pages = {073007},
}

@article{simon,
	title = {Symmetry breaking and error correction in open quantum systems},
	volume = {125},
	issn = {0031-9007, 1079-7114},
	url = {http://arxiv.org/abs/2008.02816},
	doi = {10.1103/PhysRevLett.125.240405},
	number = {24},
	urldate = {2022-08-23},
	journal = {Physical Review Letters},
	author = {Lieu, Simon and Belyansky, Ron and Young, Jeremy T. and Lundgren, Rex and Albert, Victor V. and Gorshkov, Alexey V.},
	month = dec,
	year = {2020},
	note = {arXiv:2008.02816 [cond-mat, physics:physics, physics:quant-ph]},
	pages = {240405},
}

@article{lvovsky_cat,
	title = {Engineering {Schr\"odinger} cat states with a photonic even-parity detector},
	volume = {4},
	issn = {2521-327X},
	url = {http://arxiv.org/abs/1908.10314},
	doi = {10.22331/q-2020-03-02-239},
	urldate = {2022-08-23},
	journal = {Quantum},
	author = {Thekkadath, G. S. and Bell, B. A. and Walmsley, I. A. and Lvovsky, A. I.},
	month = mar,
	year = {2020},
	note = {arXiv:1908.10314 [quant-ph]},
	pages = {239},
}

@article{alice_bob_cat,
	title = {Repetition {Cat} {Qubits} for {Fault}-{Tolerant} {Quantum} {Computation}},
	volume = {9},
	issn = {2160-3308},
	url = {http://arxiv.org/abs/1904.09474},
	doi = {10.1103/PhysRevX.9.041053},
	number = {4},
	urldate = {2022-08-23},
	journal = {Physical Review X},
	author = {Guillaud, Jérémie and Mirrahimi, Mazyar},
	month = dec,
	year = {2019},
	note = {arXiv:1904.09474 [cond-mat, physics:quant-ph]},
	pages = {041053},
}

@article{amazon_cat,
	title = {Building a fault-tolerant quantum computer using concatenated cat codes},
	volume = {3},
	issn = {2691-3399},
	url = {http://arxiv.org/abs/2012.04108},
	doi = {10.1103/PRXQuantum.3.010329},
	number = {1},
	urldate = {2022-08-23},
	journal = {PRX Quantum},
	author = {Chamberland, Christopher and Noh, Kyungjoo and Arrangoiz-Arriola, Patricio and Campbell, Earl T. and Hann, Connor T. and Iverson, Joseph and Putterman, Harald and Bohdanowicz, Thomas C. and Flammia, Steven T. and Keller, Andrew and Refael, Gil and Preskill, John and Jiang, Liang and Safavi-Naeini, Amir H. and Painter, Oskar and Brandão, Fernando G. S. L.},
	month = feb,
	year = {2022},
	note = {arXiv:2012.04108 [quant-ph]},
	pages = {010329},
}

@misc{simon_new,
	title = {Candidate for a passively-protected quantum memory in two dimensions},
	url = {http://arxiv.org/abs/2205.09767},
	doi = {10.48550/arXiv.2205.09767},
	urldate = {2022-08-23},
	publisher = {arXiv},
	author = {Lieu, Simon and Liu, Yu-Jie and Gorshkov, Alexey V.},
	month = jul,
	year = {2022},
	note = {arXiv:2205.09767 [cond-mat, physics:quant-ph]},
}

@article{cristobal_strong,
  title = {Driven Bose-Hubbard dimer under nonlocal dissipation: A bistable time crystal},
  author = {Lled\'o, C. and Mavrogordatos, Th. K. and Szyma\ifmmode \acute{n}\else \'{n}\fi{}ska, M. H.},
  journal = {Phys. Rev. B},
  volume = {100},
  issue = {5},
  pages = {054303},
  numpages = {5},
  year = {2019},
  month = {Aug},
  publisher = {American Physical Society},
  doi = {10.1103/PhysRevB.100.054303},
  url = {https://link.aps.org/doi/10.1103/PhysRevB.100.054303}
}

@article{N_scaling1,
  title = {Quantum entanglement in the spatial-symmetry-breaking phase transition of a driven-dissipative Bose-Hubbard dimer},
  author = {Casteels, Wim and Ciuti, Cristiano},
  journal = {Phys. Rev. A},
  volume = {95},
  issue = {1},
  pages = {013812},
  numpages = {5},
  year = {2017},
  month = {Jan},
  publisher = {American Physical Society},
  doi = {10.1103/PhysRevA.95.013812},
  url = {https://link.aps.org/doi/10.1103/PhysRevA.95.013812}
}

@article{N_scaling2,
  title = {Critical dynamical properties of a first-order dissipative phase transition},
  author = {Casteels, W. and Fazio, R. and Ciuti, C.},
  journal = {Phys. Rev. A},
  volume = {95},
  issue = {1},
  pages = {012128},
  numpages = {5},
  year = {2017},
  month = {Jan},
  publisher = {American Physical Society},
  doi = {10.1103/PhysRevA.95.012128},
  url = {https://link.aps.org/doi/10.1103/PhysRevA.95.012128}
}

@article{ekert_dfs,
	title = {Quantum computers and dissipation},
    author = {Palma,  G.M. and Suominen,  K. and Ekert, A.},
	volume = {452},
	issn = {1364-5021, 1471-2946},
	url = {https://royalsocietypublishing.org/doi/10.1098/rspa.1996.0029},
	doi = {10.1098/rspa.1996.0029},
	language = {english},
	number = {1946},
	urldate = {2024-06-12},
	journal = {Proceedings of the Royal Society of London. Series A: Mathematical, Physical and Engineering Sciences},
	month = dec,
	year = {1996},
	pages = {567--584},
}

@article{zanardi_noiseless_1997,
	title = {Noiseless {Quantum} {Codes}},
	volume = {79},
	url = {https://link.aps.org/doi/10.1103/PhysRevLett.79.3306},
	doi = {10.1103/PhysRevLett.79.3306},
	number = {17},
	urldate = {2024-06-12},
	journal = {Physical Review Letters},
	author = {Zanardi, P. and Rasetti, M.},
	month = oct,
	year = {1997},
	pages = {3306--3309},
}

@article{duan_preserving_1997,
	title = {Preserving {Coherence} in {Quantum} {Computation} by {Pairing} {Quantum} {Bits}},
	volume = {79},
	url = {https://link.aps.org/doi/10.1103/PhysRevLett.79.1953},
	doi = {10.1103/PhysRevLett.79.1953},
	number = {10},
	urldate = {2024-06-12},
	journal = {Physical Review Letters},
	author = {Duan, Lu-Ming and Guo, Guang-Can},
	month = sep,
	year = {1997},
	pages = {1953--1956},
}

@article{lidar_decoherence-free_1998,
	title = {Decoherence-{Free} {Subspaces} for {Quantum} {Computation}},
	volume = {81},
	url = {https://link.aps.org/doi/10.1103/PhysRevLett.81.2594},
	doi = {10.1103/PhysRevLett.81.2594},
	number = {12},
	urldate = {2024-06-12},
	journal = {Physical Review Letters},
	author = {Lidar, D. A. and Chuang, I. L. and Whaley, K. B.},
	month = sep,
	year = {1998},
	pages = {2594--2597},
}

@article{knill_theory_2000,
	title = {Theory of {Quantum} {Error} {Correction} for {General} {Noise}},
	volume = {84},
	url = {https://link.aps.org/doi/10.1103/PhysRevLett.84.2525},
	doi = {10.1103/PhysRevLett.84.2525},
	number = {11},
	urldate = {2024-06-12},
	journal = {Physical Review Letters},
	author = {Knill, Emanuel and Laflamme, Raymond and Viola, Lorenza},
	month = mar,
	year = {2000},
	pages = {2525--2528},
}

@misc{albert_lindbladians_2018,
	title = {Lindbladians with multiple steady states: theory and applications},
	shorttitle = {Lindbladians with multiple steady states},
	url = {http://arxiv.org/abs/1802.00010},
	doi = {10.48550/arXiv.1802.00010},
	urldate = {2025-10-28},
	publisher = {arXiv},
	author = {Albert, Victor V.},
	month = jan,
	year = {2018},
	note = {arXiv:1802.00010 [quant-ph]},
}

@article{liu_dissipative_2024,
	title = {Dissipative phase transitions and passive error correction},
	volume = {109},
	issn = {2469-9926, 2469-9934},
	url = {https://link.aps.org/doi/10.1103/PhysRevA.109.022422},
	doi = {10.1103/PhysRevA.109.022422},
	language = {english},
	number = {2},
	urldate = {2025-10-28},
	journal = {Physical Review A},
	author = {Liu, Yu-Jie and Lieu, Simon},
	month = feb,
	year = {2024},
	pages = {022422},
}

@article{li_perturbative_2014,
	title = {Perturbative approach to {Markovian} open quantum systems},
	volume = {4},
	issn = {2045-2322},
	url = {https://www.nature.com/articles/srep04887},
	doi = {10.1038/srep04887},
	language = {english},
	number = {1},
	urldate = {2025-10-29},
	journal = {Scientific Reports},
	author = {Li, Andy C. Y. and Petruccione, F. and Koch, Jens},
	month = may,
	year = {2014},
	pages = {4887},
}

@article{mirrahimi_dynamically_2014,
	title = {Dynamically protected cat-qubits: a new paradigm for universal quantum computation},
	volume = {16},
	copyright = {http://iopscience.iop.org/info/page/text-and-data-mining},
	issn = {1367-2630},
	shorttitle = {Dynamically protected cat-qubits},
	url = {https://iopscience.iop.org/article/10.1088/1367-2630/16/4/045014},
	doi = {10.1088/1367-2630/16/4/045014},
	number = {4},
	urldate = {2025-10-29},
	journal = {New Journal of Physics},
	author = {Mirrahimi, Mazyar and Leghtas, Zaki and Albert, Victor V and Touzard, Steven and Schoelkopf, Robert J and Jiang, Liang and Devoret, Michel H},
	month = apr,
	year = {2014},
	pages = {045014},
}

@article{lledo_driven_2019,
	title = {Driven {Bose}-{Hubbard} dimer under nonlocal dissipation: {A} bistable time crystal},
	volume = {100},
	issn = {2469-9950, 2469-9969},
	shorttitle = {Driven {Bose}-{Hubbard} dimer under nonlocal dissipation},
	url = {https://link.aps.org/doi/10.1103/PhysRevB.100.054303},
	doi = {10.1103/PhysRevB.100.054303},
	language = {english},
	number = {5},
	urldate = {2025-10-29},
	journal = {Physical Review B},
	author = {Lledó, C. and Mavrogordatos, Th. K. and Szymańska, M. H.},
	month = aug,
	year = {2019},
	pages = {054303},
}

@misc{stefanini_is_2025,
	title = {Is {Lindblad} for me?},
	copyright = {arXiv.org perpetual, non-exclusive license},
	url = {https://arxiv.org/abs/2506.22436},
	doi = {10.48550/ARXIV.2506.22436},
	urldate = {2025-10-29},
	publisher = {arXiv},
	author = {Stefanini, Martino and Ziolkowska, Aleksandra A. and Budker, Dmitry and Poschinger, Ulrich and Schmidt-Kaler, Ferdinand and Browaeys, Antoine and Imamoglu, Atac and Chang, Darrick and Marino, Jamir},
	year = {2025},
	note = {arXiv:2506.22436},
}

@article{fowler_surface_2012,
	title = {Surface codes: {Towards} practical large-scale quantum computation},
	volume = {86},
	copyright = {http://link.aps.org/licenses/aps-default-license},
	issn = {1050-2947, 1094-1622},
	shorttitle = {Surface codes},
	url = {https://link.aps.org/doi/10.1103/PhysRevA.86.032324},
	doi = {10.1103/PhysRevA.86.032324},
	language = {english},
	number = {3},
	urldate = {2025-11-24},
	journal = {Physical Review A},
	author = {Fowler, Austin G. and Mariantoni, Matteo and Martinis, John M. and Cleland, Andrew N.},
	month = sep,
	year = {2012},
	pages = {032324},
}

@article{shor_scheme_1995,
	title = {Scheme for reducing decoherence in quantum computer memory},
	volume = {52},
	copyright = {http://link.aps.org/licenses/aps-default-license},
	issn = {1050-2947, 1094-1622},
	url = {https://link.aps.org/doi/10.1103/PhysRevA.52.R2493},
	doi = {10.1103/PhysRevA.52.R2493},
	language = {english},
	number = {4},
	urldate = {2025-11-24},
	journal = {Physical Review A},
	author = {Shor, Peter W.},
	month = oct,
	year = {1995},
	pages = {R2493--R2496},
}

@article{steane_error_1996,
	title = {Error {Correcting} {Codes} in {Quantum} {Theory}},
	volume = {77},
	copyright = {http://link.aps.org/licenses/aps-default-license},
	issn = {0031-9007, 1079-7114},
	url = {https://link.aps.org/doi/10.1103/PhysRevLett.77.793},
	doi = {10.1103/PhysRevLett.77.793},
	language = {english},
	number = {5},
	urldate = {2025-11-24},
	journal = {Physical Review Letters},
	author = {Steane, A. M.},
	month = jul,
	year = {1996},
	pages = {793--797},
}

@article{gottesman_theory_1998,
	title = {Theory of fault-tolerant quantum computation},
	volume = {57},
	copyright = {http://link.aps.org/licenses/aps-default-license},
	issn = {1050-2947, 1094-1622},
	url = {https://link.aps.org/doi/10.1103/PhysRevA.57.127},
	doi = {10.1103/PhysRevA.57.127},
	language = {english},
	number = {1},
	urldate = {2025-11-24},
	journal = {Physical Review A},
	author = {Gottesman, Daniel},
	month = jan,
	year = {1998},
	pages = {127--137},
}

@article{dennis_topological_2002,
	title = {Topological quantum memory},
	volume = {43},
	issn = {0022-2488, 1089-7658},
	url = {https://pubs.aip.org/jmp/article/43/9/4452/230976/Topological-quantum-memory},
	doi = {10.1063/1.1499754},
	language = {english},
	number = {9},
	urldate = {2025-11-24},
	journal = {Journal of Mathematical Physics},
	author = {Dennis, Eric and Kitaev, Alexei and Landahl, Andrew and Preskill, John},
	month = sep,
	year = {2002},
	pages = {4452--4505},
}

@article{kelly_state_2015,
	title = {State preservation by repetitive error detection in a superconducting quantum circuit},
	volume = {519},
	issn = {0028-0836, 1476-4687},
	url = {https://www.nature.com/articles/nature14270},
	doi = {10.1038/nature14270},
	language = {english},
	number = {7541},
	urldate = {2025-11-24},
	journal = {Nature},
	author = {Kelly, J. and Barends, R. and Fowler, A. G. and Megrant, A. and Jeffrey, E. and White, T. C. and Sank, D. and Mutus, J. Y. and Campbell, B. and Chen, Yu and Chen, Z. and Chiaro, B. and Dunsworth, A. and Hoi, I.-C. and Neill, C. and O’Malley, P. J. J. and Quintana, C. and Roushan, P. and Vainsencher, A. and Wenner, J. and Cleland, A. N. and Martinis, John M.},
	month = mar,
	year = {2015},
	pages = {66--69},
}

@article{ofek_extending_2016,
	title = {Extending the lifetime of a quantum bit with error correction in superconducting circuits},
	volume = {536},
	issn = {0028-0836, 1476-4687},
	url = {https://www.nature.com/articles/nature18949},
	doi = {10.1038/nature18949},
	language = {english},
	number = {7617},
	urldate = {2025-11-24},
	journal = {Nature},
	author = {Ofek, Nissim and Petrenko, Andrei and Heeres, Reinier and Reinhold, Philip and Leghtas, Zaki and Vlastakis, Brian and Liu, Yehan and Frunzio, Luigi and Girvin, S. M. and Jiang, L. and Mirrahimi, Mazyar and Devoret, M. H. and Schoelkopf, R. J.},
	month = aug,
	year = {2016},
	pages = {441--445},
}

@article{zurek_decoherence_2003,
	title = {Decoherence, einselection, and the quantum origins of the classical},
	volume = {75},
	copyright = {http://link.aps.org/licenses/aps-default-license},
	issn = {0034-6861, 1539-0756},
	url = {https://link.aps.org/doi/10.1103/RevModPhys.75.715},
	doi = {10.1103/RevModPhys.75.715},
	language = {english},
	number = {3},
	urldate = {2025-11-24},
	journal = {Reviews of Modern Physics},
	author = {Zurek, Wojciech Hubert},
	month = may,
	year = {2003},
	pages = {715--775},
}

@article{wilczek_quantum_2012,
	title = {Quantum Time Crystals},
	volume = {109},
	url = {https://link.aps.org/doi/10.1103/PhysRevLett.109.160401},
	doi = {10.1103/PhysRevLett.109.160401},
	pages = {160401},
	number = {16},
	journal = {Physical Review Letters},
	shortjournal = {Phys. Rev. Lett.},
	publisher = {American Physical Society},
	author = {Wilczek, Frank},
	urldate = {2026-01-20},
	date = {2012-10-15},
	year = {2012},
}

@article{khemani_phase_2016,
	title = {Phase Structure of Driven Quantum Systems},
	volume = {116},
	url = {https://link.aps.org/doi/10.1103/PhysRevLett.116.250401},
	doi = {10.1103/PhysRevLett.116.250401},
	pages = {250401},
	number = {25},
	journal = {Physical Review Letters},
	shortjournal = {Phys. Rev. Lett.},
	publisher = {American Physical Society},
	author = {Khemani, Vedika and Lazarides, Achilleas and Moessner, Roderich and Sondhi, S. L.},
	urldate = {2026-01-20},
	date = {2016-06-21},
	year = {2016},
}

@article{liang_statistical_2024,
	title = {Statistical and dynamical aspects of quantum chaos in a kicked Bose-Hubbard dimer},
	volume = {109},
	url = {https://link.aps.org/doi/10.1103/PhysRevA.109.033316},
	doi = {10.1103/PhysRevA.109.033316},
	pages = {033316},
	number = {3},
	journal = {Physical Review A},
	shortjournal = {Phys. Rev. A},
	publisher = {American Physical Society},
	author = {Liang, Chenguang and Zhang, Yu and Chen, Shu},
	urldate = {2026-01-20},
	date = {2024-03-22},
	year = {2024},
}

@article{pudlik_dynamics_2013,
	title = {Dynamics of entanglement in a dissipative Bose-Hubbard dimer},
	volume = {88},
	url = {https://link.aps.org/doi/10.1103/PhysRevA.88.063606},
	doi = {10.1103/PhysRevA.88.063606},
	pages = {063606},
	number = {6},
	journal = {Physical Review A},
	shortjournal = {Phys. Rev. A},
	publisher = {American Physical Society},
	author = {Pudlik, Tadeusz and Hennig, Holger and Witthaut, D. and Campbell, David K.},
	urldate = {2026-01-20},
	date = {2013-12-04},
	year = {2013},
}

@article{unruh_maintaining_1995,
	title = {Maintaining coherence in quantum computers},
	volume = {51},
	url = {https://link.aps.org/doi/10.1103/PhysRevA.51.992},
	doi = {10.1103/PhysRevA.51.992},
	pages = {992--997},
	number = {2},
	journal = {Physical Review A},
	shortjournal = {Phys. Rev. A},
	publisher = {American Physical Society},
	author = {Unruh, W. G.},
	urldate = {2026-01-20},
	date = {1995-02-01},
	year = {1995},
}

@article{chuang_quantum_1995,
	title = {Quantum Computers, Factoring, and Decoherence},
	volume = {270},
	issn = {0036-8075, 1095-9203},
	url = {http://arxiv.org/abs/quant-ph/9503007},
	doi = {10.1126/science.270.5242.1633},
	pages = {1633--1635},
	number = {5242},
	journal = {Science},
	shortjournal = {Science},
	author = {Chuang, I. and Laflamme, Raymond and Shor, P. and Zurek, W.},
	urldate = {2026-01-20},
	date = {1995-12-08},
	year = {1995},
	eprinttype = {arxiv},
	eprint = {quant-ph/9503007},
}

@article{knill_resilient_1998,
	title = {Resilient Quantum Computation: Error Models and Thresholds},
	volume = {454},
	issn = {1364-5021, 1471-2946},
	url = {http://arxiv.org/abs/quant-ph/9702058},
	doi = {10.1098/rspa.1998.0166},
	shorttitle = {Resilient Quantum Computation},
	pages = {365--384},
	number = {1969},
	journal = {Proceedings of the Royal Society of London. Series A: Mathematical, Physical and Engineering Sciences},
	shortjournal = {Proc. R. Soc. Lond. A},
	author = {Knill, Emanuel and Laflamme, Raymond and Zurek, Wojciech H.},
	urldate = {2026-01-20},
	date = {1998-01-08},
	year = {1998},
	eprinttype = {arxiv},
	eprint = {quant-ph/9702058},
}

@article{suter_colloquium_2016,
	title = {Colloquium: Protecting quantum information against environmental noise},
	volume = {88},
	url = {https://link.aps.org/doi/10.1103/RevModPhys.88.041001},
	doi = {10.1103/RevModPhys.88.041001},
	shorttitle = {Colloquium},
	pages = {041001},
	number = {4},
	journal = {Reviews of Modern Physics},
	shortjournal = {Rev. Mod. Phys.},
	publisher = {American Physical Society},
	author = {Suter, Dieter and Álvarez, Gonzalo A.},
	urldate = {2026-01-20},
	date = {2016-10-10},
	year = {2016},
}

@article{egan_fault-tolerant_2021,
	title = {Fault-tolerant control of an error-corrected qubit},
	volume = {598},
	rights = {2021 The Author(s), under exclusive licence to Springer Nature Limited},
	issn = {1476-4687},
	url = {https://www.nature.com/articles/s41586-021-03928-y},
	doi = {10.1038/s41586-021-03928-y},
	pages = {281--286},
	number = {7880},
	journal = {Nature},
	publisher = {Nature Publishing Group},
	author = {Egan, Laird and Debroy, Dripto M. and Noel, Crystal and Risinger, Andrew and Zhu, Daiwei and Biswas, Debopriyo and Newman, Michael and Li, Muyuan and Brown, Kenneth R. and Cetina, Marko and Monroe, Christopher},
	urldate = {2026-01-20},
	date = {2021-10},
	year = {2021},
	langid = {english},
}

@article{ding_universally_2025,
	title = {Universally robust control of open quantum systems},
	rights = {2025 The Author(s)},
	issn = {2056-6387},
	url = {https://www.nature.com/articles/s41534-025-01166-y},
	doi = {10.1038/s41534-025-01166-y},
	journal = {npj Quantum Information},
	shortjournal = {npj Quantum Inf},
	publisher = {Nature Publishing Group},
	author = {Ding, Lixiang and Fan, Jingtao and Qiu, Xingze},
	urldate = {2026-01-20},
	date = {2025-12-27},
	year = {2025},
	langid = {english},
}

@article{viola_dynamical_1999,
	title = {Dynamical Decoupling of Open Quantum Systems},
	volume = {82},
	url = {https://link.aps.org/doi/10.1103/PhysRevLett.82.2417},
	doi = {10.1103/PhysRevLett.82.2417},
	pages = {2417--2421},
	number = {12},
	journal = {Physical Review Letters},
	shortjournal = {Phys. Rev. Lett.},
	publisher = {American Physical Society},
	author = {Viola, Lorenza and Knill, Emanuel and Lloyd, Seth},
	urldate = {2026-01-20},
	date = {1999-03-22},
	year = {1999},
}

@article{krinner_realizing_2022,
	title = {Realizing repeated quantum error correction in a distance-three surface code},
	volume = {605},
	rights = {2022 The Author(s), under exclusive licence to Springer Nature Limited},
	issn = {1476-4687},
	url = {https://www.nature.com/articles/s41586-022-04566-8},
	doi = {10.1038/s41586-022-04566-8},
	pages = {669--674},
	number = {7911},
	journal = {Nature},
	publisher = {Nature Publishing Group},
	author = {Krinner, Sebastian and Lacroix, Nathan and Remm, Ants and Di Paolo, Agustin and Genois, Elie and Leroux, Catherine and Hellings, Christoph and Lazar, Stefania and Swiadek, Francois and Herrmann, Johannes and Norris, Graham J. and Andersen, Christian Kraglund and Müller, Markus and Blais, Alexandre and Eichler, Christopher and Wallraff, Andreas},
	urldate = {2026-01-20},
	date = {2022-05},
	year = {2022},
	langid = {english},
}

@article{sivak_real-time_2023,
	title = {Real-time quantum error correction beyond break-even},
	volume = {616},
	rights = {2023 The Author(s), under exclusive licence to Springer Nature Limited},
	issn = {1476-4687},
	url = {https://www.nature.com/articles/s41586-023-05782-6},
	doi = {10.1038/s41586-023-05782-6},
	pages = {50--55},
	number = {7955},
	journal = {Nature},
	publisher = {Nature Publishing Group},
	author = {Sivak, V. V. and Eickbusch, A. and Royer, B. and Singh, S. and Tsioutsios, I. and Ganjam, S. and Miano, A. and Brock, B. L. and Ding, A. Z. and Frunzio, L. and Girvin, S. M. and Schoelkopf, R. J. and Devoret, M. H.},
	urldate = {2026-01-20},
	date = {2023-04},
	year = {2023},
	langid = {english},
}

@article{terhal_quantum_2015,
	title = {Quantum error correction for quantum memories},
	volume = {87},
	url = {https://link.aps.org/doi/10.1103/RevModPhys.87.307},
	doi = {10.1103/RevModPhys.87.307},
	pages = {307--346},
	number = {2},
	journal = {Reviews of Modern Physics},
	shortjournal = {Rev. Mod. Phys.},
	publisher = {American Physical Society},
	author = {Terhal, Barbara M.},
	urldate = {2026-01-20},
	date = {2015-04-07},
	year = {2015},
}

@article{osullivan_signatures_2020,
	title = {Signatures of discrete time crystalline order in dissipative spin ensembles},
	volume = {22},
	issn = {1367-2630},
	url = {https://doi.org/10.1088/1367-2630/ab9fbe},
	doi = {10.1088/1367-2630/ab9fbe},
	pages = {085001},
	number = {8},
	journal = {New Journal of Physics},
	shortjournal = {New J. Phys.},
	publisher = {{IOP} Publishing},
	author = {O’Sullivan, James and Lunt, Oliver and Zollitsch, Christoph W and Thewalt, M L W and Morton, John J L and Pal, Arijeet},
	urldate = {2026-01-20},
	date = {2020-08},
	year = {2020},
	langid = {english},
}

@article{taheri_dissipative_2022,
	title = {Dissipative discrete time crystals in a pump-modulated Kerr microcavity},
	volume = {5},
	rights = {2022 The Author(s)},
	issn = {2399-3650},
	url = {https://www.nature.com/articles/s42005-022-00926-y},
	doi = {10.1038/s42005-022-00926-y},
	pages = {159},
	number = {1},
	journal = {Communications Physics},
	shortjournal = {Commun Phys},
	publisher = {Nature Publishing Group},
	author = {Taheri, Hossein and Matsko, Andrey B. and Herr, Tobias and Sacha, Krzysztof},
	urldate = {2026-01-20},
	date = {2022-06-22},
	year = {2022},
	langid = {english},
}

@misc{verstraete_quantum_2008,
	title = {Quantum computation, quantum state engineering, and quantum phase transitions driven by dissipation},
	url = {http://arxiv.org/abs/0803.1447},
	doi = {10.48550/arXiv.0803.1447},
	number = {{arXiv}:0803.1447},
	publisher = {{arXiv}},
	author = {Verstraete, Frank and Wolf, Michael M. and Cirac, J. Ignacio},
	urldate = {2026-01-20},
	date = {2008-03-12},
	year = {2008},
	eprinttype = {arxiv},
	eprint = {0803.1447 [quant-ph]},
}

@article{lescanne_exponential_2020,
	title = {Exponential suppression of bit-flips in a qubit encoded in an oscillator},
	volume = {16},
	rights = {2020 The Author(s), under exclusive licence to Springer Nature Limited},
	issn = {1745-2481},
	url = {https://www.nature.com/articles/s41567-020-0824-x},
	doi = {10.1038/s41567-020-0824-x},
	pages = {509--513},
	number = {5},
	journal = {Nature Physics},
	shortjournal = {Nat. Phys.},
	publisher = {Nature Publishing Group},
	author = {Lescanne, Raphaël and Villiers, Marius and Peronnin, Théau and Sarlette, Alain and Delbecq, Matthieu and Huard, Benjamin and Kontos, Takis and Mirrahimi, Mazyar and Leghtas, Zaki},
	urldate = {2026-01-20},
	date = {2020-05},
	year = {2020},
	langid = {english},
}

@article{grimm_stabilization_2020,
	title = {Stabilization and operation of a Kerr-cat qubit},
	volume = {584},
	rights = {2020 The Author(s), under exclusive licence to Springer Nature Limited},
	issn = {1476-4687},
	url = {https://www.nature.com/articles/s41586-020-2587-z},
	doi = {10.1038/s41586-020-2587-z},
	pages = {205--209},
	number = {7820},
	journal = {Nature},
	publisher = {Nature Publishing Group},
	author = {Grimm, A. and Frattini, N. E. and Puri, S. and Mundhada, S. O. and Touzard, S. and Mirrahimi, M. and Girvin, S. M. and Shankar, S. and Devoret, M. H.},
	urldate = {2026-01-20},
	date = {2020-08},
	year = {2020},
	langid = {english},
}

@article{leghtas_confining_2015,
	title = {Confining the state of light to a quantum manifold by engineered two-photon loss},
	volume = {347},
	issn = {0036-8075, 1095-9203},
	url = {http://arxiv.org/abs/1412.4633},
	doi = {10.1126/science.aaa2085},
	pages = {853--857},
	number = {6224},
	journal = {Science},
	shortjournal = {Science},
	author = {Leghtas, Zaki and Touzard, Steven and Pop, Ioan M. and Kou, Angela and Vlastakis, Brian and Petrenko, Andrei and Sliwa, Katrina M. and Narla, Anirudh and Shankar, Shyam and Hatridge, Michael J. and Reagor, Matthew and Frunzio, Luigi and Schoelkopf, Robert J. and Mirrahimi, Mazyar and Devoret, Michel H.},
	urldate = {2026-01-20},
	date = {2015-02-20},
	year = {2015},
	eprinttype = {arxiv},
	eprint = {1412.4633 [quant-ph]},
}

@article{puri_bias-preserving_2020,
	title = {Bias-preserving gates with stabilized cat qubits},
	volume = {6},
	issn = {2375-2548},
	url = {http://arxiv.org/abs/1905.00450},
	doi = {10.1126/sciadv.aay5901},
	pages = {eaay5901},
	number = {34},
	journal = {Science Advances},
	shortjournal = {Sci. Adv.},
	author = {Puri, Shruti and St-Jean, Lucas and Gross, Jonathan A. and Grimm, Alexander and Frattini, N. E. and Iyer, Pavithran S. and Krishna, Anirudh and Touzard, Steven and Jiang, Liang and Blais, Alexandre and Flammia, Steven T. and Girvin, S. M.},
	urldate = {2026-01-20},
	date = {2020-08-21},
	year = {2020},
	eprinttype = {arxiv},
	eprint = {1905.00450 [quant-ph]},
}

@article{chamberland_building_2022,
	title = {Building a fault-tolerant quantum computer using concatenated cat codes},
	volume = {3},
	issn = {2691-3399},
	url = {http://arxiv.org/abs/2012.04108},
	doi = {10.1103/PRXQuantum.3.010329},
	pages = {010329},
	number = {1},
	journal = {{PRX} Quantum},
	shortjournal = {{PRX} Quantum},
	author = {Chamberland, Christopher and Noh, Kyungjoo and Arrangoiz-Arriola, Patricio and Campbell, Earl T. and Hann, Connor T. and Iverson, Joseph and Putterman, Harald and Bohdanowicz, Thomas C. and Flammia, Steven T. and Keller, Andrew and Refael, Gil and Preskill, John and Jiang, Liang and Safavi-Naeini, Amir H. and Painter, Oskar and Brandão, Fernando G. S. L.},
	urldate = {2026-01-20},
	date = {2022-02-23},
	year = {2022},
	eprinttype = {arxiv},
	eprint = {2012.04108 [quant-ph]},
}

@article{heugel_role_2023,
	title = {The role of fluctuations in quantum and classical time crystals},
	volume = {6},
	issn = {2666-9366},
	url = {https://www.scipost.org/SciPostPhysCore.6.3.053?acad_field_slug=physics},
	doi = {10.21468/SciPostPhysCore.6.3.053},
	pages = {053},
	number = {3},
	journal = {{SciPost} Physics Core},
	author = {Heugel, Toni Louis and Eichler, Alexander and Chitra, Ramasubramanian and Zilberberg, Oded},
	urldate = {2026-01-20},
	date = {2023-08-10},
	year = {2023},
	langid = {english},
}

@article{li_quantum_2023,
	title = {Quantum nonstationary phenomena of spin systems in collision models},
	volume = {107},
	url = {https://link.aps.org/doi/10.1103/PhysRevA.107.042205},
	doi = {10.1103/PhysRevA.107.042205},
	pages = {042205},
	number = {4},
	journal = {Physical Review A},
	shortjournal = {Phys. Rev. A},
	publisher = {American Physical Society},
	author = {Li, Yan and Li, Xingli and Jin, Jiasen},
	urldate = {2026-02-05},
	date = {2023-04-07},
	year = {2023},
}

@misc{zhao_quantum_2025,
	title = {Quantum Synchronization of Perturbed Oscillating Coherences},
	url = {http://arxiv.org/abs/2510.11601},
	doi = {10.48550/arXiv.2510.11601},
	number = {{arXiv}:2510.11601},
	publisher = {{arXiv}},
	author = {Zhao, Yi J. and Moore, Joel E. and Thingna, Juzar and Wächtler, Christopher W.},
	urldate = {2026-02-05},
	date = {2025-10-13},
	year = {2025},
	eprinttype = {arxiv},
	eprint = {2510.11601 [quant-ph]},
}

@misc{dutta_introduction_2025,
	title = {An introduction to Markovian open quantum systems},
	url = {http://arxiv.org/abs/2510.26530},
	doi = {10.48550/arXiv.2510.26530},
	number = {{arXiv}:2510.26530},
	publisher = {{arXiv}},
	author = {Dutta, Shovan},
	urldate = {2026-02-05},
	date = {2025-11-04},
	year = {2025},
	eprinttype = {arxiv},
	eprint = {2510.26530 [quant-ph]},
}

@article{pi_dynamics_2024,
	title = {The dynamics of an open {Bose}–{Hubbard} dimer with effective asymmetric coupling},
	volume = {97},
	issn = {1434-6036},
	url = {https://doi.org/10.1140/epjb/s10051-024-00661-z},
	doi = {10.1140/epjb/s10051-024-00661-z},
	number = {3},
	urldate = {2026-02-06},
	journal = {The European Physical Journal B},
	author = {Pi, Jinghui and Chen, Feng and Liu, Qi and You, Li and Lü, Rong},
	month = mar,
	year = {2024},
	pages = {26},
}

@article{nielsen_fighting_2010,
	title = {Fighting decoherence in a continuous two-qubit odd- or even-parity measurement with a closed-loop setup},
	volume = {81},
	url = {https://link.aps.org/doi/10.1103/PhysRevA.81.012307},
	doi = {10.1103/PhysRevA.81.012307},
	number = {1},
	urldate = {2026-02-06},
	journal = {Physical Review A},
	publisher = {American Physical Society},
	author = {Nielsen, Anne E. B.},
	month = jan,
	year = {2010},
	pages = {012307},
}

@article{minganti2018spectral,
  title={Spectral theory of Liouvillians for dissipative phase transitions},
  author={Minganti, Fabrizio and Biella, Alberto and Bartolo, Nicola and Ciuti, Cristiano},
  journal={Physical Review A},
  volume={98},
  number={4},
  pages={042118},
  year={2018},
  publisher={APS}
}

@article{johansson2012qutip,
  title   = {QuTiP: An open-source Python framework for the dynamics of open quantum systems},
  author  = {Johansson, J. R. and Nation, P. D. and Nori, F.},
  journal = {Computer Physics Communications},
  volume  = {183},
  pages   = {1760--1772},
  year    = {2012}
}

@book{lehoucq1998arpack,
  title     = {ARPACK Users' Guide: Solution of Large-Scale Eigenvalue Problems with Implicitly Restarted Arnoldi Methods},
  author    = {Lehoucq, Richard B. and Sorensen, Danny C. and Yang, Chao},
  year      = {1998},
  publisher = {SIAM},
  address   = {Philadelphia}
}

@misc{yoshida_theory_2026,
	title = {Theory of {Steady} {States} for {Lindblad} {Equations} beyond {Time}-{Independence}: {Classification}, {Uniqueness} and {Symmetry}},
	shorttitle = {Theory of {Steady} {States} for {Lindblad} {Equations} beyond {Time}-{Independence}},
	url = {http://arxiv.org/abs/2602.13095},
	doi = {10.48550/arXiv.2602.13095},
	urldate = {2026-02-21},
	publisher = {arXiv},
	author = {Yoshida, Hironobu and Hamazaki, Ryusuke},
	month = feb,
	year = {2026},
	note = {arXiv:2602.13095 [quant-ph]},
}

@article{barends_superconducting_2014,
	title = {Superconducting quantum circuits at the surface code threshold for fault tolerance},
	volume = {508},
	copyright = {2014 Springer Nature Limited},
	issn = {1476-4687},
	url = {https://www.nature.com/articles/nature13171},
	doi = {10.1038/nature13171},
	language = {english},
	number = {7497},
	urldate = {2026-02-21},
	journal = {Nature},
	publisher = {Nature Publishing Group},
	author = {Barends, R. and Kelly, J. and Megrant, A. and Veitia, A. and Sank, D. and Jeffrey, E. and White, T. C. and Mutus, J. and Fowler, A. G. and Campbell, B. and Chen, Y. and Chen, Z. and Chiaro, B. and Dunsworth, A. and Neill, C. and O’Malley, P. and Roushan, P. and Vainsencher, A. and Wenner, J. and Korotkov, A. N. and Cleland, A. N. and Martinis, John M.},
	month = apr,
	year = {2014},
	pages = {500--503},
}

@article{chen_exponential_2021,
	title = {Exponential suppression of bit or phase errors with cyclic error correction},
	volume = {595},
	copyright = {2021 The Author(s)},
	issn = {1476-4687},
	url = {https://www.nature.com/articles/s41586-021-03588-y},
	doi = {10.1038/s41586-021-03588-y},
	language = {english},
	number = {7867},
	urldate = {2026-02-21},
	journal = {Nature},
	publisher = {Nature Publishing Group},
	author = {Chen, Zijun and Satzinger, Kevin J. and Atalaya, Juan and Korotkov, Alexander N. and Dunsworth, Andrew and Sank, Daniel and Quintana, Chris and McEwen, Matt and Barends, Rami and Klimov, Paul V. and Hong, Sabrina and Jones, Cody and Petukhov, Andre and Kafri, Dvir and Demura, Sean and Burkett, Brian and Gidney, Craig and Fowler, Austin G. and Paler, Alexandru and Putterman, Harald and Aleiner, Igor and Arute, Frank and Arya, Kunal and Babbush, Ryan and Bardin, Joseph C. and Bengtsson, Andreas and Bourassa, Alexandre and Broughton, Michael and Buckley, Bob B. and Buell, David A. and Bushnell, Nicholas and Chiaro, Benjamin and Collins, Roberto and Courtney, William and Derk, Alan R. and Eppens, Daniel and Erickson, Catherine and Farhi, Edward and Foxen, Brooks and Giustina, Marissa and Greene, Ami and Gross, Jonathan A. and Harrigan, Matthew P. and Harrington, Sean D. and Hilton, Jeremy and Ho, Alan and Huang, Trent and Huggins, William J. and Ioffe, L. B. and Isakov, Sergei V. and Jeffrey, Evan and Jiang, Zhang and Kechedzhi, Kostyantyn and Kim, Seon and Kitaev, Alexei and Kostritsa, Fedor and Landhuis, David and Laptev, Pavel and Lucero, Erik and Martin, Orion and McClean, Jarrod R. and McCourt, Trevor and Mi, Xiao and Miao, Kevin C. and Mohseni, Masoud and Montazeri, Shirin and Mruczkiewicz, Wojciech and Mutus, Josh and Naaman, Ofer and Neeley, Matthew and Neill, Charles and Newman, Michael and Niu, Murphy Yuezhen and O’Brien, Thomas E. and Opremcak, Alex and Ostby, Eric and Pató, Bálint and Redd, Nicholas and Roushan, Pedram and Rubin, Nicholas C. and Shvarts, Vladimir and Strain, Doug and Szalay, Marco and Trevithick, Matthew D. and Villalonga, Benjamin and White, Theodore and Yao, Z. Jamie and Yeh, Ping and Yoo, Juhwan and Zalcman, Adam and Neven, Hartmut and Boixo, Sergio and Smelyanskiy, Vadim and Chen, Yu and Megrant, Anthony and Kelly, Julian and {Google Quantum AI}},
	month = jul,
	year = {2021},
	pages = {383--387},
}

@misc{Esencan2026GitHub,
  author       = {Mert Esencan},
  title        = {GitHub: Dissipative Time Crystals as Passively Protected Oscillating Qubits, Numerical Data and Simulation Code},
  year         = {2026},
  howpublished = {\url{https://github.com/mertesencan/time-crystals-oscillating-qubits}},
}

@article{mamaev_dissipative_2018,
	title = {Dissipative stabilization of entangled cat states using a driven {Bose}-{Hubbard} dimer},
	volume = {2},
	url = {https://quantum-journal.org/papers/q-2018-03-27-58/},
	doi = {10.22331/q-2018-03-27-58},
	language = {english},
	urldate = {2026-02-22},
	journal = {Quantum},
	publisher = {Verein zur Förderung des Open Access Publizierens in den Quantenwissenschaften},
	author = {Mamaev, M. and Govia, L. C. G. and Clerk, A. A.},
	month = mar,
	year = {2018},
	pages = {58},
}

@article{suri_uniformly_2024,
	title = {Uniformly decaying subspaces for error-mitigated quantum computation},
	volume = {110},
	url = {https://link.aps.org/doi/10.1103/PhysRevA.110.042621},
	doi = {10.1103/PhysRevA.110.042621},
	number = {4},
	urldate = {2026-02-22},
	journal = {Physical Review A},
	publisher = {American Physical Society},
	author = {Suri, Nishchay and Saied, Jason and Venturelli, Davide},
	month = oct,
	year = {2024},
	pages = {042621},
}

@article{buca_algebraic_2022,
  title = {Algebraic theory of quantum synchronization and limit cycles under dissipation},
  author = {Bu{\v c}a, Berislav and Booker, Cameron and Jaksch, Dieter},
  journal = {SciPost Physics},
  volume = {12},
  pages = {097},
  year = {2022},
  doi = {10.21468/SciPostPhys.12.3.097},
  eprint = {2103.01808},
  archivePrefix = {arXiv},
  primaryClass = {quant-ph}
}

@article{carusotto_quantum_2013,
  title = {Quantum fluids of light},
  author = {Carusotto, Iacopo and Ciuti, Cristiano},
  journal = {Rev. Mod. Phys.},
  volume = {85},
  pages = {299--366},
  year = {2013},
  doi = {10.1103/RevModPhys.85.299}
}

@article{bogoliubov_superfluidity_1947,
	title = {On the theory of superfluidity},
	volume = {11},
	pages = {23--32},
	journal = {Journal of Physics (USSR)},
	author = {Bogoliubov, N. N.},
	year = {1947}
}

@book{pitaevskii_stringari_2016,
	title = {Bose-Einstein Condensation and Superfluidity},
	publisher = {Oxford University Press},
	address = {Oxford},
	author = {Pitaevskii, Lev and Stringari, Sandro},
	year = {2016}
}

\clearpage
\onecolumngrid
\begin{center}
\textbf{End Matter}
\end{center}
\vspace{-0.5em}
\twocolumngrid
\hypersetup{hidelinks}

\renewcommand{\topfraction}{0.8}
\renewcommand{\textfraction}{0.05}
\setcounter{equation}{0}
\renewcommand{\theequation}{A\arabic{equation}}
\renewcommand{\theHequation}{A\arabic{equation}}

\begin{figure}[b]
\centering
\includegraphics[width=0.96\linewidth]{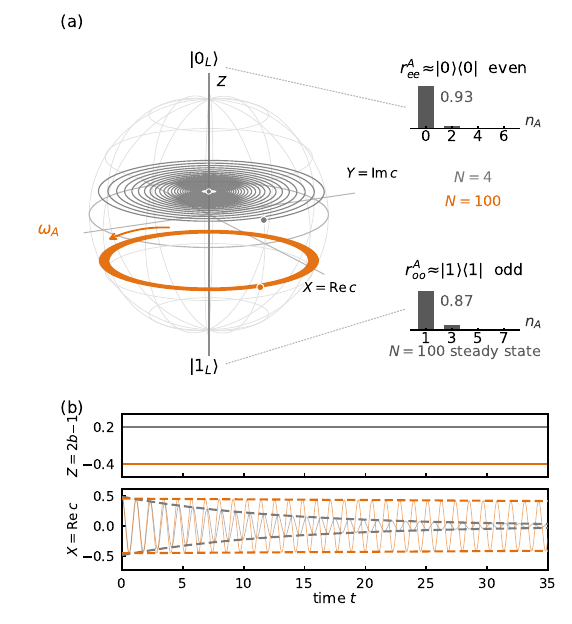}
\caption{Logical qubit at $\tilde F=1.8$, in Bloch coordinates $X=\mathrm{Re}\,c$, $Y=\mathrm{Im}\,c$, $Z=2b-1$. (a)~Trajectories of two maximal-coherence states, $b=0.3$ at $N=100$ (orange) and $b=0.6$ at $N=4$ (grey), each with $|c(0)|=\sqrt{b(1-b)}$, i.e.\ $0.46$ and $0.49$. The plotted coherence is $c(t)=\langle\!\langle l_{eo}|\rho(t)\rangle\!\rangle$, with the measured slow-branch eigenvalues $\lambda_{eo}=-0.0027+5.506i$ and $-0.077+5.389i$. Bar charts: Fock-state probabilities of the even and odd parity sectors of the antibonding-mode reduction of the $N=100$ steady state. (b)~Readout of the same two states: the parity channel $Z$ and the coherence quadrature $X=\mathrm{Re}\,c$, the latter inside the $\pm|c(t)|$ envelope (dashed).}
\label{fig:em_bloch}
\end{figure}

\vspace{0.45em}
\noindent\textit{Qubit visualization.---}
As stated in the main text, the qubit is a superposition of the even and odd subspaces of the Hilbert space. The bar charts in Fig.~\ref{fig:em_bloch}(a) show the logical states in the Fock basis of the antibonding mode. with the logical 0 and 1 being dominated by the vacuum and single-photon states with the probabilities $0.93$ and $0.87$, respectively (for $\tilde F=1.8, N=100$). 

The qubit evolution consists in precession of the Bloch vector around the $Z$ axis at the time-crystal frequency. The coherence $c(t)=\langle\!\langle l_{eo}|\rho(t)\rangle\!\rangle$ quantifies the superposition of the logical states: its quadratures serve as $X$ and $Y$, oscillating at the time-crystal frequency and with decaying slower for larger $N$s [Fig.~\ref{fig:em_bloch}(a,b)], as evidenced by the trajectories for $N=4$ (gray) and $N=100$ (orange).

\vspace{0.45em}
\noindent\textit{Unbalanced loss.---}
Consider a loss channel affecting the two modes in a slightly unbalanced fashion: $L=a_B+\eta\,a_A$, deviating from the ideal collective channel at $\eta=0$ to fully site-local loss at $\eta=1$. This imperfection violates the parity symmetry: $[L,P]\neq0$ for any $\eta\neq0$ and compromises the exact block structure that defines the Liouvillian. We show here that the cost is of order $\eta^{2}$, by a parity selection rule.

\begin{figure}[t]
\centering
\includegraphics[width=0.74\linewidth]{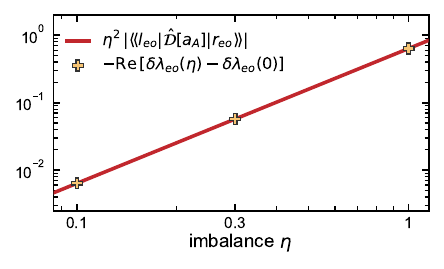}
\caption{Coherence eigenvalue decay rate at $\tilde F=1.8$, per unit loss rate, against the imbalance $\eta$ for $N=100$, showing quadratic dependence.}
\label{fig:em_asym}
\end{figure}

The dissipator is quadratic in its jump operator:
\begin{equation}
\hat{\mathcal{D}}[L]=\hat{\mathcal{D}}[a_B]+\eta^{2}\hat{\mathcal{D}}[a_A]+\eta\,\hat{\mathcal{C}},
\label{eq:end_cross}    
\end{equation}
where
\begin{equation}
\hat{\mathcal{C}}\rho = a_B\rho\, a_A^{\dagger}+a_A\rho\, a_B^{\dagger}
-\tfrac12\{a_B^{\dagger}a_A+a_A^{\dagger}a_B,\rho\},
\label{eq:end_crossterm}
\end{equation}
which
is of the first order with respect to the imbalance. However, each term in $\hat{\mathcal{C}}$ contains only one operator $a_A$, meaning that $\hat{\mathcal{C}}$ will map the density operator from subspaces $eo$ and $oe$ to subspaces $ee$ and $oo$ of the operator space and vice versa. As a result, $\langle\!\langle l_{ee}|\hat{\mathcal{C}}|r_{ee}\rangle\!\rangle=\langle\!\langle l_{oo}|\hat{\mathcal{C}}|r_{oo}\rangle\!\rangle=\langle\!\langle l_{eo}|\hat{\mathcal{C}}|r_{eo}\rangle\!\rangle=0$, so the $\hat{\mathcal{C}}$ term does not contribute to the decoherence rate. This result is confirmed by the numerical calculation in Fig.~\ref{fig:em_asym}, showing quadratic dependence of the real part of the time-crystal state eigenvalue on $\eta$. We also note that the second term in Eq.~\eqref{eq:end_cross} does not diverge in the thermodynamic limit because the antibonding mode remains microscopic as $N\to\infty$.

\noindent\textit{Mechanism of the dephasing protection.---} Figure 8 shows the photon number differences $\Delta n_A$ and $\Delta n_B$ between the even and odd subspaces of the antibonding and bonding modes, as well as their sum $\Delta n_{\rm tot}=\Delta n_A+\Delta n_B$, representing the entire state. As discussed in the main text, the dephasing of the qubit is proportional to the latter number. In the bright state, the bonding mode has more photons in the even-even than the odd-odd state, canceling most of the antibonding difference. The cancellation becomes optimal at $\tilde{F}\approx1.22$ where $\Delta n_A=-\Delta n_B$, and an independent ab initio Bogoliubov theory in the SM predicts $\tilde F^{*}=1.217$ as $N\to\infty$ \cite{supplements}. The same imbalance sets the response to a phase kick: a collective kick of angle $\phi$ rotates the logical coherence through $\approx\phi\,|\Delta n_{\rm tot}|$ at leading order, vanishing at the same $\tilde F=\tilde F^{*}$ \cite{supplements}.

Interestingly, an imperfect coupling can also be protected by tuning the drive. For $L_d^{\eta}=n_B+\eta\,n_A$ the leading-order decay vanishes wherever $\Delta n_B=-\eta\,\Delta n_A$, so each $\eta$ from $0.8$ to $1.15$ has its own decoupling drive $\tilde F^{*}(\eta)$: for example, our representative $\tilde F=1.8$ fully corrects $\eta^{*}\approx1.15$. Further explanation and a broader category of error channels, including the lab-basis loss and the corresponding frequency shifts, are given in the SM \cite{supplements}.

\begin{figure}[t]
\centering
\includegraphics[width=0.74\linewidth]{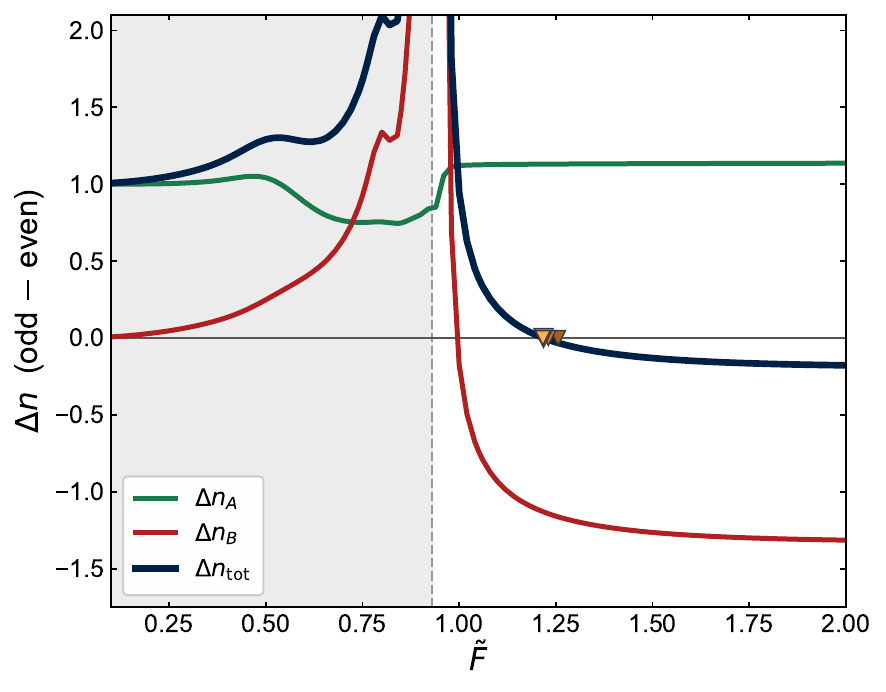}
\caption{Even-odd photon-number differences vs drive at $N=40$ (odd minus even). Triangles on the zero line mark the drives where $\Delta n_A=-\Delta n_B$, i.e.\ $\Delta n_{\rm tot}=0$, for $N=10$--$50$; with growing $N$ they converge on $\tilde F^{*}\to1.217$. Shading: no DNS below $\tilde F_c$.}
\label{fig:em_dntot}
\end{figure}

  \end{document}


\title{Supplemental Material for ``Dissipative Time Crystals as Passively Protected Oscillating Qubits"}

\author{Mert Esencan}
\affiliation{Clarendon Laboratory, University of Oxford, Parks Road, OX1 3PU, Oxford, United Kingdom}

\author{A. I. Lvovsky}
\affiliation{Clarendon Laboratory, University of Oxford, Parks Road, OX1 3PU, Oxford, United Kingdom}

\author{Berislav Bu\v ca}
\affiliation{Universit\' e Paris-Saclay, CNRS, LPTMS, 91405, Orsay, France}
\affiliation{Niels Bohr Institute, University of Copenhagen, Blegdamsvej 17, 2100, Copenhagen, Denmark}
\affiliation{Clarendon Laboratory, University of Oxford, Parks Road, OX1 3PU, Oxford, United Kingdom}

\maketitle

\section{What the Qubit Is: A Logical Subsystem}\label{sec:dressed}

\begin{figure}[!h]
    \centering
    \includegraphics[width=0.5\linewidth]{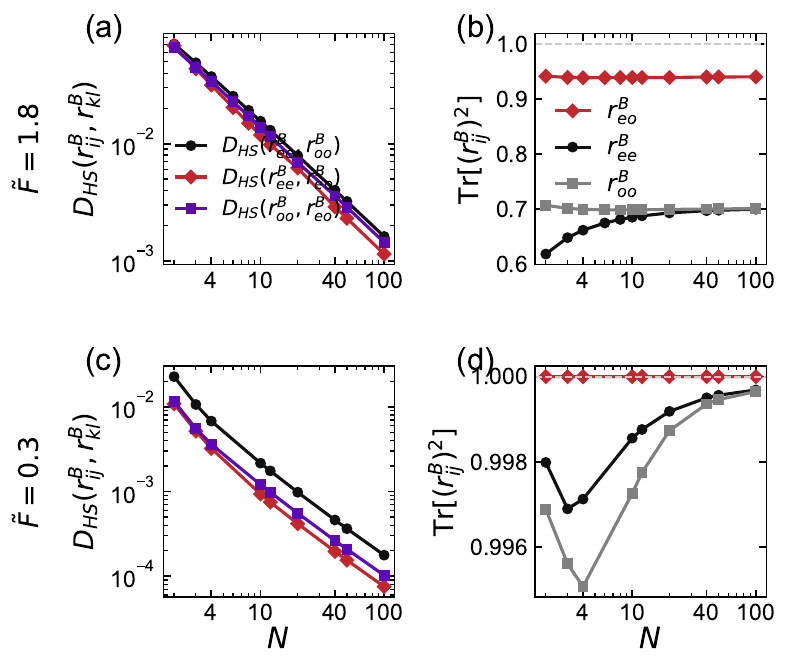}
    \caption{Bonding-mode convergence at both drives, $N=2$ to $100$. $\Ft=1.8$ (top row) and $\Ft=0.3$ (bottom row). (a,c) The normalized HS distances of Eq.~(7) of the main text between the bonding reductions of the eigenoperators fall with $N$ in both phases. (b,d) Purities of the same reductions. Above the transition the shared bonding state is mixed at an intensive $\mathrm{Tr}[(r^B)^2]\approx0.70$, the coherence factor near $0.94$. Below the transition the shared factor is nearly pure, $\mathrm{Tr}[(r^B)^2]\to1$.}
    \label{fig:sfig_bondconv}
\end{figure}

\paragraph{The bonding mode behaving as a common macroscopic background.---}\!\!\!\!\!\!\! The bonding-mode reductions $r^B_{ij}$ become increasingly similar with $N$, consistent with the bonding mode behaving as a common macroscopic background. This is quantified by the decreasing Hilbert--Schmidt distances in Fig.~\ref{fig:sfig_bondconv}(a,c). The mean photon numbers of these reductions nonetheless differ by about one photon at every size but this effect is minimal: a one-photon shift of a bonding mode holding $3N$ photons lies far below its number spread of order $\sqrt{N}$, so the mean difference persists while the distinguishability of the reductions falls as $1/N$.

Fig.~\ref{fig:sfig_bondconv}(b,d) adds a measurement of the purities of the same bonding reductions. The two steady-state bonding states are mixed and stay mixed for the high drive, their purity settling at an intensive value of $\approx0.70$, independent of $N$ over the measured range. This can be understood as dissipation causing decoherence, as all the dissipation lives in the common bonding background, which is why it never purifies, while its distinguishability across the sectors falls as $1/N$. But at $\Ft=0.3$ the bonding mode is nearly pure, $\mathrm{Tr}[(r^B)^2]\ge0.995$ at every size and approaching one with $N$. Thus the mixedness is a property of the bright phase. Above the transition the common factor is macroscopic and carries all the dissipation, while below it photon occupation is low and purity is high.

\begin{figure}[!h]
    \centering
    \includegraphics[width=0.69\linewidth]{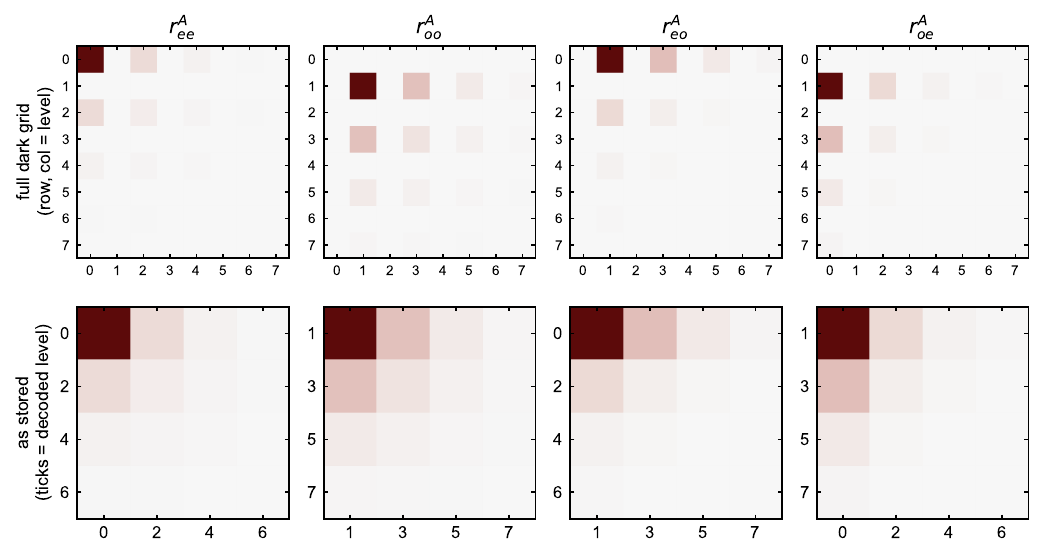}
    \caption{Matrix color plots of the four antibonding blocks ($N=100$, $\Ft=1.8$, $K_A=8$). Top row: the antibonding reductions of the four eigenoperators on the full $8\times8$ grid with each sector populating only its own sublattice. Bottom row: the same matrices with the empty sites stripped away, dense $4\times4$, ticks giving the level each stored index decodes to.}
    \label{fig:dark-sublattice}
\end{figure}

\paragraph{The noiseless subsystem compared to the antibonding mode.---}\!\!\!\!\!\!\! Let us now investigate the common structure of the four antibonding blocks. Within a fixed parity sector each block occupies just one sublattice of $K_A/2=4$ drawn on the full $K_A=8$ grid: $r^A_{ee}$ the even-even sites, $r^A_{oo}$ the odd-odd, and the coherences the two mixed sublattices [Fig.~\ref{fig:dark-sublattice}, top row]. We can remove the empty sites and collapse them onto dense $4\times4$ matrices [bottom row], mapping even sectors to $0,2,4,6$, odd to $1,3,5,7$. Thus the sectors differ by the \emph{parity address} of the sublattice they live on and otherwise look alike.

\begin{figure}[b!]
    \centering
    \includegraphics[width=0.5\linewidth]{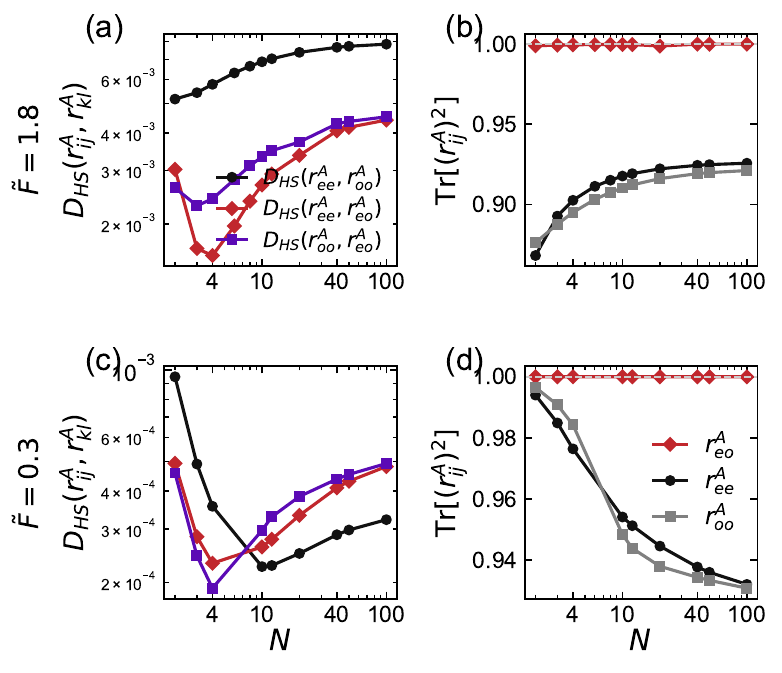}
    \caption{Antibonding-mode convergence and purities on the stored $4\times4$ blocks (the sublattice from Fig.~\ref{fig:dark-sublattice}), at $\Ft=1.8$ (top row) and $\Ft=0.3$ (bottom row), $N=2$ to $100$. (a,c) Normalized HS distances between the stored antibonding blocks. They do not fall with $N$ as there is a residue from the intensive dressing. (b,d) Purities of the same blocks. The coherence block is exactly pure at both drives, while the sector blocks settle near $0.9$ for both drives.}
    \label{fig:sfig_red}
\end{figure}

How alike are the sublattice blocks? In Fig.~\ref{fig:sfig_red} we show the same convergence analysis as Fig.~\ref{fig:sfig_bondconv}, now on the stored $4\times4$ antibonding blocks. (On the $8\times8$ grids the sectors occupy different sublattices and are orthogonal by construction, so only the sublattices can be compared at all.) The distances sit at the half-percent level at $\Ft=1.8$ and a few $10^{-4}$ at $\Ft=0.3$, and they do \emph{not} fall with $N$.
Thus the antibonding-mode reductions do not all become pure: $r^A_{eo}$ approaches a pure coherence, whereas $r^A_{ee}$ and $r^A_{oo}$ remain mixed [Fig.~\ref{fig:sfig_red}(b)]. The antibonding mode is dark only to the direct loss channel and remains strongly coupled to the dissipating bonding mode. The noiseless subsystem is therefore not the antibonding mode alone. Instead, it is a noiseless subsystem encoding \cite{zanardi_noiseless_1997, knill_theory_2000} whose logical computational-basis labels are the Liouvillian strong-parity sectors \cite{beri_symmetry, simon}.

The physical states realizing the two labels are \emph{dressed} and spread over both modes: the odd sector differs from the even sector by a single dressed quasiparticle which we will formally define in Sec.~\ref{sec:qf}. The bar charts in Fig.~6(a) of the main text illustrate this dressing, identifying the poles with the logical basis, the even (odd) sector being $|0\rangle$ ($|1\rangle$) dominated with probability $0.93$ ($0.87$) for $\tilde F = 1.8, N=100$.

\section{Semiclassical Dynamics}\label{app:scconv}

\begin{figure}[b]
    \centering
    \includegraphics[width=0.72\linewidth]{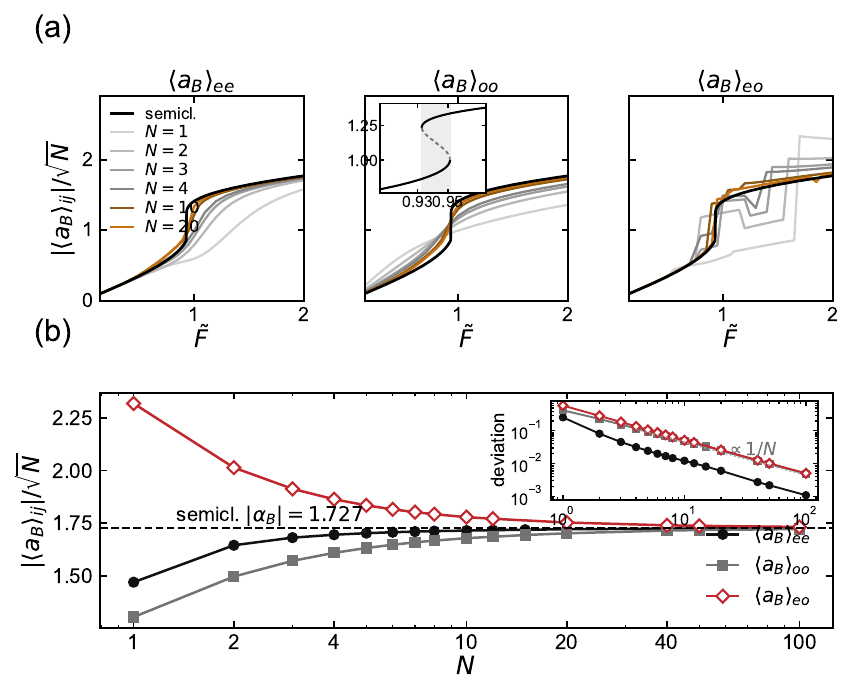}
    \caption{Convergence of the bonding amplitude to the semiclassical fixed point. (a) $|\langle a_B\rangle_{ij}|/\sqrt N$ versus drive for the sizes with dense drive sweeps, against the semiclassical amplitude (black) with the branch switch at $\tilde F\approx0.93$. Inset: the three real roots of the semiclassical cubic near the transition, stable branches solid and the unstable middle branch dashed, with the three-root window $\tilde F\in[0.933,\,0.951]$ shaded. (b) The same quantity at $\tilde F=1.8$ versus $N$, per sector, with the semiclassical value dashed. Inset: absolute deviation from the semiclassical amplitude versus $N$.}
    \label{fig:sfig_scconv}
\end{figure}

\noindent\textit{Semiclassical dynamics.---}
We describe here semiclassical calculation leading to Eq.~(5) of the main text. The bonding mode approaches a driven-dissipative state, whereas a small nonzero antibonding amplitude undergoes persistent oscillations.

The fixed point with $\alpha_A=0$ is reached only when the antibonding mode is initialized exactly at zero. For a small nonzero $\alpha_A$, that mode instead undergoes persistent oscillations. For the dynamics we use \cite{cristobal_strong, cristobal}
\begin{equation}
\frac{d}{dt}\langle a_k\rangle=\operatorname{Tr}\!\big(a_k\,\hat{\mathcal L}\rho\big)= i\langle[H,a_k]\rangle+\gamma\Big\langle L^{\dagger}a_kL-\tfrac12\{L^{\dagger}L,a_k\}\Big\rangle
\label{eq:sm_heis}
\end{equation}
and replace the mode operators by coherent amplitudes $a_k\rightarrow\alpha_k\simeq\langle a_k\rangle$:

\begin{align}
\dot{\alpha}_B
&=(i\Delta+iJ-\gamma)\alpha_B-\sqrt{2}iF
\nonumber\\[-0.25em]
&\quad-iU\!\left(
\alpha_B|\alpha_B|^2+2|\alpha_A|^2\alpha_B
+\alpha_A^2\alpha_B^*\right),
\label{eq:end_sc_B}\\[-0.1em]
\dot{\alpha}_A
&=i(\Delta-J)\alpha_A
\nonumber\\[-0.25em]
&\quad-iU\!\left(
\alpha_B^2\alpha_A^*+2|\alpha_B|^2\alpha_A
+\alpha_A|\alpha_A|^2\right).
\label{eq:end_sc_A}
\end{align}
Setting $\dot\alpha_B=0$ with $\alpha_A=0$ gives the fixed point in closed form,
\begin{equation}
\alpha_B=\frac{\sqrt{2}\,i\tilde F}{\,i\!\left(\Delta+J-U|\alpha_B|^{2}\right)-\gamma\,},
\label{eq:sm_sc_fixedpoint}
\end{equation}
a self-consistent equation whose modulus obeys the cubic $|\alpha_B|^{2}[\gamma^{2}+(\Delta+J-U|\alpha_B|^{2})^{2}]=2\tilde F^{2}$. The key feature is the abrupt branch jump near $\Ft\approx0.93$. As $\Ft$ is increased continuously, the amplitudes do not evolve smoothly through this point, but switch to a different solution branch. For $\Ft$ between $0.9326$ and $0.9513$, the cubic admits three real solutions (bistable points at the edges), with the middle branch unstable and both outer branches stable [inset of Fig.~\ref{fig:sfig_scconv}(a)]. The jump happens at the lower bistable point of an S-shaped curve, and these S-shaped curves with multiple solutions arise in driven Kerr resonators. We chose our parameters to sit close to the monostable regime, so the window of drives with multiple solutions is narrow. (Same parameters were used in \cite{cristobal} for a different model.) In practice the second stable solution plays no role in our sweeps, and we follow the branch continuously connected to the unique high-drive solution, which is also the branch the quantum steady state follows. For $J=1.1$, $\Delta=0.8$, $U=1$, $\tilde F=1.8$ and $\gamma=1$ the stable solution is
\begin{equation}
\alpha_B=-1.269-1.172\,i,\qquad |\alpha_B|=1.727,
\label{eq:sm_sc_value}
\end{equation}
with $\alpha_A=0$, corresponding to $\langle n_B\rangle \approx |\alpha_B|^{2}N\approx3N$, which sets the bonding-mode Fock cutoffs used numerically, elaborated in Sec.~\ref{app:computational}. The same equations follow after the thermodynamic rescaling of $F$ and $U$ \cite{N_scaling1, N_scaling2}. The black
curves in Fig.~1(c) of the main text show the corresponding solutions, for which the selected branch changes discontinuously near $\tilde F\simeq0.93$. We extend those calculations here to each individual parity sector (h.c. of $ r_{eo}$ is $r_{oe}$ and is omitted as usual) in Figure~\ref{fig:sfig_scconv}. Panel (a) shows $|\langle a_B\rangle_{ij}|/\sqrt N = |\mathrm{Tr}(a_B\,r^B_{ij})|/\sqrt N$ against the drive for the sizes with dense drive sweeps, on top of the semiclassical fixed-point amplitude $|\alpha_B(\tilde F)|$, including its discontinuous branch switch at $\tilde F\approx0.93$. Panel (b) fixes $\tilde F=1.8$ and follows the same quantity across every available size, $N=1$ to $100$: the diagonal-sector amplitudes approach $|\alpha_B|=1.7271$ from below and the coherence-sector amplitude from above, with the deviation falling algebraically in $N$ (inset).

To investigate the oscillations, we hold $\alpha_B$ fixed and linearize (remove all quadratic terms in $a_A$) Eq.~\eqref{eq:end_sc_A} to get
\begin{equation}
i\,\frac{d}{dt}
\begin{pmatrix}
\alpha_A\\[0.1em]
\alpha_A^*
\end{pmatrix}
=V
\begin{pmatrix}
\alpha_A\\[0.1em]
\alpha_A^*
\end{pmatrix}.
\label{eq:end_linearized_A}
\end{equation}
Writing $V=(V_{mn})$, its nonzero entries are
\begin{align}
V_{11}&=(J-\Delta)+2U|\alpha_B|^2,
&V_{12}&=U\alpha_B^2,\nonumber\\[-0.2em]
V_{21}&=-U(\alpha_B^*)^2,
&V_{22}&=-V_{11}.
\label{eq:end_linearized_entries}
\end{align}
The eigenvalues of $V$ are $\pm\omega_A$, giving the oscillation frequency in Eq.~(5) of the main text, and its branch change near $\tilde F\simeq0.93$ corresponds to the dissipative transition.

\section{Beyond Semiclassics: Quantum Fluctuations}\label{sec:qf}

\begin{figure}[t]
    \centering
    \includegraphics[width=0.94\linewidth]{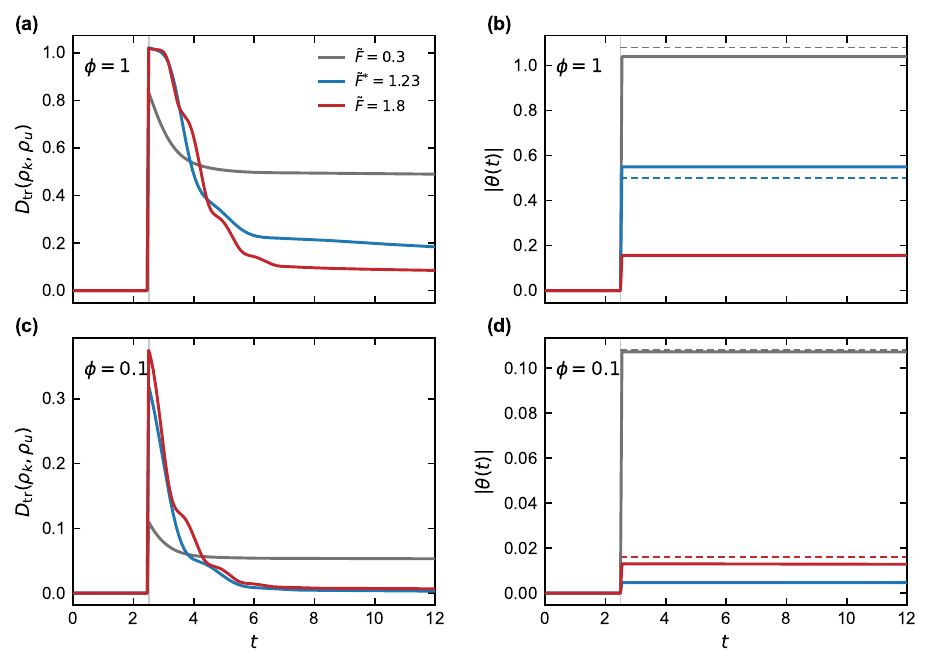}
    \caption{Single-kick response by drive ($N=20$, kick at $t=2.5$, grey vertical): $\Ft=0.3$ (grey), $\Ft^{*}=1.23$ (blue), $\Ft=1.8$ (red). (a, c) Distance between the kicked and unkicked states. (b, d) Logical phase $|\theta(t)|$. Dashed guides: the theoretical predictions.}
    \label{fig:sfig_kickdrives}
\end{figure}

\begin{figure*}[t]
    \centering
    \begin{subfigure}[t]{0.47\textwidth}
        \centering
        \includegraphics[width=\linewidth]{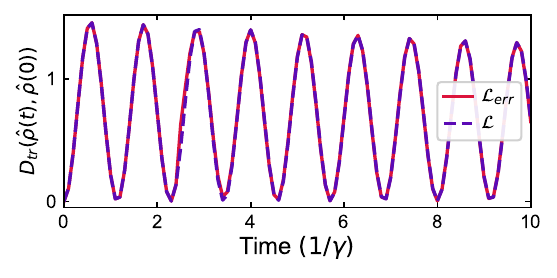}
        \caption{$\phi=0.1$.}
    \end{subfigure}\hfill
    \begin{subfigure}[t]{0.47\textwidth}
        \centering
        \includegraphics[width=\linewidth]{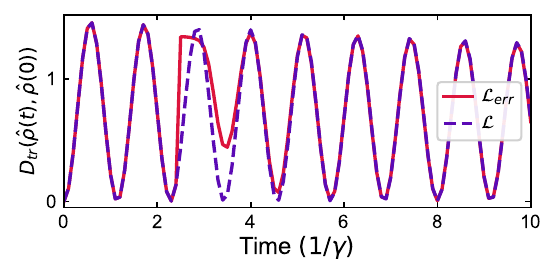}
        \caption{$\phi=0.5$.}
    \end{subfigure}

    \vspace{0.35em}
    \begin{subfigure}[t]{0.47\textwidth}
        \centering
        \includegraphics[width=\linewidth]{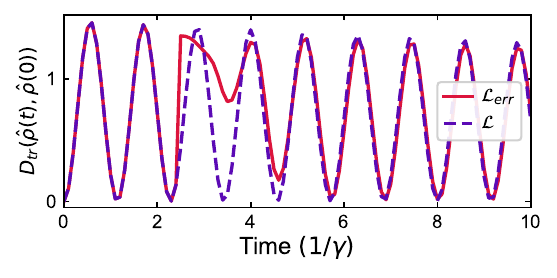}
        \caption{$\phi=1.0$ (same kick scale as main text).}
    \end{subfigure}\hfill
    \begin{subfigure}[t]{0.47\textwidth}
        \centering
        \includegraphics[width=\linewidth]{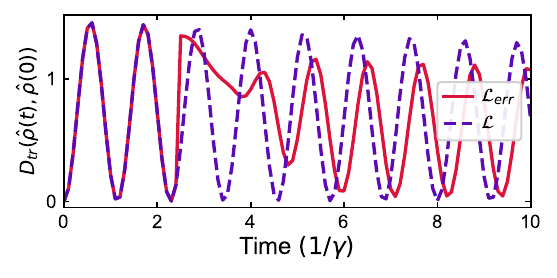}
        \caption{$\phi=2.0$.}
    \end{subfigure}
    \caption{Kick-strength scan for the same trajectory diagnostic used in the main text. Curves show $D_{\mathrm{tr}}(\hat\rho(t),\hat\rho(0))$ for $\mathcal{L}_{\mathrm{err}}$ (red) and $\mathcal{L}$ (dashed purple) at fixed $\tilde F=1.8$, $N=20$, with a kick at $t=2.5$.}
    \label{fig:sfig_kick_strengths}
\end{figure*}

\paragraph{Anatomy of a phase kick.---}\!\!\!\!\!\!\! Now let us develop the theory behind the phase-kick numerics of the main text. We first describe the anatomy of a kick and then derive the kick response from a Bogoliubov expansion \cite{bogoliubov_superfluidity_1947, pitaevskii_stringari_2016, carusotto_quantum_2013} around the semiclassical fixed point, relaxing our linearization condition above and allowing for quadratic terms in $\langle a_A\rangle$. The theory predicts a ``magic'' drive strength $\Ft^{*}$ at which the linear response to collective phase errors vanishes. The perturbation studied in the main text is an instantaneous collective phase kick,
\begin{equation}\label{eq:kick}
\rho \;\to\; e^{i\phi\,\ntot}\,\rho\, e^{-i\phi\,\ntot},
\qquad
\ntot = a_B^\dagger a_B + a_A^\dagger a_A,
\end{equation}
which rotates both modes by the common angle $\phi$. Numerically the kick is exact and instantaneous: $\ntot$ is diagonal in the Fock basis, so the conjugation multiplies each matrix element of $\rho$ by $e^{i\phi(m_A+m_B-n_A-n_B)}$ between two Lindblad evolutions. Following from the strong symmetry of the system, $\ntot$ commutes with $P$, so the kick acts within each parity sector separately, thus the logical populations $b$ and $1-b$ of Eq.~(4) of the main text are exactly preserved. The response on the other hand separates into a physical and a logical part. The physical part is the recovery seen in Fig.~5 of the main text. The bonding mode is pulled back to its unique dissipative attractor within a few time steps (in units of $1/\gamma$), and through the cross-phase modulation this recovery also restores the phase of the antibonding mode. The kicked state rejoins the unkicked trajectory, the attractor restoring the overall (gauge) phase that the kick had displaced. What survives is a permanent rotation of the logical coherence, $c(t)\to c(t)\,e^{i\theta(\phi)}$. Although $\theta(\phi)$ is a deterministic function of the kick angle, a known kick is calibratable and can be undone, we are interested in random kicks to better understand dephasing. We will see later those are suppressed for small $\phi$ when quantum fluctuations are accounted for.

Figure~\ref{fig:sfig_kickdrives} runs the same protocol from Fig~5 of main text at the three representative drives, now tracking both the physical distance and the logical (qubit) phase through the biorthogonal projection of the two branches onto the slow coherence mode, $\theta(t)=\arg c_k(t)-\arg c_u(t)$. At every drive $\theta$ jumps at the kick instant and is afterwards constant to numerical precision: the logical rotation is imprinted by the kick itself, and the recovery that follows is only in the physical state. The size of the jump follows the photon-counting prediction (to be derived below): $|\theta|=0.16$ at $\Ft=1.8$ (the linear response $s\phi$), $0.55$ at $\Ft^{*}=1.23$ (the quadratic residual of a macroscopic kick is larger for the operating drive), and $1.04$ at $\Ft=0.3$. Repeating the protocol at a small kick, $\phi=0.1$ [panel (c)] inverts the ordering: the operating drive is now the best protected, $|\theta|=0.005$ against $0.013$ at $\Ft=1.8$, while in the non-NS phase ($\Ft=0.3$) the effect is an order of magnitude larger with $|\theta|=0.107$.

To emphasize our point, there are two distinct categories of observables one can read after a kick. The physical one is the antibonding mode itself, $\arg\langle a_A\rangle$. The logical one is the coefficient of the coherence, $\arg c$ with $c=\dbra{l_{eo}}\rho\rangle\rangle$. The semiclassical description of $\langle a_A \rangle$ does not capture the mechanism of the qubit protection. The antibonding mode never damps, since the loss enters only the bonding equation, so every kick residue rings forever and an ensemble read this way spreads at every drive with no magic point to cancel it out. This is why for the purposes of memory information is read out from the biorthogonal projection $c$, not from $\arg\langle a_A\rangle$. Now let us look at macroscopic kicks to illustrate further the recovery of the physical modes.

\paragraph{Different kick strengths.---}\!\!\!\!\!\!\! In the main text we show phase-kick recovery for $\phi=1$ by direct master-equation evolution \cite{johansson2012qutip} [Fig.~5 of the main text]. Here we repeat the same study for additional kick strengths for $\tilde{F}=1.8$. Figure~\ref{fig:sfig_kick_strengths} compares the trajectories under the kicked Liouvillian ($\mathcal{L}_{\mathrm{err}}$, red) and the unkicked Liouvillian ($\mathcal{L}$, dashed purple). $D_{\mathrm{tr}}$ plotted there is the same distance used in the main text. We classify recovery by whether the kicked trajectory returns close to and tracks the unkicked oscillatory orbit after the kick. The additional scans show that for small kicks ($\phi=0.1$ and $0.5$) the perturbed trajectory quickly returns to the same oscillatory orbit and remains close to the reference dynamics. For $\phi=1.0$, which is the same case shown in the main text, the trajectory still displays substantial recovery and follows the same long-time oscillatory structure, although with a visible residual mismatch. For larger kicks, the recovery becomes incomplete on the simulated timescale. In particular, at $\phi=2.0$ the kicked trajectory does not fully relock to the reference orbit. Thus passive correction is supported for small (and, in practice, moderate) kicks, while very large kicks are only partially corrected in this finite-time evolution.

\paragraph{Kick-size response, and dephasing as accumulated kicks.---}\!\!\!\!\!\!\!
The permanent rotation is the phase of the projection of the kicked state onto the slow coherence mode. Expanding the kick as nested commutators, $e^{i\phi\ntot}\rho\,e^{-i\phi\ntot}=\rho+i\phi[\ntot,\rho]+\tfrac{(i\phi)^{2}}{2}[\ntot,[\ntot,\rho]]+\ldots$, and projecting onto the left eigenoperator $\dbra{l_{eo}}$ of the main text gives
\begin{equation}\label{eq:theta}
\theta(\phi) = \phi\, \mathrm{Re}\, T_1 - \tfrac{\phi^2}{2}\, \mathrm{Im}\!\left(T_2 - T_1^2\right) + O(\phi^3),
\end{equation}
with the matrix elements
\begin{equation}\label{eq:T12}
T_1 = \dbra{l_{eo}}[\ntot, r_{eo}]\rangle\rangle,
\qquad
T_2 = \dbra{l_{eo}}[\ntot,[\ntot, r_{eo}]]\rangle\rangle.
\end{equation}
We write the linear coefficient as $s \equiv \mathrm{Re}\,T_1$. Whenever $s$ is nonzero the small-kick response is linear, $\theta = s\phi$. At $\Ft = 1.8$ we measure $s = 0.145$ for the $\phi=1$ kick of Fig.~5 ($N=20$, while the $\phi \to 0$ slope evaluated from the matrix element of Eq.~\eqref{eq:T12} is $0.155$ at $N = 40$), meaning that a macroscopic collective phase error reaches the logical subspace at the $15\%$ level. At the magic drive $\Ft^{*}$ derived below the linear term vanishes, and the measured response is quadratic, $\theta \approx 0.5\,\phi^2$. 

\begin{figure}[t]
    \centering
    \includegraphics[width=0.62\linewidth]{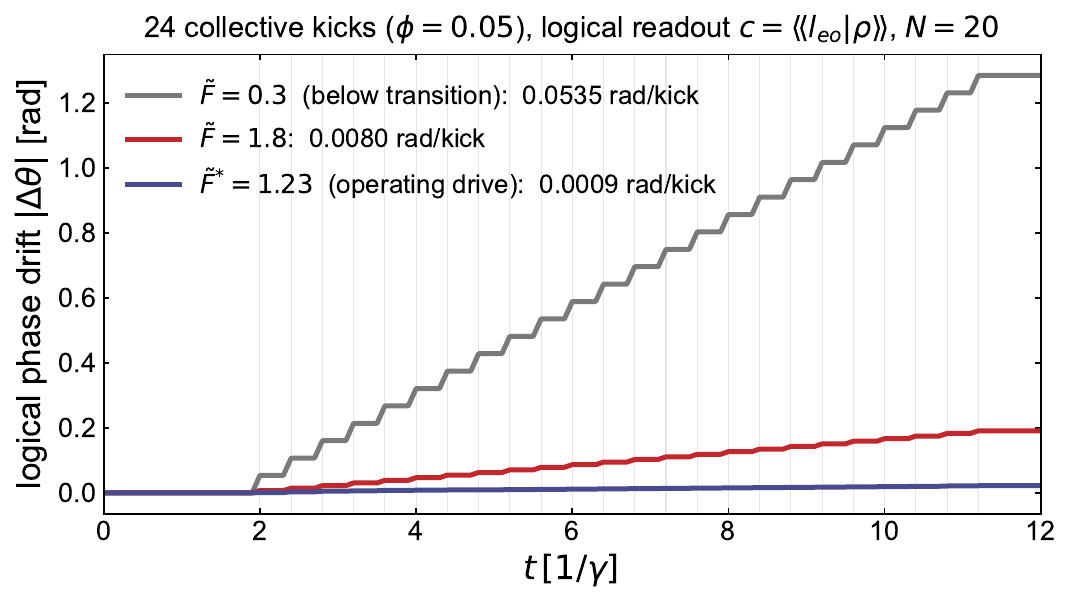}
    \caption{Kick response by drive ($N=20$): a train of 24 collective kicks ($\phi=0.05$, grey verticals). Accumulated logical phase drift $|\Delta\varphi|$: below the transition ($\tilde F=0.3$) the qubit takes essentially the full kick, $\approx0.054$~rad per kick. Above the transition at $\tilde F=1.8$ the phase drifts by $\approx0.008$~rad per kick, while at the operating drive $\tilde F^{*}=1.23$ the residual drift is $\approx0.0009$~rad per kick, an order of magnitude smaller since the linear kick response, and with it the leading dephasing coupling vanishes.}
    \label{fig:sfig_kickdrive}
\end{figure}

The coefficient $s$ is also the first-order effect of collective dephasing. The dephasing dissipator $\hat{\mathcal D}[L_d]$ of the main text is equivalent to a Gaussian accumulation of infinitesimal collective kicks. In the number basis a kick multiplies $\rho_{mn}$ by $e^{i\phi(m-n)}$, and averaging over a Gaussian angle of variance $\gamma_d t$ gives $\rho_{mn}\,e^{-\gamma_d t(m-n)^{2}/2}$, exactly the solution of $\dot\rho=\gamma_d\hat{\mathcal D}[\ntot]\rho$ \cite{open_quantum}. Each infinitesimal kick rotates the logical phase through the same factor $s$, so the phase noise of a collective bath reaches the qubit only to the extent that a single kick does. The same quantity therefore governs the protection against a macroscopic kick and against continuous collective dephasing, as we show next. Figure~\ref{fig:sfig_kickdrive} demonstrates the cancellation directly in the time domain: under a train of small kicks the accumulated logical phase drifts steadily below the phase transition at $\Ft=0.3$ but only residually above at $\Ft=1.23$ and $\Ft=1.8$. At the operating drive $\Ft=1.23$ the drift is minimized.

\paragraph{Origin of the decoupling point.---}\!\!\!\!\!\!\!
The collective bath couples only to $\ntot$. In the thermodynamic limit, where the four sector eigenoperators share a common factor [Eq.~(8) of the main text], the matrix element in Eq.~\eqref{eq:T12} reduces to a difference of sector expectation values, $\langle\cdot\rangle_{ii}\equiv\mathrm{Tr}[\,\cdot\;r_{ii}]$ with the trace-normalized sector states, with $\Delta n_{\mathrm{tot}}$ in the main-text convention (odd minus even),
\begin{equation}\label{eq:penetration}
s = \mathrm{Re}\,T_1 \;\xrightarrow{\;N\to\infty\;}\; \langle \ntot\rangle_{ee} - \langle \ntot\rangle_{oo} = -\Delta n_{\mathrm{tot}}:
\end{equation}
the leading-order effect of a collective phase error equals, up to sign, the total-photon-number imbalance between the two logical sectors. To see why the limit produces sector expectation values, split $\ntot$ into its sector means and the fluctuation about them using the parity operator $P$ of the main text, $\ntot=\bar n+\tfrac12\bigl(\langle\ntot\rangle_{ee}-\langle\ntot\rangle_{oo}\bigr)P+\delta n$, with $\bar n$ the average of the two sector values. The $\bar n$ term commutes with everything, and since $P\,r_{eo}=r_{eo}$ while $r_{eo}\,P=-r_{eo}$, the $P$ term gives the exact identity
\begin{equation}\label{eq:T1split}
[\ntot,\,r_{eo}]=\bigl(\langle\ntot\rangle_{ee}-\langle\ntot\rangle_{oo}\bigr)\,r_{eo}+[\delta n,\,r_{eo}].
\end{equation}
$T_1$ is the sector imbalance plus the matrix element of the fluctuation term. As $N\to\infty$ the four eigenoperators share one common factor [Eq.~(8) of the main text], so $\delta n$ acts only inside that shared factor and its matrix element reduces to the trace of a commutator, which vanishes. Only the imbalance survives. The bonding mode's $\approx3N$ photons are common to both sectors and cancel in the difference. What survives is an imbalance of about one photon, with two competing contributions: the photon content $\Delta n_A$ of the extra antibonding quasiparticle carried by the odd-odd sector, and the compensating response $\Delta n_B$ of the bonding mode, the same ``guardian'' response that restores the state in Fig.~5. If the bonding mode's compensation cancels the excitation's photon content, $\Delta n_B=-\Delta n_A$, the qubit decouples from collective phase noise at first order.

\paragraph{Bogoliubov theory of the sector photon numbers.---}\!\!\!\!\!\!\! We now restore the discarded orders one at a time, and each order adds an additional physical effect to be investigated. The goal of this is to construct the qubit's computational basis states step by step, which will turn out to be the eigenstates of a dressed quasiparticle. We will see that the quadratic terms dress the excitation and set its photon content, the cubic terms govern the response of the bonding mode, and the balance of the two determines the protection of the qubit against the collective phase error. But first, to do the full analysis, let's start by removing $\alpha_A$ from the differential equations altogether. Many times we have reported the antibonding mode to be small compared to the bonding mode, so this is a natural first step. All of our steps are listed below.

\begin{enumerate}
\item Remove $\alpha_A$ altogether: the classical bonding mode alone. This step captures the guardian's recovery, and it predicts no qubit protection at all.
\item Keep $\alpha_A$ to linear order: this gives us the oscillation frequency $\omega_A$.
\item Promote the antibonding amplitude back to an operator, $\alpha_A\to a_A$: the mode gets dressed, and its photon content becomes $\Delta n_A$. This is the step where the quantum enters.
\item Add all the higher order terms: the bonding mode responds with the screened compensation $\Delta n_B$. The balance of the last two steps decides the protection and cancels at $\Ft^{*}$.
\end{enumerate}

Step 1: We set $\alpha_A=0$, such that the problem collapses to the single bonding mode, where the recovery is fully visible and the protection is described fully classically.
The single-mode mean-field equation of motion for the bonding amplitude is
\begin{equation}\label{eq:naive-eom}
\dot\alpha = -i(\Delta+J)\,\alpha - \gamma\,\alpha - iU|\alpha|^{2}\alpha - \sqrt2\,iF,
\end{equation}
together with its complex conjugate. Two exact consequences follow. First, the modulus obeys
\begin{equation}\label{eq:naive-modulus}
\frac{d|\alpha|^{2}}{dt} = -2\gamma\,|\alpha|^{2} + \sqrt2\,iF\,(\alpha-\alpha^{*}),
\end{equation}
so the drive is the only source of photons and the loss the only drain. Second, the accumulated phase can be calculated without solving the dynamics. Consider the two integrals
\begin{equation}
A_{\alpha}=\int \frac{\dot\alpha}{\alpha}\,dt,
\qquad
B_{\alpha}=\int \frac{\dot\alpha^{*}}{\alpha^{*}}\,dt .
\end{equation}
We write $\alpha=re^{i\varphi}$, $A_{\alpha}=\ln(r_f/r_i)+i\Delta\varphi$ and $B_{\alpha}=\ln(r_f/r_i)-i\Delta\varphi$, so their difference gives the phase,
\begin{equation}\label{eq:naive-phase}
A_{\alpha}-B_{\alpha}=2i\,\Delta\varphi .
\end{equation}
Evaluated instead from the equation of motion, every term is explicit,
\begin{equation}
A_{\alpha}-B_{\alpha} = -2i(\Delta+J)\,t - 2iU\!\int\!|\alpha|^{2}\,dt - \sqrt2\,iF\!\int\!\left(\frac{1}{\alpha}+\frac{1}{\alpha^{*}}\right)dt ,
\end{equation}
and since $1/\alpha+1/\alpha^{*}=2\,\mathrm{Re}\,\alpha/|\alpha|^{2}$, comparing the kicked trajectory $\alpha$ with the unkicked reference $\alpha_{0}$ cancels the detuning term and leaves the permanent relative phase as a functional of the two trajectories alone,
\begin{equation}\label{eq:naive-boxed}
\Delta\Phi = -\,U\!\int\!\left(|\alpha|^{2}-|\alpha_{0}|^{2}\right)dt
\;-\;\sqrt2\,F\!\int\!\left(\frac{\mathrm{Re}\,\alpha}{|\alpha|^{2}}-\frac{\mathrm{Re}\,\alpha_{0}}{|\alpha_{0}|^{2}}\right)dt .
\end{equation}

Thus the entire mean-field kick response accumulates during the recovery alone, from how much brighter the kicked trajectory momentarily is than the reference (the Kerr term) and from how much it is misaligned with the drive (the drive term). The bonding phase itself always recovers, as the drive locks it back to the fixed point. The two integrals total exactly $-\phi$ and the recovery generates the compensating phase that undoes the kick.

Step 2: We allow up to linear terms now. We work with the intensive ($N$-independent) amplitude $\beta=\langle a_B\rangle/\sqrt{N}$, $x=|\beta|^{2}$, and the effective detuning $\delta=\Delta+J$. Keeping the dark amplitude to linear order, the two mean-field equations read
\begin{equation}\label{eq:linear-eom}
\begin{aligned}
\dot\alpha_B &= -i(\Delta+J)\,\alpha_B - \gamma\,\alpha_B - iU|\alpha_B|^{2}\alpha_B - \sqrt2\,iF,\\
\dot\alpha_A &= -i\,w_0\,\alpha_A - i\,U\alpha_B^{2}\,\alpha_A^{*},
\qquad w_0 = J-\Delta+2U|\alpha_B|^{2}.
\end{aligned}
\end{equation}

In the linear order the antibonding mode has no effect on the bonding mode. Every antibonding term in the bonding equation is quadratic, so the bonding amplitude stays frozen at the fixed point of Step 1. The antibonding amplitude, coupled to its own conjugate, oscillates at the time crystal frequency $\omega_A=\sqrt{w_0^2-(\tilde Ux)^2}$ of Eq.~(5) of the main text as in the derivation from Sec.~\ref{app:scconv}. This is a bare harmonic ripple of the antibonding mode on a frozen bonding state, with no photon content and no response from the bonding mode.

Step 3: Now we move from classical to quantum as the antibonding amplitude is promoted back to an operator, $\alpha_A\to a_A$, and the calculation of Sec.~\ref{app:scconv} is repeated at the operator level, the same quadratic problem but with $a_A$ and $a_A^{\dagger}$ kept as operators rather than replaced by amplitudes.

\begin{equation}\label{eq:step3-eom}
\begin{aligned}
\dot{\alpha}_B &= -i(\Delta+J)\,\alpha_B - \gamma\,\alpha_B - iU|\alpha_B|^{2}\alpha_B - \sqrt{2}\,iF,\\
\dot{a}_A &= -i\,w_0\,a_A - i\,U\alpha_B^{2}\,a_A^{\dagger}.
\end{aligned}
\end{equation}

The frequency will necessarily be the same, but now we get the photon content of the vacuum and of the excitation because of the nonzero commutator. The semiclassical fixed point [Eq.~\eqref{eq:sm_sc_fixedpoint} of Sec.~\ref{app:scconv}] satisfies the cubic
\begin{equation}\label{eq:cubic}
x\left[\gamma^2 + (\delta - \Ut x)^2\right] = 2\Ft^2,
\end{equation}
of which we take the stable branch. Displacing $a_B \to \sqrt{N}\beta + \delta a_B$ in Eq.~(1) of the main text and keeping the terms quadratic in the antibonding mode gives the effective antibonding-mode Hamiltonian,
\begin{equation}\label{eq:HA2}
H_A^{(2)} = w_0\, a_A^\dagger a_A + \tfrac{\Ut}{2}\!\left(\beta^2 a_A^{\dagger 2} + \mathrm{h.c.}\right),
\quad
w_0 = J - \Delta + 2\Ut x.
\end{equation}
The discarded higher order terms are small and become zero in the thermodynamic limit. The antibonding self-Kerr is quartic with coefficient $\Ut/N$ acting on states of about one antibonding photon at most, and the mixed terms with one bonding-mode fluctuation carry $1/\sqrt N$. These produce the small back-action on the bonding mode and will be explained below. A Bogoliubov rotation \cite{bogoliubov_superfluidity_1947, pitaevskii_stringari_2016} introduces the quasiparticle annihilation operator
\begin{equation}\label{eq:bogq}
q=\cosh(r)\,a_A+e^{i\theta}\sinh(r)\,a_A^{\dagger},
\qquad \tanh 2r = \Ut x/w_0,
\end{equation}
with the phase $\theta$ set by the bonding phase $\arg\beta^{2}$, which diagonalizes Eq.~\eqref{eq:HA2} into a plain oscillator, $H_A^{(2)}=\omega_A\,q^{\dagger}q$ up to a constant, with the frequency
\begin{equation}\label{eq:omegaA}
\omega_A = \sqrt{w_0^2 - (\Ut x)^2},
\end{equation}
which reproduces Eq.~(5) of the main text. For a quadratic Hamiltonian the Bogoliubov excitation frequency coincides with the classical linear-response frequency obtained from the linearization of Sec.~\ref{app:scconv} (with the same $\omega_A$). Because the loss channel never acts on $a_A$ and $H_A^{(2)}$ commutes with $q^{\dagger}q$, the quasiparticle number cannot change at this order, and the two logical sectors are the quasiparticle vacuum in the even-even sector and the single-quasiparticle state in the odd-odd one.

The rotation of Eq.~\eqref{eq:bogq} dresses the quasiparticle, mixing the bare photon with its conjugate, so the stationary states are not bare Fock states but squeezed combinations, with the bonding amplitude supplying the mixing through the $\beta^{2}a_A^{\dagger2}$ term of Eq.~\eqref{eq:HA2}. Inverting the rotation, $a_A=\cosh(r)\,q-e^{i\theta}\sinh(r)\,q^{\dagger}$, and substituting into $n_A=a_A^{\dagger}a_A$ gives
\begin{equation}
n_A=\cosh 2r\;q^{\dagger}q+\sinh^{2}r-\tfrac{1}{2}\sinh 2r\left(e^{i\theta}q^{\dagger 2}+\mathrm{h.c.}\right),
\end{equation}
so a state of $m$ quasiparticles carries $\langle n_A\rangle_m=\cosh 2r\,m+\sinh^{2}r$ bare photons, and the difference between the one and zero quasiparticle states is
\begin{equation}\label{eq:qA}
\Delta n_A = \cosh 2r = \frac{w_0}{\omega_A} = 1.137,
\end{equation}
rather than one bare photon. We measure this difference to be $+1.1355$ at $\Ft=1.8$, $N=40$, and it is nearly flat in $\Ft$ and $N$ because $\cosh 2r$ is a ratio of two intensive quantities.

The quasiparticle also shifts $\langle a_A^{2}\rangle$ as each excitation changes it by $-\Ut\beta^{2}/\omega_A$, and its phase is fixed to that of $\beta^{2}$, since in the steady state the equation of motion for $\langle a_A^{2}\rangle$ is sourced only by the $\beta^{2}$ term.

The quasiparticle operator $q$ of Eq.~\eqref{eq:bogq} is the antibonding mode mixed with its own conjugate, so that $q^{\dagger}$ creates one dressed quasiparticle. Put simply $q$ is to the quasiparticle what $a$ is to a photon. Within the truncation where the Hamiltonian is $H_A^{(2)}$ alone, $[H_A^{(2)}, q^{\dagger}]=\omega_A\,q^{\dagger}$, and $[L, q^{\dagger}]=[L^{\dagger}, q^{\dagger}]=0$ holds exactly at every $N$, since the loss lives on the bonding mode and $q$ is built from only the antibonding mode. Now apply $q^{\dagger}$ to the steady state. It adds one dark quantum, so the ket side flips to odd parity while the bra side stays even. $q^{\dagger}\rho_{ss}$ lands in the even-odd block, exactly where the logical coherence $c$ lives. At finite $N$ the discarded $1/\sqrt N$ terms break the Hamiltonian half of the algebra, which is why the coherence decays at $\propto1/N$ instead of living forever, and we investigate this effect next.

Step 4: Here we add all the terms. The two mean-field equations of Eq.~\eqref{eq:linear-eom} become
\begin{equation}\label{eq:full-eom}
\begin{aligned}
\dot\alpha_B &= -i(\Delta+J)\,\alpha_B - \gamma\,\alpha_B - iU|\alpha_B|^{2}\alpha_B - \sqrt2\,iF - 2iU\,a_A^{\dagger}a_A\,\alpha_B - iU\alpha_B^{*}a_A^{2},\\
\dot{a}_A &= -i\,w_0\,a_A - i\,U\alpha_B^{2}\,a_A^{\dagger} - iU\,a_A^{\dagger}a_A\,a_A.
\end{aligned}
\end{equation}
The cubic ones bring in the back-action. Both shifts act back on the bonding mode through the cross-Kerr and pair-exchange terms of Eq.~(1) of the main text, producing a source smaller by $1/N$ in the equation of motion for $\beta$. The phase locking makes the source exactly detuning-like as the equation of motion for $\beta$ acquires the term $\frac{i}{N}\, c_{\mathrm{eff}}\,\beta$, where
\begin{equation}\label{eq:ceff}
c_{\mathrm{eff}} = 2\cosh 2r - \sinh 2r = \frac{0.6 + 3x}{\omega_A},
\end{equation}
so the odd-odd bonding state is the even-even bonding state at a detuning shifted by $-c_{\mathrm{eff}}/N$, and the response collapses to the static compressibility of the cubic \eqref{eq:cubic},
\begin{equation}\label{eq:dnB}
\Delta n_B = -c_{\mathrm{eff}}\, \frac{dx}{d\delta}.
\end{equation}
The $\langle a_A^{2}\rangle$ shift is essential here. The occupation enters the bonding equation through the $2U a_A^{\dagger}a_A\,\alpha_B$ term of Eq.~\eqref{eq:full-eom} and alone would replace the bare coupling $2$ by $2\cosh 2r = 2.27$, an overshoot. The $\langle a_A^{2}\rangle$ shift enters through the $U\alpha_B^{*}a_A^{2}$ term and reduces the sum to $c_{\mathrm{eff}} = 2\cosh 2r - \sinh 2r = 1.73$, turning the naive mean-field prediction $|\Delta n_B| = 1.50$ (at $\Ft = 1.8$) into $1.297$ (which we measure to be $1.306$ at $N = 100$).

\begin{figure}[!tb]
\centering
\includegraphics[width=0.52\linewidth]{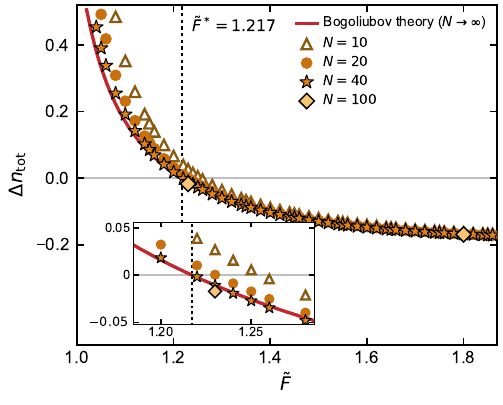}
\caption{Sector photon-number imbalance $\Delta n_{\mathrm{tot}}$ as a function of the drive. The ab initio theory of Eq.~\eqref{eq:dntot} (line) against the quantum eigenoperators (symbols). Inset: the decoupling region magnified. The zero crossing $\Ft^{*}=1.217$ (dotted line) is the decoupling point at which the linear response to collective phase errors is tuned through zero.}
\label{fig:sweetspot}
\end{figure}

Collecting the two contributions,
\begin{equation}\label{eq:dntot}
\Delta n_{\mathrm{tot}}(\Ft) = \Delta n_A - c_{\mathrm{eff}}\,\frac{dx}{d\delta},
\end{equation}
with $x(\Ft)$ from Eq.~\eqref{eq:cubic}. This function, shown in Fig.~\ref{fig:sweetspot}, crosses zero at
\begin{equation}\label{eq:fstar}
\Ft^{*} = 1.217,
\end{equation}
inside the window $1.21$ to $1.24$ where the quantum crossings are found. At $\Ft^{*}$ the guardian's screened compensation exactly balances the dressed quasiparticle's photon content, and by Eq.~\eqref{eq:penetration} the linear response to any collective phase error is tuned through zero.

We note that to obtain the Bogoliubov prediction we approximated the bonding mode as a macroscopic bonding state with small fluctuations. This holds only for the high-drive phase and in the thermodynamic limit as $N\to\infty$, as no macroscopic bonding state exists below the transition. The eigenoperators we got from diagonalizing the Liouvillian thus cover a wider range of drives in addition to finite size fluctuations, and match Bogoliubov theory in the NS phase.

The quantum fluctuations around the bonding mode also have an effect. The quasiparticle couples to the damped fluctuation $\delta a_B$ through the mixed terms discarded in Step 3, and so inherits a damping of its own, as whatever mixes into $\delta a_B$ is carried away by the loss. This is the quasiparticle damping $\kappa(N)$ of Sec.~\ref{sec:coh}, quadratic in the $1/\sqrt{N}$ coupling and hence falling as $1/N$. The steady state of the driven, lossy bonding mode leaves $\delta a_B$ with nonzero occupation, so the same coupling also works in reverse and feeds quanta into the quasiparticle mode. The population settles where this feed balances with the damping. Both rates fall as $1/N$ together, so the balance point does not change with $N$. This is why the purities of the antibonding reductions saturate [Fig.~\ref{fig:sfig_red}(b)] instead of approaching one.

\section{Treating Dephasing as a Perturbation in the Leading Order}\label{sec:deph}

\begin{figure}[!t]
    \centering
    \includegraphics[width=0.98\linewidth]{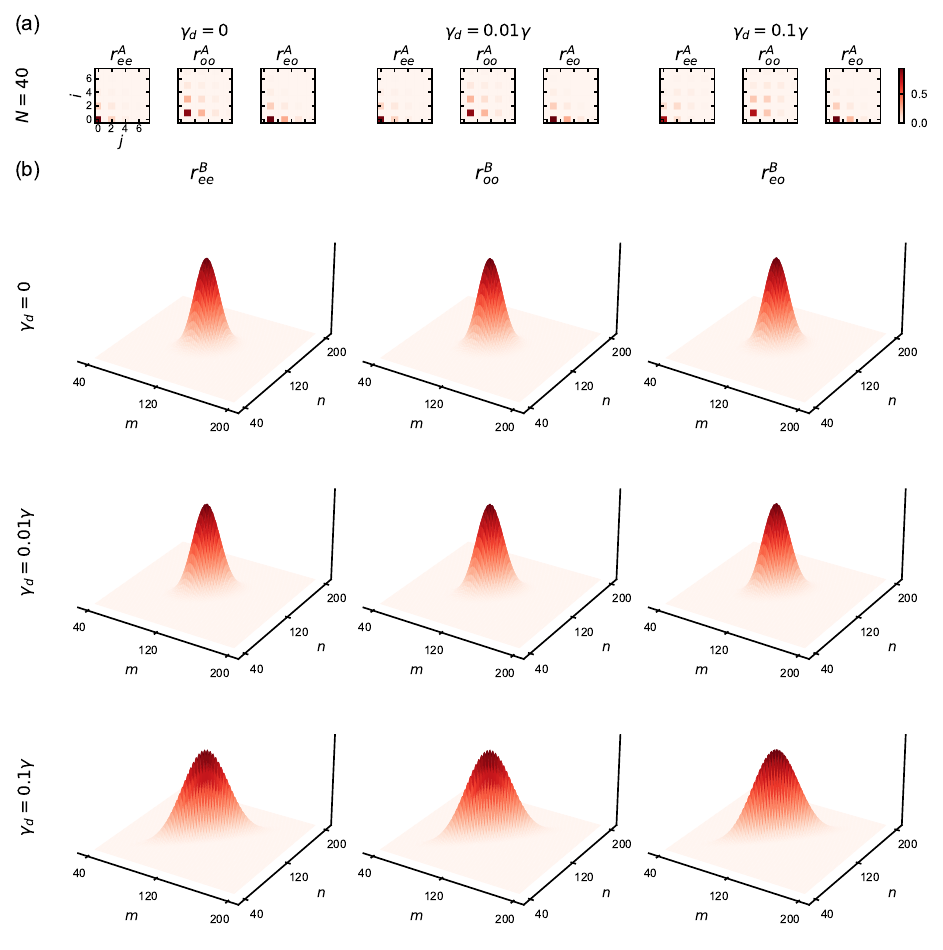}
    \caption{Mode reductions of all the sectors ($N=40$, $\tilde F=1.8$) for no dephasing, for the rate in the main text $\gamma_d=0.01\gamma$, and $\gamma_d=0.1\gamma$.
    \textbf{(a)} Antibonding reductions $|r^{A}_{ij}|$ on a shared color scale, barely affected by dephasing.
    \textbf{(b)} Bonding reductions $|r^{B}_{ij}|$. Within every row the three sectors carry the same macroscopic bonding state, peaked at $n_B\approx119$. Increasing the dephasing rate broadens that bonding-mode band along the Fock diagonal without disturbing the sector structure. The broadening is also what makes large-$N$ dephased computations costly, and at $N=40$ the distribution still remains well inside the $K_B=200$ window.}
    \label{fig:sfig3}
\end{figure}

To estimate the effect of collective dephasing in the leading order, we apply perturbation theory \cite{li_perturbative_2014, open_quantum}. We consider the perturbed Liouvillian of the form
\begin{equation}
    \mathcal{L} = \mathcal{L}_0 + \alpha\,\mathcal{L}_1,
\end{equation}
where $\alpha \ll 1$ is a small parameter.
Right and left eigenvectors of $\mathcal{L}_0$ obey
\begin{equation}
    \mathcal{L}_0 | r_j^{(0)} \rangle\!\rangle
        = \lambda_j^{(0)} | r_j^{(0)} \rangle\!\rangle,
    \qquad
    \mathcal{L}_0^\dagger | l_j^{(0)} \rangle\!\rangle
        = \lambda_j^{(0)\,*} | l_j^{(0)} \rangle\!\rangle,
\end{equation}
with Hilbert--Schmidt normalization
\begin{equation}
    \langle\!\langle l_j^{(0)} | r_k^{(0)} \rangle\!\rangle
    \equiv \mathrm{Tr}\!\left[(l_j^{(0)})^\dagger r_k^{(0)}\right]
    = \delta_{jk}.
\end{equation}
We now look for perturbed eigenvalues and eigenvectors of $\mathcal{L}$ in the form of power series in~$\alpha$:
\begin{align}
    \lambda_j &= \lambda_j^{(0)} + \alpha\,\lambda_j^{(1)} + O(\alpha^2),\\
    | r_j \rangle\!\rangle
        &= | r_j^{(0)} \rangle\!\rangle
         + \alpha\,| r_j^{(1)} \rangle\!\rangle
         + O(\alpha^2).
\end{align}
Inserting this expansion into the perturbed eigenvalue equation and keeping terms up to first order in~$\alpha$ gives

\begin{equation}
\begin{split}
    &(\mathcal{L}_0 + \alpha\,\mathcal{L}_1)
    \bigl(| r_j^{(0)} \rangle\!\rangle
        + \alpha\,| r_j^{(1)} \rangle\!\rangle\bigr) =
    \bigl(\lambda_j^{(0)} + \alpha\,\lambda_j^{(1)}\bigr)
    \bigl(| r_j^{(0)} \rangle\!\rangle
        + \alpha\,| r_j^{(1)} \rangle\!\rangle\bigr)
        + O(\alpha^2).
\end{split}
\end{equation}

Expanding and grouping terms by powers of $\alpha$ yields in the zeroth order:

\begin{equation}
    \mathcal{L}_0 | r_j^{(0)} \rangle\!\rangle
    = \lambda_j^{(0)} | r_j^{(0)} \rangle\!\rangle,
\end{equation}
which is just the unperturbed eigenvalue equation. The terms proportional to $\alpha$ on the other hand give us in first (leading) order
\begin{equation}
    \mathcal{L}_0 | r_j^{(1)} \rangle\!\rangle
    + \mathcal{L}_1 | r_j^{(0)} \rangle\!\rangle
    =
    \lambda_j^{(0)} | r_j^{(1)} \rangle\!\rangle
    + \lambda_j^{(1)} | r_j^{(0)} \rangle\!\rangle.
\end{equation}
Rearranging, we get
\begin{equation}
    (\mathcal{L}_0 - \lambda_j^{(0)}) | r_j^{(1)} \rangle\!\rangle
    =
    -\bigl(\mathcal{L}_1 - \lambda_j^{(1)}\bigr)
     | r_j^{(0)} \rangle\!\rangle.
    \label{eq:first_order_vector_equation}
\end{equation}
To extract $\lambda_j^{(1)}$ we now left--multiply
Eq.~\eqref{eq:first_order_vector_equation} by $\langle\!\langle l_j^{(0)}|$.
The left-hand side of Eq.~\eqref{eq:first_order_vector_equation} vanishes,
and we obtain
\begin{equation}
    0
    = - \langle\!\langle l_j^{(0)} |
        \bigl(\mathcal{L}_1 - \lambda_j^{(1)}\bigr)
        | r_j^{(0)} \rangle\!\rangle,
\end{equation}
and therefore
\begin{equation}
    \langle\!\langle l_j^{(0)} | \mathcal{L}_1 | r_j^{(0)} \rangle\!\rangle
    = \lambda_j^{(1)}
      \langle\!\langle l_j^{(0)} | r_j^{(0)} \rangle\!\rangle.
\end{equation}
Using the normalization
$\langle\!\langle l_j^{(0)} | r_j^{(0)} \rangle\!\rangle = 1$,
we finally obtain the first--order correction
\begin{equation}
    \lambda_j^{(1)} =
    \langle\!\langle l_j^{(0)} | \mathcal{L}_1 | r_j^{(0)} \rangle\!\rangle
    = \mathrm{Tr}\!\left[
        (l_j^{(0)})^\dagger\,\mathcal{L}_1(r_j^{(0)})
    \right].
\end{equation}
This is the Liouvillian analogue of first-order perturbation theory for Hamiltonian eigenvalues. The derivation assumes the unperturbed eigenvalue $\lambda_j^{(0)}$ is nondegenerate and that the biorthogonal eigenbasis is complete \cite{minganti2018spectral}, both hold for the isolated slow eigenvalues used here. This condition does not hold for the diagonal sectors as their eigenvalues are degenerate ($\lambda_{ee} = \lambda_{oo} = 0 $), which we investigate in the next section.

For the collective dephasing of the main text the perturbation is $\mathcal L_1=\hat{\mathcal D}[\ntot]$, with $\ntot=a_A^{\dagger}a_A+a_B^{\dagger}a_B$ the total photon number, and the small parameter is the dephasing rate, $\alpha=\gamma_d$. The real and imaginary parts of $\lambda^{(1)}$ have clear physical meanings: $ -\mathrm{Re}\{\,\lambda^{(1)}\}$ is the additional decay rate and  $\mathrm{Im}\,\{\lambda^{(1)}\}$ is the frequency shift due to dephasing. The states themselves, at the largest solved size and under dephasing, are shown in Fig.~\ref{fig:sfig3}. In the main text, Fig.~4(a) reports the matrix element $\lambda^{(1)}$ itself, the shift per unit dephasing rate.

\section{Asymmetric and Non-Collective Errors}\label{sec:asym}

\paragraph{Taxonomic classification of errors.---}\!\!\!\!\!\!\! We investigated the protection of the qubit against collective and strong symmetric errors, specifically in the form of photon loss (dissipation) and dephasing. However experiments in the real world usually do not have such amiable error structures. This section describes a more general taxonomy of the qubit. We investigate error channels fixed by three binary choices. Whether it is a loss (the jump is an annihilation operator) or a dephasing (the jump is a number operator), acts in the beam basis or in the lab basis, or is collective or local. Collective and local are properties of the dissipator, not of the operators. A collective channel is a single jump into one shared bath, a local channel is a sum of independent dissipators, one bath per mode. Table~\ref{tab:errorset} lists the resulting families, with the asymmetry $\eta$ interpolating within each.

\begin{table}[h]
\centering
\small
\begin{tabular}{llcc}
\hline
 & & collective (one bath) & local (independent baths) \\
\hline
loss & beam & $\mathcal{D}[a_B{+}\eta a_A]$ & $\mathcal{D}[a_B]+\mathcal{D}[\eta a_A]$ \\
 & lab & $\mathcal{D}[a_1{+}\eta a_2]$ & $\mathcal{D}[a_1]+\mathcal{D}[\eta a_2]$ \\
dephasing & beam & $\mathcal{D}[n_B{+}\eta n_A]$ & $\mathcal{D}[n_B]+\mathcal{D}[\eta n_A]$ \\
 & lab & $\mathcal{D}[n_1{+}\eta n_2]$ & $\mathcal{D}[n_1]+\mathcal{D}[\eta n_2]$ \\
\hline
\end{tabular}
\caption{The error set at the dissipator level.}
\label{tab:errorset}
\end{table}

\paragraph{Imperfect collective dephasing.---}\!\!\!\!\!\!\! The dephasing coupling of Fig.~4(a) of the main text was computed not only at the balanced point but also for $\eta=0.9$. To avoid confusion, we call this an imbalance rather than asymmetry as any $\eta$ in this form still leads to strong symmetric errors. For every $\eta$ the coupling $L_d^{\eta}$ is built from the two number operators, so it still commutes with the parity $P$. Thus the exact bit is untouched, and only the weight of the noise on the two modes becomes realistic.

This realistic form also brings a freedom for the experimentalist along with this imbalance. The decoupling condition depends on the drive, so instead of demanding that the hardware deliver $\eta=1$, one can measure the asymmetry in the experiment and choose the drive that decouples it.

For the asymmetric collective coupling $L_d^{\eta}=n_B+\eta\,n_A$ the leading-order decay is proportional to $\Delta n_B+\eta\,\Delta n_A$, which is tuned through zero at
\begin{equation}
\eta^{*}(\Ft)=-\frac{\Delta n_B(\Ft)}{\Delta n_A(\Ft)}.
\end{equation}

\begin{figure}[h]
    \centering
    \includegraphics[width=0.58\linewidth]{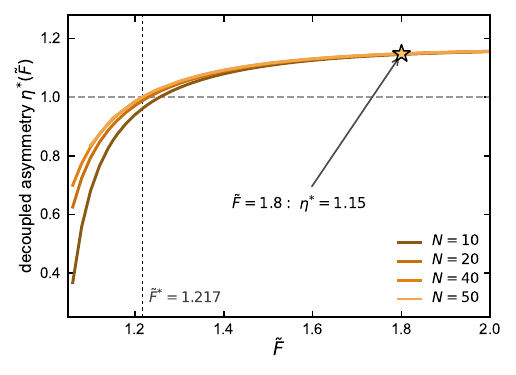}
    \caption{The decoupling asymmetry $\eta^{*}(\tilde F)=-\Delta n_B/\Delta n_A$ for $N=10$ to $50$: the drive that cancels the leading-order dephasing decay of $L_d^{\eta}$. $\eta^{*}=1$ at the symmetric operating drive. The star marks $\tilde F=1.8$, which corrects $\eta^{*}=1.15$.}
    \label{fig:sfig_etastar}
\end{figure}

Figure~\ref{fig:sfig_etastar} shows the decoupling map for $N=10,20,40,50$ (the sector photon differences themselves, with the $\Delta n_{\rm tot}$ crossings converging onto Eq.~\eqref{eq:fstar}, are plotted in the End Matter of the main text). These four sizes collapse onto one curve, which rises from the transition, crosses the symmetric point $\eta^{*}=1$ at the operating drive, and saturates near $1.16$. Thus every asymmetry in the band $\eta\approx0.8$ to $1.15$ has its own perfect-decoupling drive in $\Ft\in[1.1,2.0]$, and the operating point can be matched to the hardware's coupling asymmetry rather than the reverse giving rise to more experimental bandwidth. For the purpose of our analysis we chose the magic drive to be $F^*=1.23$ as $N=20$ crosses the $\eta^{*}=1$ at that point.

\paragraph{Bit protection with degenerate perturbation theory.---}\!\!\!\!\!\!\! We mostly focused on protecting the coherence, taking the bit protection for granted (and exact) due to the strong symmetry. But now let us look at the bit. Its protection cannot be checked with the perturbation theory of Sec.~\ref{sec:deph}, because the bit lives in a degenerate pair of exact zero modes thus we need degenerate perturbation theory. We show that against the unbalanced drain the bit too is protected at first order, by the same parity selection rule, and bounded at second order.

The perturbation theory of Sec.~\ref{sec:deph} does not apply to the logical populations, because the unperturbed eigenvalue is doubly degenerate. $r_{ee}$ and $r_{oo}$ are both exact zero modes, so every mixture $b\,r_{ee}+(1-b)\,r_{oo}$ is stationary and the perturbation itself decides how the pair splits. The correct tool is degenerate perturbation theory in the zero subspace, the eigenvalues of the $2\times2$ matrix
\begin{equation}
M_{ij}=\langle\langle l_i|\hat{\mathcal{L}}_1|r_j\rangle\rangle,
\qquad i,j\in\{ee,oo\},
\label{eq:sm_degpt}
\end{equation}
per unit channel rate, with $\hat{\mathcal{L}}_1$ the perturbing dissipator of the expansion in Sec.~\ref{sec:deph}, and where the left zero modes $l_{ee}$ and $l_{oo}$ are exact.\footnote{They are simply the parity projectors, the identity restricted to each sector. Since every jump operator commutes with $P$, each sector's probability is separately conserved, and a conserved observable is a left zero mode. This makes the left eigenoperators exact at every $N$, in contrast to the right zero modes, the sector steady states, which carry all the $N$-dependent structure.} Acting on an operator they return the trace of its even or odd parity block, because the evolution preserves the trace within each sector. For every parity-preserving (strong-symmetric) channel, collective loss and collective dephasing (for dephasing $L_d^{\eta}$ at any imperfection $\eta$ included), $\hat{\mathcal{L}}_1$ maps each diagonal block to itself and acts trace-preservingly there, so all four entries are zero and the bit is untouched at leading order.

The unbalanced drain (asymmetric loss) of the End Matter, $L=a_B+\eta\,a_A$, is different. Its entries of first order in $\eta$ are still zero by the parity selection rule of Eq.~(A2) of the End Matter, every term of the cross term $\hat{\mathcal{C}}$ moving the diagonal blocks into the coherence blocks, whose sector traces are zero. At order $\eta^{2}$, however, the jump $a_A\rho\,a_A^{\dagger}$ has a finite trace. Physically this removes one dark photon, thus it flips the antibonding parity moving a state from the even sector to the odd one and vice-versa.

To evaluate Eq.~\eqref{eq:sm_degpt} we need two numbers: the dark photon occupations of the two sectors. We measure them from the $N=100$ sector steady states at $\Ft=1.8$:
\begin{equation}
\langle n_A\rangle_e\approx0.16,
\qquad
\langle n_A\rangle_o\approx1.30.
\label{eq:sm_occupations}
\end{equation}
Their difference, $1.30-0.16=1.14$, is the quasiparticle photon content $\cosh 2r$ derived with the Bogoliubov theory of Sec.~\ref{sec:qf}. The photon loss rate in a sector is its antibonding occupation,
\begin{equation}
\mathrm{rate}(even\to odd)=\gamma\eta^{2}\langle n_A\rangle_e,
\qquad
\mathrm{rate}(odd\to even)=\gamma\eta^{2}\langle n_A\rangle_o,
\label{eq:sm_rates}
\end{equation}
and Eq.~\eqref{eq:sm_degpt} assembles the two rates into
\begin{equation}
M=\eta^{2}\begin{pmatrix}-\langle n_A\rangle_e & +\langle n_A\rangle_o\\[2pt] +\langle n_A\rangle_e & -\langle n_A\rangle_o\end{pmatrix}
\label{eq:sm_bitmatrix}
\end{equation}
a Markov matrix for plain classical probabilities in units of the loss rate $\gamma$. Column by column, probability leaves a sector at that sector's loss rate and lands in the other, so each column sums to $0$. The zero eigenvalue (not to be confused with our usual eigenoperators, these are just the old-school eigenvalues of $M$) is the conservation of total probability, and its eigenvector is the new steady state. The logical weight $b$ of Eq.~(4) of the main text, exactly frozen until now, acquires the steady value $b_{\rm ss}=\langle n_A\rangle_o/(\langle n_A\rangle_e+\langle n_A\rangle_o)\approx0.89$. The drain empties the antibonding mode, so it slowly resets the qubit towards the vacuum-like logical $|0\rangle$, emptying the odd sector about eight times faster than the even one. The second eigenvalue is the rate at which any imbalance relaxes to that value, as the bit decays at $\Gamma_{\rm bit}=\gamma\eta^{2}(\langle n_A\rangle_e+\langle n_A\rangle_o)\approx1.46\,\gamma\eta^{2}$. Thus at first order in the imbalance there is no effect on the bit, but at second order the bit (slowly) decays suppressed by $\eta^{2}$, and at a rate comparable to the coherence floor $\gamma\eta^{2}|t_A|$, with $t_A=\langle\!\langle l_{eo}|\mathcal{D}[a_A]|r_{eo}\rangle\!\rangle$ the penalty per unit of antibonding-mode loss. Against that coherence floor this is the relation $T_2\approx2T_1$ of a biased amplitude-damping channel.

\paragraph{General asymmetric and non-collective errors.---}\!\!\!\!\!\!\! The channels so far were written in the beam basis. Physical errors act on the two sites, so we re-introduce our Hamiltonian in the lab basis,
\begin{equation}
    \hat{H} = F (\hat{a}_1 + \hat{a}_2 + \hat{a}^\dagger_1 + \hat{a}^\dagger_2) -\Delta\hat{a}^\dagger_1\hat{a}_1 -\Delta\hat{a}^\dagger_2\hat{a}_2 - J(\hat{a}^\dagger_1\hat{a}_2 +\hat{a}_1\hat{a}^\dagger_2) + \sum_{1,2}(U \hat{a}^\dagger_i \hat{a}^\dagger_i \hat{a}_i \hat{a}_i)
\end{equation}
for modes $a_1$ and $a_2$, related to the beam modes by
\begin{equation}
a_1=\frac{a_B+a_A}{\sqrt{2}},\qquad a_2=\frac{a_B-a_A}{\sqrt{2}}.
\end{equation}

A sum of dissipators is invariant under the beam-splitter rotation, $\mathcal{D}[a_1]+\mathcal{D}[a_2]=\mathcal{D}[a_B]+\mathcal{D}[a_A]$, so local lab loss is exactly collective bonding loss from our original system, plus antibonding-mode loss. Dephasing does bring new channels, because number operators do not rotate linearly. The sum is basis invariant, $n_1+n_2=n_A+n_B$, but the difference is the hopping $n_1-n_2=a_B^{\dagger}a_A+a_A^{\dagger}a_B$, which are the cross terms that perturb the strong parity symmetry.

\begin{figure}[t]
    \centering
    \includegraphics[width=\linewidth]{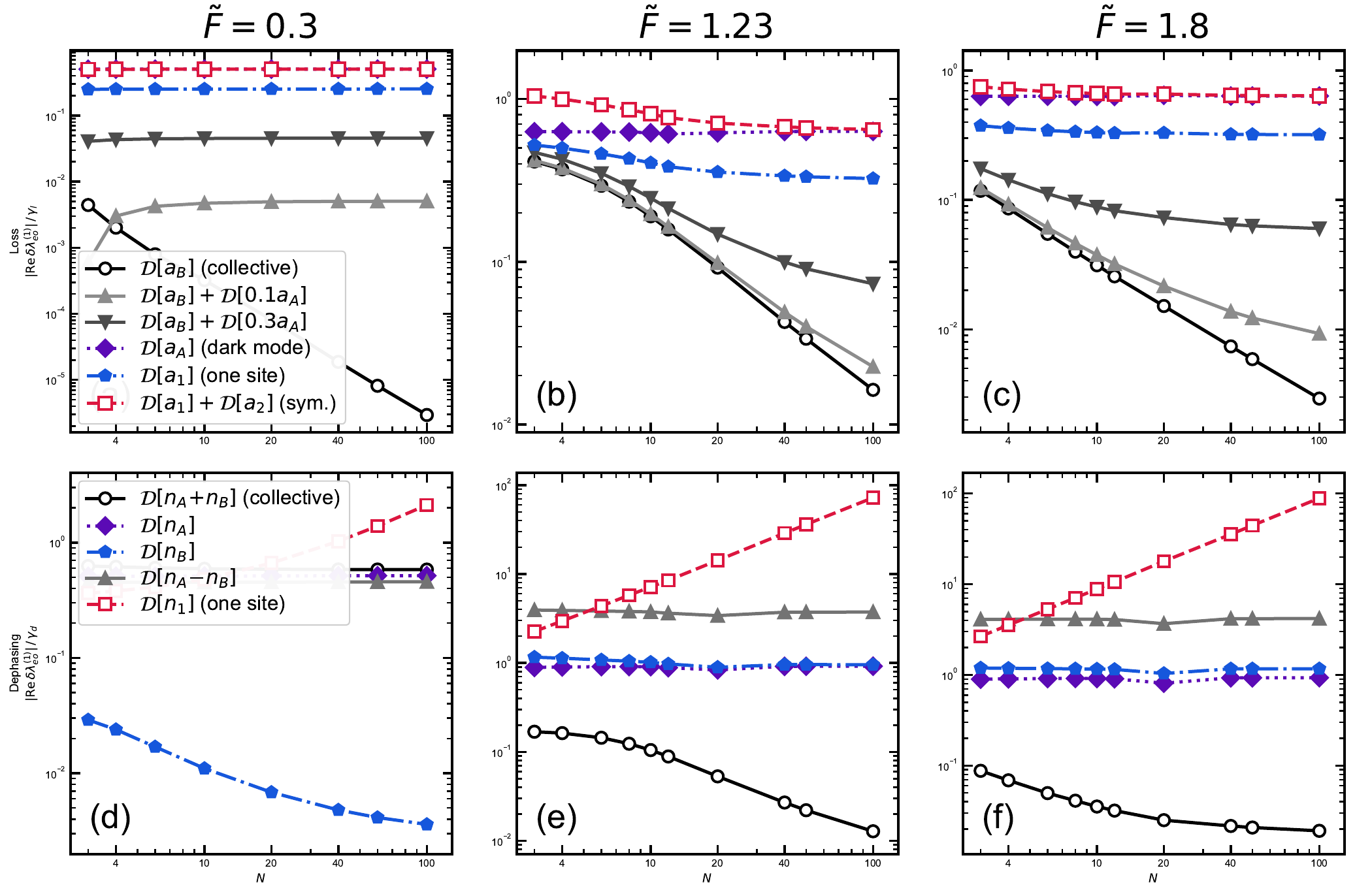}
    \caption{First-order coherence decay $-\mathrm{Re}\,\delta\lambda_{eo}$ (per unit rate) for the full list of loss and dephasing channels versus $N$, at $\tilde F=0.3$ (data to $N=200$), $1.23$ and $1.8$ (to $N=100$), classes as in Table~\ref{tab:channels}. Protected channels fall with $N$, beam-local channels are flat, and site-local dephasing $\mathcal{D}[n_1]$ grows $\propto N$.}
    \label{fig:sfig_channels}
\end{figure}

We evaluate below the first-order coherence decay $-\mathrm{Re}\,\delta\lambda_{eo}$ per unit channel rate for each error. Figure~\ref{fig:sfig_channels} shows all of them. We allocate the error channels into three classes according to how their cost behaves towards the thermodynamic limit: (i) \emph{protected:} the cost falls with $N$, (ii) \emph{bounded:} settles onto an $N$-independent floor, or (iii) \emph{extensive:} grows with $N$. Throughout, $t_B=\langle\!\langle l_{eo}|\mathcal{D}[a_B]|r_{eo}\rangle\!\rangle$ and $t_A=\langle\!\langle l_{eo}|\mathcal{D}[a_A]|r_{eo}\rangle\!\rangle$ are the bonding and antibonding mode loss matrix elements as defined before. The non-strongly symmetric loss case is derived in full in the End Matter of the main text.

\emph{(i) Protected: the cost falls with $N$.} Two channels we already investigated are passively protected. Engineered collective loss $\mathcal{D}[a_B]$, whose cost is $t_B$ itself, falls with $N$ at every drive. The antibonding mode does not dissipate under $a_B$, and the attractor restores the displaced bonding state. Collective dephasing $\mathcal{D}[n_B+\eta n_A]$ is protected too at the leading order, but only above the transition for a range of $\eta$, where its cost falls as $1/N$ due to the cancellation with the photon-counting compensation. The leading order effect here is small and can be made zero by tuning $F$.

\emph{(ii) Bounded: an $N$-independent floor.} For this category, let us first investigate non-collective (local) errors in the beam basis. For these errors, the first-order cost stops falling and settles at a constant, so increasing $N$ no longer helps \cite{buca_algebraic_2022}. We use as reference the antibonding mode loss $\mathcal{D}[a_A]$, costing $|t_A|=0.63$ per unit rate at both drives above the transition ($0.636$ at $\Ft=1.8$, $0.632$ at $\Ft=1.23$) and $0.51$ at $\Ft=0.3$, flat in $N$. The beam-asymmetric loss $a_B+\eta a_A$ inherits that scale with a quadratic prefactor, its excess over the symmetric point being exactly $\eta^{2}|t_A|$, because parity forbids the cross terms as we've seen before.

Balanced local loss, two independent site drains at equal rate (the $\eta=1$ member of the local lab family $\mathcal{D}[a_1]+\mathcal{D}[\eta a_2]$), costs $|t_B+t_A|\approx|t_A|$. The two site drains decompose exactly into bonding plus antibonding mode loss. Since the bonding part is the protected one, the antibonding mode element is all that is left as the cost of the error at large $N$.

Site-local loss $\mathcal{D}[a_1]$ costs about half as much, $\tfrac12|t_B+t_A|\approx\tfrac12|t_A|\approx0.3$. The basis rotation splits it evenly between the two beam channels, and since bonding part again is the protected one, $|t_B|\ll|t_A|$, only the dark half dominates the error. The decomposition behind both entries is
\begin{align}
\bigl(\mathcal{D}[a_1]+\mathcal{D}[\eta a_2]\bigr)\rho
&=\frac{1+\eta^2}{2}\bigl(\mathcal{D}[a_B]+\mathcal{D}[a_A]\bigr)\rho\notag\\
&\quad+\frac{1-\eta^2}{2}\,\hat{\mathcal{C}}\rho,
\end{align}
with $\hat{\mathcal{C}}$ the cross dissipator of Eq.~(A2) of the End Matter, whose first-order matrix element on the coherence is zero by the same parity selection rule. Independent site baths pay $(1+\eta^{2})/2$ of bonding plus antibonding mode loss, the $\eta=0$ special case as the site-local one and $\eta=1$ the balanced local one.

Beam-local dephasing sits at the same order-one scale ($\approx0.9$, $1.2$ and $4.2$ for $\mathcal{D}[n_A]$, $\mathcal{D}[n_B]$ and $\mathcal{D}[n_A-n_B]$). The jump operator of any dephasing channel is Hermitian, $L_d=L_d^{\dagger}$. For a Hermitian jump the dissipator collapses exactly to a double commutator, $\mathcal{D}[L_d]\rho=-\tfrac12[L_d,[L_d,\rho]]$, so the first-order cost is the mean squared difference of the coupling across the two logical sectors. Every curve therefore has a floor $\tfrac12(\Delta L_d)^{2}$ set by the sector contrast of the coupling, plus a variance part that falls away as the two sectors converge onto the shared macroscopic bonding state: every curve in Fig.~\ref{fig:sfig_channels} flattens onto its floor ($\tfrac12(1.01)^{2}=0.51$ for $n_A$ and $\tfrac12(1.08)^{2}=0.59$ for $n_A{+}n_B$ below the transition), and $\mathcal{D}[n_B]$ keeps falling there only because its floor, $\tfrac12(0.07)^{2}$, is invisibly small. Collective dephasing joins this class below the transition, flat at $0.58$ from $N=3$ to $N=200$, just above the bare $\tfrac12$ an undressed $\ketbra{0}{1}$ coherence pays under number dephasing, and in fact more than $\mathcal{D}[n_A]$ alone ($0.58$ against $0.51$), the $n_B$ noise simply adding. Without a macroscopic bonding state there is no counter-response, so the compensation switches on exactly at the transition [Fig.~\ref{fig:sfig_channels}(a) and (d)].

\begin{table*}[t]
\caption{Error-channel taxonomy for the oscillating qubit: which channels are passively corrected, which are merely bounded, and why. Rates are the first-order coherence decay $-\mathrm{Re}\,\delta\lambda_{eo}$ per unit channel rate at the parameters from the main text (Fig.~\ref{fig:sfig_channels}), and ``transition'' refers to $\tilde F_c\approx0.93$. The symmetry column states whether the channel preserves the strong parity symmetry (every jump commutes with $P$), only the weak one (the dissipator maps each parity block onto a parity block), or neither.}
\label{tab:channels}
\begin{ruledtabular}
\begin{tabular}{p{0.14\textwidth}p{0.155\textwidth}p{0.085\textwidth}p{0.1\textwidth}p{0.4\textwidth}}
Channel & Dissipator & Symmetry & Decay & Mechanism \\
\hline
\multicolumn{5}{l}{\emph{Protected: cost falls with $N$ (passively corrected)}}\\[0.1em]
Collective loss & $\mathcal{D}[a_B]$ & strong & falls with $N$ at every drive & the dark qubit mode never jumps, the attractor restores the displaced bonding state (guardian response) \\
Collective dephasing (above transition) & $\mathcal{D}[n_A{+}n_B]$ & strong & $\propto1/N$ at $\tilde F^{*}\!\approx\!1.23$, where it crosses zero, elsewhere the floor $\tfrac12\Delta n_{\rm tot}^{2}$ & photon-counting cancellation: quasiparticle photon content screened by the bonding-mode counter-response (Sec.~\ref{sec:qf}) \\
\hline
\multicolumn{5}{l}{\emph{Bounded: flat, $N$-independent floor (not corrected, set by asymmetry or rate)}}\\[0.1em]
Dark-mode loss & $\mathcal{D}[a_A]$ & weak & $0.63$ above transition, $0.505$ below & the sectors differ by one dark photon, so even the absence of a jump tells them apart \\
Beam-asymmetric loss & $\mathcal{D}[a_B{+}\eta a_A]$ & none & excess exactly $\eta^{2}\times0.63$ & parity forbids cross terms, quadratic in the asymmetry \\
Lab-symmetric loss & $\mathcal{D}[a_1]{+}\mathcal{D}[a_2]$ & weak & $0.63$ & decomposes exactly into collective plus dark-mode loss \\
Site-local loss & $\mathcal{D}[a_1]$ & none & $\approx0.3$ & half-rate dark-mode loss plus a parity-projected cross term \\
Beam-local dephasing & $\mathcal{D}[n_A]$, $\mathcal{D}[n_B]$, $\mathcal{D}[n_A{-}n_B]$ & strong & $0.9$, $1.2$, $4.2$ & floor set by each coupling's sector contrast, no bonding-mode cancellation \\
Collective dephasing (below transition) & $\mathcal{D}[n_A{+}n_B]$ & strong & $0.58$ & no macroscopic bonding state, hence no counter-response: the protection requires the bright phase \\
\hline
\multicolumn{5}{l}{\emph{Extensive: cost grows $\propto N$ (must be excluded architecturally)}}\\[0.1em]
Site-local dephasing & $\mathcal{D}[n_1]$ & none & $\propto N$ & the bonding mode is the bright arm of an interferometer amplifying which-site noise \\
Parity-breaking dephasing & $\mathcal{D}[n_1{-}n_2]$ & weak & $3.5N$ & hopping term couples the $O(N)$ bonding occupation to the dark photon and mixes the parity sectors \\
\end{tabular}
\end{ruledtabular}
\end{table*}

\emph{(iii) Extensive: the cost grows with $N$.} The macroscopic bonding occupation that protects the collective channels becomes an error amplifier as soon as the environment can tell the two sites apart. Written in the beam basis, the error-causing operator is the difference $V=n_1-n_2=a_B^{\dagger}a_A+a_A^{\dagger}a_B$: the hopping term, which moves a single photon between the bonding mode and the antibonding mode. Every hop borrows the bonding mode's macroscopic amplitude, so which-site noise is amplified by the same bonding state that erased the which-path record for collective noise. The parity selection rule shapes the response as before. Each term of $V$ flips the parity of whichever side it acts on, so the jump term leaves the coherence sector exactly and only the no-jump term survives,
\begin{equation}
\delta\lambda_{eo}=-\tfrac12\langle\!\langle l_{eo}|\{V^{2},\,\cdot\,\}|r_{eo}\rangle\!\rangle,
\label{eq:sm_paritybreak}
\end{equation}
and $V^{2}$ contains the photon number occupation of the bonding mode $\langle n_B\rangle$, which grows with $N$. Single-site dephasing $\mathcal{D}[n_1]=\mathcal{D}[n_2]$, which contains half a unit of $V$, costs $\approx0.9N$ per unit rate at $\Ft=1.8$, the difference channel four times that, and below the transition the growth survives but is roughly forty times weaker. (We note that even outside the bright phase the driven occupation is still extensive but small [Fig.~\ref{fig:sfig_channels}].) Thus under which-site dephasing the coherence survives only for $\gamma_d\ll\gamma/N$. An experimental device must be wired so that its dephasing is common mode. The same hopping also flips the bit. Its jump term, projected out of the coherence, converts one sector into the other at first order in the channel rate, again bonding-amplified.

In practice the protection therefore presupposes the engineered collective loss, the drain coupled to the bright mode $a_B$ alone. Generic unengineered local loss is bounded but never corrected, at $\approx0.3$ per unit rate for all $\eta\le0.1$ and essentially independent of $N$ [Fig.~\ref{fig:sfig_channels}(a--c), blue pentagons], so it removes the coherence within a few oscillation periods and residual dark-mode loss sets the lifetime budget.

The taxonomy leads to three points for a potential experiment. First, the loss channel must be collective, since all the protection we discussed relies on it. An experimentalist can build this with standard tools from optics by interfering the two physical outputs on a beam splitter, so that the two hybrid outputs are the bonding and antibonding combinations, and by coupling a controlled dissipative port to the bright output alone \cite{leghtas_confining_2015, lescanne_exponential_2020}. The residual coupling of the antibonding output would be the imbalance $\eta$ of the End Matter. Second, the dephasing must be common mode, since site-resolved dephasing is the one channel whose cost grows with $N$. The device must be built with one master oscillator and shared bias lines to achieve common mode dephasing, so that no noise source exists that can address one site alone. Third, once both channels are collective, the only error left is the residual antibonding-mode loss, whose flat floor of $0.63$ per unit rate sets the qubit lifetime [Fig.~\ref{fig:sfig_channels}]. This residual noise damps the odd sector about eightfold faster than the even, exactly the biased structure that cat-qubit architectures are built to exploit \cite{puri_bias-preserving_2020, chamberland_building_2022}.

Throughout this section we only investigated the real part of the first-order shift $\delta\lambda_{eo}$. We can justify only looking at the first order as we assume smaller errors compared to other experimental parameters. But the first order effect is a complex number, and its imaginary part is a frequency change. An error channel might not only damp the clock of the qubit, but also change its rate. Yet from the error-correction standpoint this change of rate does not matter. We saw in Sec.~\ref{sec:deph} that the dephased Liouvillian hosts a noiseless subsystem of its own at a slightly shifted frequency [Fig.~4(b) of the main text]. This is still a qubit, just oscillating at a different frequency. An error is continuous and that is exactly what a dissipator models, so the frequency shift amounts to one constant recalibration of the frequency, not a decoherence mechanism. The frequency change does matter in another domain, where it is indeed useful rather than harmful. Turning a channel on for a while changes the frequency for a while, and a temporary frequency change is an accumulated phase inducing a phase rotation of the qubit potentially useful for quantum information processing.

\section{Justification for the Choice of Coherence Eigenstates}\label{sec:coh}

The main text selects the eigenoperator whose eigenvalue has real part closest to zero. This selection rule is seemingly arbitrary to how we construct the qubit, other than minimizing the decay rate. In the thermodynamic limit more than one even-odd and odd-even eigenvalue approach the imaginary axis, and near the transition the tracked coherence is not even the closest one to the axis, which is why we select the eigenoperators by their overlaps with operators in close higher or lower drives rather than by eigenvalue as explained in Sec.~\ref{app:computational}. The choice of these two specific eigenoperators is necessary though, as we now show.

\begin{figure*}[!t]
    \centering
    \includegraphics[width=0.92\linewidth]{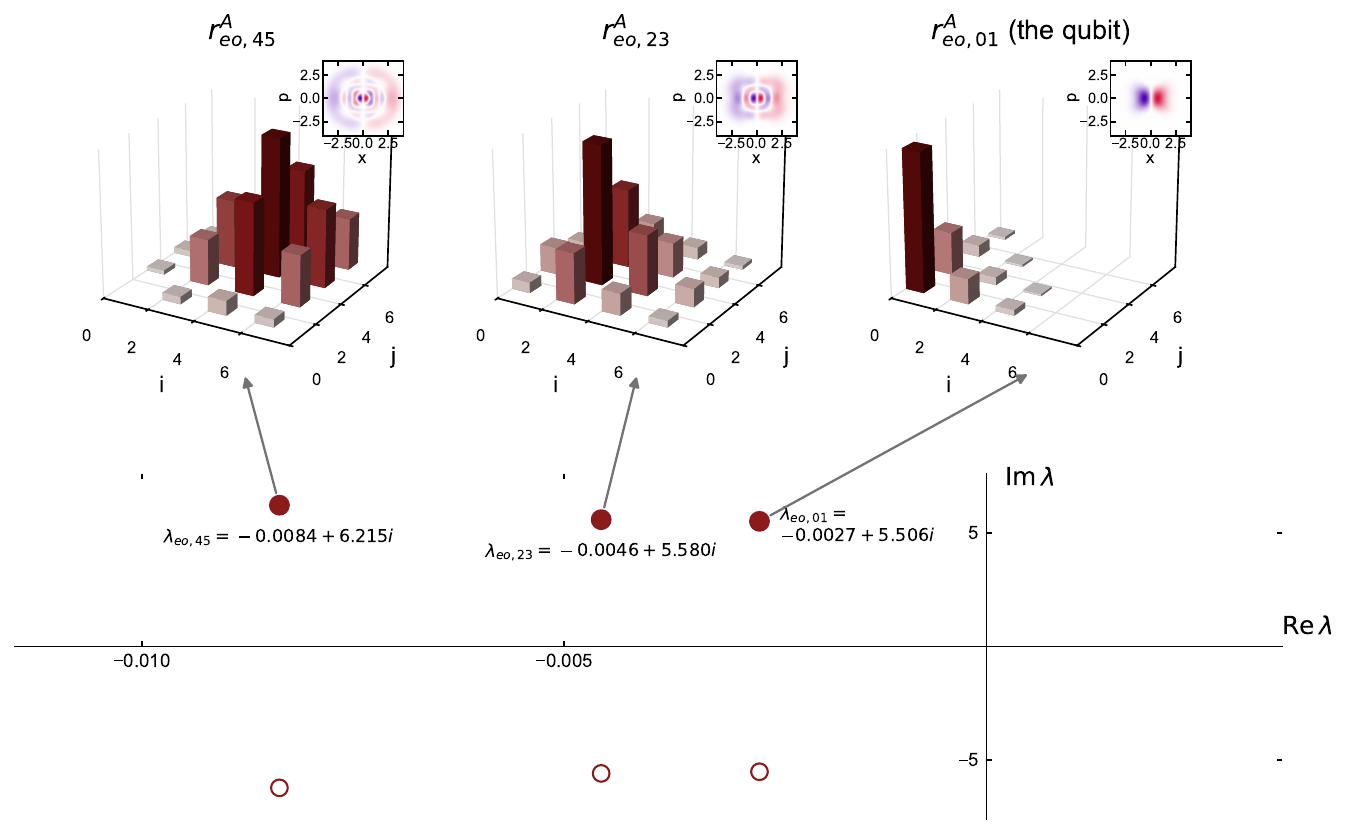}
    \caption{The slowest decaying even-odd eigenoperators at $N=100$, $\tilde F=1.8$, shown as the three even-odd eigenvalues in the complex plane (open circles: their odd-even conjugates) and the antibonding blocks of the corresponding eigenoperators, with the real part of the Wigner function as insets.}
    \label{fig:sfig_tower}
\end{figure*}

If several coherences reached the axis together, there might be many choices for us to construct the qubit. And in fact many slow even-odd decay rates close as $1/N$, our chosen pair with the slowest decay rate (for $\Ft=1.8$ the fastest dissipative gap closes at about $0.27/N$ with the next one about $0.46/N$ [Fig.~\ref{fig:sfig_tower}]). Thus what creates the qubit is the choice of the initial state. We initialize our state with the even-odd and odd-even eigenoperators and those will be preserved by the linearity of the Liouvillian. But again, why not initialize with a different set, or a mixture of many of these?

We showed with the Bogoliubov description of Sec.~\ref{sec:qf} that the antibonding mode is a quasiparticle oscillator decoupled from the loss at leading order, so that its damping comes only from the residual coupling to the lossy bonding mode and falls as $1/N$. Here we model it as a linearly damped oscillator with an effective rate $\kappa(N)=2|\mathrm{Re}\,\lambda_{eo}|$, for which the Liouvillian spectrum is known in closed form \cite{open_quantum, albert_lindbladians_2018},
\begin{equation}
\lambda_{mn}=-\tfrac{\kappa}{2}\,(m+n)-i\,(m-n)\,\omega_A
\label{eq:ladder}
\end{equation}
labelled by the coherence $\ketbra{m}{n}$ between quasiparticle Fock states. The parity grading places the members with even $m$ and odd $n$ in the even-odd sector, so the ladder of single-quantum coherences, $0\!\leftrightarrow\!1$, $2\!\leftrightarrow\!3$, $4\!\leftrightarrow\!5,\dots$, becomes slowly decaying as $N\to\infty$, at frequencies near $\omega_A$ [Fig.~\ref{fig:sfig_tower}]. Higher-orders such as $0\!\leftrightarrow\!3$, which Eq.~\eqref{eq:ladder} would place at $3\kappa/2$ and $3\omega_A$ that decay much faster than the eigenoperators reported [Table~\ref{tab:eo-tower-fates}].

The single rate $\kappa$ of Eq.~\eqref{eq:ladder} is itself an idealization as each member inherits its own effective damping from the bonding mode at the frequency it oscillates at, so $\kappa$ differs from state to state. The dominant element of $r^{A}_{eo,23}$ sits at $(2,3)$ with half the block weight and that of $r^{A}_{eo,45}$ at $(4,5)$. Pairings such as $0\!\leftrightarrow\!2$ are excluded because the even-odd block holds only even-to-odd coherences. The coherences we used to construct the qubit lie near $\omega_A$ as $|m-n|=1$ requires while a $0\!\leftrightarrow\!3$ member would oscillate near $3\omega_A$. Every other even-odd mode lies beyond $\mathrm{Re}\,\lambda=-1$. We choose the $0\!\leftrightarrow\!1$ eigenoperators because of two properties.

\begin{table}[!t]
    \centering
    \small
    \begin{tabular}{@{}p{4.3cm}p{2.1cm}p{2.6cm}p{4.4cm}@{}}
    \hline
    coherence $\ketbra{m}{n}$ & frequency & dissipative gap & mechanism \\
    \hline
    $|m-n|$ even (e.g.\ $0\!\leftrightarrow\!2$) & -- & excluded & the block holds only even-to-odd coherences \\
    $|m-n|=1$ (e.g.\ $0\!\leftrightarrow\!1$, $2\!\leftrightarrow\!3$) & $\approx\omega_A$ & closes as $1/N$ & dynamical-symmetry protection, no fast modes nearby \\
    $|m-n|\geq3$ (e.g.\ $0\!\leftrightarrow\!3$) & $\geq3\omega_A$ & stays at order $\gamma$ & inherit order-$\gamma$ decay from the bonding mode\\
    \hline
    \end{tabular}
    \caption{The quasiparticle coherences in the even-odd block. Only the single-quantum case becomes slow with $N$.}
    \label{tab:eo-tower-fates}
\end{table}

First, the specific coherence we need to choose is decided by our steady states $r_{ee}$ and $r_{oo}$. As a reminder, our qubit is two stationary populations together with the coherences connecting them, as shown in Eq.~(8) of the main text. The coherence connecting our two steady states is the $0\!\leftrightarrow\!1$ coherence, and this is also why it decays the slowest. In Eq.~\eqref{eq:ladder} we can see that its rate $\kappa/2$ lies strictly below others. A candidate qubit built on, say, the $2\!\leftrightarrow\!3$ coherence would need the excited population $\ketbra{2}{2}$ and $\ketbra{3}{3}$ as its poles, and that is not stationary as it relaxes into the sector steady state. So even if such a qubit existed it would decay through its even and odd diagonal eigenoperators, where its coherence operators might not.

Second, the rest of the tower cannot mix with the off-diagonal eigenoperators we have chosen. Under the unperturbed dynamics each slow mode evolves independently (and linearly), carrying its own coefficient. The eigenmodes are biorthogonal, $\langle\!\langle l_j|r_k\rangle\!\rangle=\delta_{jk}$, so the qubit coefficient evolves as $c(t)=c(0)\,e^{\lambda_{eo}t}$.

\section{Computational Methods and Cutoff Convergence}\label{app:computational}

\paragraph{Exact diagonalization.---}\!\!\!\!\!\!\! In bosonic systems the Hilbert space should be of infinite size. However to be able to perform any numerical analysis, we need to work with finite matrices. As such we truncate these systems in the Fock basis. If we keep the $K$ Fock states $\ket{0},\dots,\ket{K-1}$ in each mode, then our coupled state of two modes is a column vector with $K^2$ elements, its corresponding density matrix is $K^2$ by $K^2$, and in the doubled Hilbert space is a column vector of size $K^4$. Thus the Liouvillian we are trying to diagonalize is a $K^4$ by $K^4$ matrix. As one can see, the computational requirements increase dramatically with the Fock truncation limit, severely limiting one in studying large systems. For the quantum problem, earlier studies of this dimer stopped near effective sizes of $N\approx5$ \cite{lledo_driven_2019, cristobal}, and the pipeline described in this section takes the same system to $N=100$.

The most direct way to obtain the spectrum is to build $\hat{\mathcal{L}}$ explicitly as a matrix in this space and diagonalize it. A better way is to split it into its four symmetry blocks (each parity sector evolves independently, as derived in the main text), and hand each block to a dense eigensolver, which returns every eigenvalue and eigenoperator of the block at once. The computational cost grows as the cube of the block dimension, and in practice dense diagonalization stops at block dimensions of a few times $10^4$, which corresponds to $N\lesssim4$.

Thus we performed computations for $N\leq4$ with exact diagonalization of the parity-resolved Liouvillian blocks, where all eigenstates were found (in Figs.~1 and 3 of the main text we use dense $\tilde F$-sweeps with the full spectrum retained at every point.)

\paragraph{Sparse diagonalization at large $N$.---}\!\!\!\!\!\!\! Past $N\approx4$ working directly with dense solvers is not feasible. Since we need only a few eigenvalues, use sparse iterative solvers to reach large $N$. For $N=10$ to $100$ we used sparse parity-block diagonalization with shift-invert ARPACK methods \cite{johansson2012qutip, lehoucq1998arpack}, computing the leading eigenpairs of each block together with the corresponding left eigenoperators. The largest calculations ($N=100$: each parity subspace holds $(K_A/2)\times K_B=4\times400=1600$ states, so each Liouvillian block has dimension $1600^{2}=2.56\times10^{6}$) are dominated by the linear solves of the shift-invert step. Each Krylov iteration solves $(\hat{\mathcal L}-\sigma)x=b$ at the shift $\sigma$, which is done by factoring the sparse matrix once into a product of a lower and an upper triangular matrix (the $LU$ decomposition, $L$ for the lower and $U$ for the upper factor, Gaussian elimination stored in reusable form), so that every subsequent solve reduces to two triangular substitutions. The factors fill in to roughly $10^{11}$ nonzero entries, approximately $1.5$~TB of memory and more entries than a 32-bit integer can index, so the solver must be built with 64-bit indexing. We note that time-domain results (the evolutions) require storing and propagating the full two-mode density matrix rather than a few eigenoperators, which limits them to $N=20$.

A more efficient pipeline is to never build the Liouvillian. In this case, the Hamiltonian and the jump operator are restricted to the even and odd sectors first, and each parity block of $\hat{\mathcal{L}}$ is assembled directly in sparse form from Kronecker products of these small sub-blocks. The terabyte-scale construction disappears, and memory is set by the eigensolver instead. For each operator we also knew where to look as the diagonal blocks for their smallest-magnitude eigenvalues (the steady state and the slowest relaxations), the coherence block for its largest-real-part eigenvalues. The Krylov dimension (ARPACK's NCV) is tuned per block rather than left at defaults, and every returned eigenvector is scaled to unit norm with its first nonzero component rotated to be real and positive, a fixed convention that lets runs at neighbouring parameters be identified with one another and feeds the branch tracking below. For the left eigenoperators, instead of a second run on $\hat{\mathcal{L}}^\dagger$, the left basis can be constructed from the right eigenoperators through the inverse of their Gram matrix, the biorthogonal dual. With this pipeline the $N=100$ blocks, of dimension $2.56\times10^6$, fit on a $3$~TB node. Convergence was verified by increasing both Fock cutoffs and the Krylov dimension until the relevant eigenvalues and observables were stable, and by checking the continuity of eigenoperators with respect to small changes of $\tilde F$.

\paragraph{Basis choice and Fock cutoffs.---}\!\!\!\!\!\!\! We use an antibonding cutoff $K_A=8$ for all data unless noted otherwise, after the convergence analysis summarized below showed that $K_A=6$, although sufficient for eigenvalues and for all quantities at $N\lesssim20$, artificially limits the smallest Hilbert-Schmidt distances at large $N$. We justify this cutoff by direct convergence tests, collected in Table~\ref{tab:cutoffs}. The candidate cutoffs are small because the antibonding mode stays microscopic. The Fock representations of the antibonding reductions [Fig.~2(b--d) of the main text] have no visible weight beyond $n_A\approx4$ even at $N=100$, so $K_A=6$ or $8$ already contains all the occupied states. The tests cover all four parity blocks, $r_{ee}$, $r_{oo}$ and $r_{eo}$ with $r_{oe}=r_{eo}^{\dagger}$, together with the coherence eigenvalue $\lambda_{eo}$. For the representative system $\tilde F=1.8$ (the drive used throughout the main text), at $N=10$ we compare $K_A=6$ against $8$ and $K_B=50$ against $60$ with the same NS-basis diagnostics as in the main analysis. Every diagnostic is stable, and $K_A=6$ stays sufficient for eigenvalues and for every quantity we report at $N\lesssim20$. The exception appears at larger $N$ in the smallest Hilbert-Schmidt distances, for which the sector distance sits on a $K_A=6$ floor near $1.2\times10^{-5}$ from $N=20$ on, an order of magnitude above its converged value, so we need a larger antibonding truncation and switch to $K_A=8$ generally. The $N=40$ rows of the table show $K_A=8$ converges (switching to $K_A=10$ is a $2\times10^{-7}$ effect).

Within a fixed sector only every other dark level is populated, so every sector block is assembled and stored on a half-size dark ladder of $K_A/2$ levels, with the decoding rule that stored label $k$ means level $2k$ in an even sector and $2k+1$ in an odd one, the picture is drawn in Fig.~\ref{fig:dark-sublattice} of Sec.~\ref{sec:dressed}.

\begin{table}[h]
\centering
\footnotesize
\setlength{\tabcolsep}{3pt}
\begin{tabular}{ccccccccccc}
\hline
$N$ & $K_A$ & $K_B$ & $\langle n_B\rangle_{r_{ee}}$ & $\langle n_B\rangle_{r_{oo}}$ & $\langle n_B\rangle_{r_{eo}}$ & $D_{\mathrm{HS}}(r_{ee},r_{oo})$ & $D_{\mathrm{HS}}(r_{ee},r_{eo})$ & $D_{\mathrm{HS}}(r_{eo},r_{oe})$ & $\mathrm{Re}\,\lambda_{eo}$ & $\mathrm{Im}\,\lambda_{eo}$ \\
\hline
10 & 6 & 50 & 29.9156 & 28.6180 & 29.4351 & $3.2115\times10^{-5}$ & $2.6828\times10^{-3}$ & $8.1411\times10^{-3}$ & $-0.02860$ & $5.4683$ \\
10 & 6 & 60 & 29.9324 & 28.6218 & 29.4364 & $3.4095\times10^{-5}$ & $2.5887\times10^{-3}$ & $8.1269\times10^{-3}$ & $-0.02847$ & $5.4685$ \\
10 & 8 & 50 & 29.9478 & 28.6718 & 29.4336 & $3.5775\times10^{-5}$ & $2.5122\times10^{-3}$ & $8.1425\times10^{-3}$ & $-0.02842$ & $5.4586$ \\
10 & 8 & 60 & 29.9635 & 28.6749 & 29.4348 & $3.1780\times10^{-5}$ & $2.4329\times10^{-3}$ & $8.1282\times10^{-3}$ & $-0.02829$ & $5.4588$ \\
\hline
40 & 6 & 200 & 119.4177 & 118.0833 & 118.9214 & $1.163\times10^{-5}$ & $9.753\times10^{-4}$ & $1.960\times10^{-3}$ & $-0.00683$ & $5.5099$ \\
40 & 8 & 200 & 119.4520 & 118.1486 & 118.9240 & $2.136\times10^{-6}$ & $8.399\times10^{-4}$ & $1.945\times10^{-3}$ & $-0.00678$ & $5.4983$ \\
40 & 10 & 200 & 119.4608 & 118.1651 & 118.9198 & $1.895\times10^{-6}$ & $8.058\times10^{-4}$ & $1.985\times10^{-3}$ & $-0.00676$ & $5.4971$ \\
\hline
\end{tabular}
\caption{Cutoff convergence at $\tilde F=1.8$: sector occupations, the three block distances, and the coherence eigenvalue. At $N=10$ every diagnostic is stable across all four $(K_A,K_B)$ settings. At $N=40$, with $K_B=200$ fixed by the five-standard-deviation rule, the sector distance requires $K_A=8$: at $K_A=6$ it sits a factor of five too high, while the step from $K_A=8$ to $10$ moves it by only $2\times10^{-7}$. The eigenvalues are insensitive to the cutoff throughout.}
\label{tab:cutoffs}
\end{table}

The bonding-mode cutoff must grow with the bonding photon number. We use $K_B\gtrsim\langle n_B\rangle+5\sqrt{\langle n_B\rangle}$, a five-standard-deviation margin over the Poissonian spread of the near-coherent bonding state. At $\tilde F=1.8$, where $\langle n_B\rangle\approx|\alpha_B|^{2}N\approx3N$, this reads $K_B\gtrsim3N+5\sqrt{3N}$, giving $K_B=60,100,200,240,400$ for $N=10,20,40,50,100$. At the smaller bonding occupation of $\tilde F=1.23$, $K_B=350$ suffices at $N=100$. The margin can be checked directly on the stored $N=100$ state as the even-sector bonding-mode distribution at $\tilde F=1.8$ has mean $298$ and carries less than $10^{-10}$ of its weight in the top ten Fock levels of the $K_B=400$ window. The dephased runs of Fig.~4(b) of the main text use these same cutoffs at every $N$. At $\gamma_d=0.01\gamma$ the dephased state is $\approx10\%$ broader along the Fock diagonal [Fig.~\ref{fig:sfig3}(b)], which raises the requirement at $N=40$ from $K_B\ge187$ to $K_B\ge195$, still inside the $K_B=200$ used but makes larger $N$ computations infeasible.

\paragraph{Tracking eigenvalue branches.---}\!\!\!\!\!\!\! Near the transition, simply selecting the eigenvalue with real part closest to zero is not enough to identify the same eigenoperator branch across all \(\tilde F\). Near the level crossing, the branch that is slowest-decaying on one side becomes subleading on the other side. We therefore track branches by eigenoperator overlap with \emph{two anchors}. The low-drive branch is anchored deep in the low-drive phase and walked upward in $\tilde F$, while the high-drive branch is anchored deep in the bright phase and walked downward. At each step we select the candidate eigenoperator with the largest normalized Hilbert-Schmidt overlap with the previous point on the same branch. For $N=4$ the full exact spectrum is available at every point, so both branches are tracked throughout, including after the crossing region. For $N=10,20,40$ the sparse solver retains the several modes with smallest dissipative gaps of each block at every grid point, which is sufficient for the same two-anchor walk up to the transition. The resulting dense lines are shown in Fig.~1 of the main text and Fig.~\ref{fig:sweetspot}. Drive sweeps use a base grid $\Delta\tilde F=0.02$ over $\tilde F=0.1$ to $2.0$, with additional points including the operating drive $\tilde F^{*}=1.23$, and a coarser $\Delta\tilde F=0.05$ grid over $0.2$ to $2.0$ at $N=50$.

\bibliography{citations}